\documentclass[a4paper,12pt,onecolumn]{article}
\usepackage[english]{babel}
\usepackage{fancyhdr} % um Header und Footer zu designen
\usepackage[pdftex]{graphicx} %notwendig, wenn Grafiken geladen werden sollen
\usepackage{url} % zum korrekten Einf\"ugen von URL mit dem \url Befehl, das Package hyperref erzeugt dazu noch Links
\usepackage{caption} % um Bilder neben Bildern oder Tabellen neben Tabellen zu erzeugen
\usepackage{subcaption} % um Bilder neben Bildern oder Tabellen neben Tabellen zu erzeugen
 \usepackage{natbib} %erlaubt es, andere Zitatdarstellungen vorzunehmen als numerische
\usepackage{multirow} %erlaubt es, in einer Tabelle mehrerer Zeilen zu einer zusammenzufassen
\usepackage{float} % notwendig, um Abbildungen an der Stelle zu platzieren, an der sie im Quelltext stehen
\usepackage{amssymb}
\usepackage[intlimits]{amsmath} % beides sind Mathematikpakete
\usepackage{amsthm} % Um Theoreme, Beweise etc. zu verwalten
\usepackage[german]{nomencl}
\usepackage{geometry}
\usepackage{enumitem}
\usepackage{bbm}
\usepackage{setspace}
\usepackage{array}
\makenomenclature
\usepackage[pdfborder={000},colorlinks=true,linkcolor=blue,citecolor=blue]{hyperref}
\usepackage{listings}
\usepackage[table]{xcolor}
\usepackage{fancybox}

\usepackage{blindtext}
\usepackage{tikz}
\usetikzlibrary{shapes,arrows,shadows, positioning}
\usepackage{bm,times}
\usepackage{smartdiagram}
\usepackage{attachfile}

\definecolor{b}{rgb}{0,0,.8}	%%omega-blau
\definecolor{g}{rgb}{0,.6,0}	%%Tau-grün
\definecolor{n}{rgb}{0,0,0}	%%normal-schwarz
\definecolor{h}{rgb}{0.4,0.2,0.2}	%%hint
\definecolor{v}{rgb}{0.2,0.6,0}

\newtheorem{lemma}{Lemma}

\newtheorem{theorem}{Theorem}

\newcommand{\A}{{\mathbb A}}
\newcommand{\B}{{\mathbb B}}
\newcommand{\C}{{\mathbb C}}

\newcommand{\E}{{\mathbb E}}

\newcommand{\R}{{\mathbb R}}

\newcommand{\Z}{{\mathbb Z}}

\renewcommand{\AA}{{\mathcal{A}}}

\newcommand{\MM}{{\mathcal{M}}}
\newcommand{\NN}{{\mathcal{N}}}

\newcommand{\UU}{{\mathcal{U}}}

\newcommand{\XX}{{\mathcal{X}}}

\newcommand{\bsa}{\boldsymbol a}

\newcommand{\bse}{\boldsymbol e}
\newcommand{\bsf}{\boldsymbol f}

\newcommand{\bsu}{\boldsymbol u}

\newcommand{\bsx}{\boldsymbol x}
\newcommand{\bsy}{\boldsymbol y}
\newcommand{\bsz}{\boldsymbol z}

\newcommand{\bsC}{\boldsymbol C}

\newcommand{\bsF}{\boldsymbol F}

\newcommand{\bsI}{\boldsymbol I}

\newcommand{\bsM}{\boldsymbol M}

\newcommand{\bsR}{\boldsymbol R}
\newcommand{\bsS}{\boldsymbol S}

\newcommand{\bsU}{\boldsymbol U}

\newcommand{\bsW}{\boldsymbol W}
\newcommand{\bsX}{\boldsymbol X}
\newcommand{\bsY}{\boldsymbol Y}
\newcommand{\bsZ}{\boldsymbol Z}
\newcommand{\bsone}{\boldsymbol 1}

\newcommand{\bsnull}{\boldsymbol 0}

\newcommand{\bsmu}{\boldsymbol \mu}

\newcommand{\bsvarphi}{\boldsymbol \varphi}

\newcommand{\bsSigma}{\boldsymbol \Sigma}
\newcommand{\bsPi}{\boldsymbol \Pi}

\newcommand{\eps}{{\varepsilon}}

\DeclareMathOperator{\cov}{\C ov}

\newcommand{\ov}\overline
\newcommand{\what}{\widehat}
\newcommand{\wtilde}{\widetilde}

\newcommand{\rig}\right
\newcommand{\lef}\left
\newcommand{\nf}\normalfont

\definecolor{dcyan}{rgb}{0,0.5,.5}

\definecolor{dgreen}{rgb}{0,0.7,0}
 
\definecolor{dgrey}{rgb}{0.6,0.6,.6}

\newcommand{\MS}{\text{MS}} 
\newcommand{\CS}{\text{CS}} 
\newcommand{\MCS}{\text{MCS}} 
\newcommand{\ES}{\text{ES}} 
\newcommand{\VS}{\text{VS}} 
\newcommand{\SC}{\text{SC}} 
 
\newcommand{\RelCh}{\text{RelCh}} 
 
\newcommand{\DSS}{\text{DSS}} 
\newcommand{\CRPS}{\text{CRPS}} 
\newcommand{\bsCRPS}{\text{\textbf{CRPS}}} 
\newcommand{\CES}{\text{CES}} 
\newcommand{\CVS}{\text{CVS}} 
\newcommand{\CDSS}{\text{CDSS}}

 \title{%(Strictly) Proper (multivariate?) Scoring Rules/Forecasting Evaluation Using the Marginal-Copula-Scores
Multivariate Forecasting Evaluation:
On Sensitive and Strictly Proper Scoring Rules
}  
 \author{Florian Ziel, {\normalsize  University of Duisburg-Essen, Germany, \texttt{florian.ziel@uni-due.de}  }
 \and Kevin Berk,  {\normalsize University of Siegen,  \texttt{berk@mathematik.uni-siegen.de} } }

\begin{document}
\maketitle
\lhead{\nouppercase{\leftmark}}
\begin{abstract}
In recent years, probabilistic forecasting is an emerging topic, which is why there is a growing need of suitable methods for the evaluation of multivariate predictions. We analyze the sensitivity of the most common scoring rules, especially regarding quality of the forecasted dependency structures. Additionally, we propose scoring rules based on the copula, which uniquely describes the dependency structure for every probability distribution with continuous marginal distributions. Efficient estimation of the considered scoring rules and evaluation methods such as the Diebold-Mariano test are discussed. In detailed simulation studies, we compare the performance of the renowned scoring rules and the ones we propose. Besides extended synthetic studies based on recently published results we also consider a real data example. We find that the energy score, which is probably the most widely used multivariate scoring rule, performs comparably well in detecting forecast errors, also regarding dependencies. This contradicts other studies. The results also show that a proposed copula score provides very strong distinction between models with correct and incorrect dependency structure. We close with a comprehensive discussion on the proposed methodology.
\end{abstract}


% $\mbox{MAE} \equiv \frac{1}{\cal DH} \sum_{d=1}^{\cal D} \sum_{h=1}^{\cal H} |e_{d,h}|$ ?
% then $N= {\cal DH}$

% \begin{enumerate}
%  \item Introduction
%    [Diskussion \"uber Vorteile erg\"anzen]
%  \item Scoring rules
%  \begin{enumerate}
%   \item Notations and Basics
%   \item CRPS
%   \item Energy score
%   \item Variogram score
%   \item Dawid-Sebastiani score
%   \item Marginal-copula score
%   \begin{enumerate}
%   \item Marginal score
%   \item Copula score
%   \item Combining marginal and copula score
%   \item Copula energy score
%   \item Copula variogram score
%   \item Copula DSS
%   \end{enumerate}
%   \end{enumerate}
% \item Reporting Forecasts / study design
% \item Estimating the score
% \begin{itemize}
%  \item energy score
%  \item variagram score
%  \item DS score
%  \item Copula observations
%  \item Marginal-Copula Score [problem mit covariance]
%  \end{itemize}
% \item Evaluating the score
% 
% relative Change in score/ sensitive: Tatsu-Pinson study
% 
% DM-test
% 
% \item Simulation studies
% 
% \begin{enumerate}
%  \item  Tatsu-Pinson
%  \item  Bivariate Normal (angelehnt an variogram)
%  \item 3-d-Fall, peak-study. (motivated from demand forecasting)
%  \item role M
%  \item Study with increasing dimension ?
% \end{enumerate}
% 
% \item Real data example(?)
% 
% \item Conclusion
% 
%  
% \end{enumerate}

% $\mbox{MAE}_h = \frac{1}{\cal D} \sum_{d=1}^{\cal D} |e_{d,h}|$ 
% and 
% $\mbox{MAE} = \frac{1}{\cal DH} \sum_{d=1}^{\cal D} \sum_{h=1}^{\cal H} |e_{d,h}| $ 

\section{Introduction} \label{Introduction}

Forecasting evaluation is still highly discussed in different forecasting communities, like in meteorology, energy, economics, business or social and natural sciences. Also, in recent years, the term \emph{probabilistic forecasting} (essentially density forecasting or quantile forecasting) became popular across all areas of application (\cite{gneiting2014probabilistic}, \cite{reich2015probabilistic}, \cite{hong2016probabilistic}, \cite{lerch2017forecaster}). It is clear that the evaluation of forecasts is crucial for assessing the quality of different forecasts.
However, for 1-step ahead forecasts the discussion is much more comprehensive than for multiple step ahead predictions, say $H$-step ahead forecasts. 

%FZ I think is to too energy community specific
% In order to create a common platform for all dedicated forecasters, on which they are able to compare models from very different fields and techniques, the Global Energy Forecasting Competition was initiated in 2012. While the initial competition started with the evaluation of point forecasts, the \cite{gefcom} was the first competition to focus on models for probabilistic forecasting (one month ahead, hourly electricity data) and their evaluation. The scoring rule used in this case was the so-called quantile score. The major drawback of this method is that it only evaluates the forecasts with respect to the marginal distributions. 

% Univariate scoring rules and t
The evaluation of forecasted marginal distributions of $H$-step ahead forecasts have been treated in literature using so called \emph{univariate scoring rules}. Essentially it breaks down to the 1-step ahead forecasting case for each marginal distribution. In contrast, the (fully) multivariate case where the complete $H$-dimensional distribution is usually not covered, in the forecasting itself such as in the evaluation. Obviously, this causes problems because the multivariate $H$-step ahead distribution is not fully explained by reporting $H$-step ahead marginal distributions or quantiles. 
A main reason for these circumstances is the lack of suitable evaluation measures that can judge the quality of forecasts adequately.
% For instance, in energy load forecasting, one knows that the daily volume does not exceed a certain threshold for a given day. The probabilistic forecast of hourly load values can be very accurate in marginal distributions, but for a particular simulation of the model, the daily volume could be much higher than whats acceptable since the intra-day correlation is not scored at all. 
We require approaches to assess not only the performance of a prediction with respect to the marginals but also to the dependency structures, we need suitable \emph{multivariate scoring rules}.

The most well known multivariate scoring rule, the energy score, was introduced by \cite{gneiting2007strictly}. For instance, in \cite{sari2016statistical, spath2015time, gneiting2008assessing, pinson2012adaptive, berrocal2008probabilistic, pinson2012evaluating, moller2013multivariate, baran2015joint, yang2015multi, moller2015spatially} the energy score is used for evaluation. This is mainly in the area of meteorologic forecasting, primarily wind speed and wind power forecasting. In the wind power forecasting review of \cite{zhang2014review} it is the only mentioned multivariate forecasting evaluation method. 
%In \cite{moller2015spatially}, the DM-test is applied to multivariate forecasts using the energy score framework. 

The energy score is a strictly proper scoring, meaning that only the true model optimizes the corresponding score. 
However, in many empirical applications the energy score is still avoided. In the study of \cite{pinson2013discrimination}, the discrimination ability of the energy score is checked in a simulation study while considering the bivariate normal distribution. They conclude that the energy score can not separate differences in the dependency structure well, which made the researchers and practitioners skeptical in using it. In this paper, we will discuss this study in detail and draw somewhat different conclusions.

In contrast, another multivariate scoring rule were introduced to overcome the reported problem in energy score.
\cite{scheuerer2015variogram} developed the variogram score, that is sensitive to changes in the correlation. However, it is only a proper scoring rule, but not a strictly proper one. Thus, it can not identify the true underlying model.
% This is due to the fact that it only measures pairwise distances of the marginals. 
% Hence, two forecasts with a different mean, but all other characteristics identical will get the same score, which is impractical for proper forecasting evaluation.
Another reported plausible candidate is the multivariate log-score. However, it requires that we have multivariate density forecasts which is not the case in many applications. 
As discussed in \cite{lerch2017forecaster}, this density may be approximated. But these approximation methods suffer efficiency in higher dimensions. The results depend crucially on the chosen approximation method.
Additionally, we have the Dawid-Sebastiani score (see e.g. \cite{gneiting2007strictly}), which evaluates the mean and covariance matrix of the distributions
and corresponds to the log-score in the multivariate Gaussian settings. Moreover, a characterization only by the first two moments is not sufficient for many applications.
%this corresponds to the log-score. However, in general a characterization only by the first two moments does not suit many real applications.

With all these considerations in mind, we introduce a new scoring rule that is sensitive to dependency changes in the distribution using copula theory. By Sklar's theorem, we are able to extract the dependency structure of our forecast (in the form of the copula) and construct a more sensitive measure for evaluating the dependencies. Afterwards, we combine the score of the copula with a score of the marginal distributions to obtain a proper and even strictly proper scoring rules. The approach is somehow flexible with respect to the copula and marginal scores that are used.
% Hence, we conjecture the statement of \cite{scheuerer2015variogram} that \textit{'($\ldots$) it is unlikely, however, that there exists a single multivariate score that serves all purposes.'} to some extent.

We start with the introduction of the aforementioned famous scoring rules in section~\ref{sec:scores}. Afterwards, we introduce our concept of marginal-copula scores. In section~\ref{sec:reporting} we give a short overview on how to report multivariate forecasts. Efficient estimation of scores is crucial in practice, we describe some approaches for different scoring rules in section~\ref{sec:estim}. Since the sensitivity of a score value alone does not indicate very much about the goodness-of-fit or the discrimination ability, we introduce the concept of the Diebold-Mariano test in section~\ref{sec:eval}. In section~\ref{sec:studies}, we apply the renowned scoring rules and the marginal-copula scores in three different synthetic case studies, compare and discuss the results. A real data example using airline passenger data is discussed in section~\ref{sec:realdata}. We conclude the results in section~\ref{sec:conclusion}.% and summarize some guidelines for forecasters.

\section{Scoring rules} \label{sec:scores}

In this section we are going to introduce a concept of scoring rules to adequately assess the goodness-of-fit of multivariate probabilistic forecasts. In order to do that, we are going to introduce some well known scoring rules in a first step. Additionally, we present a new score which combines univariate scoring of the forecasted marginal distributions as well as multivariate scoring for the forecasted copula in a way that desirable preferences like (strict) propriety are preserved. We start with some notations and basics on scoring rules.

\subsection{Notations and basics}

Let $\bsY = (Y_1,\ldots, Y_H)'$ be an $H$-dimensional random variable with multivariate (cumulative) distribution function $\bsF_{\bsY}$ which we want to forecast. The most standard example would be 
that $\bsY$ is a $H$-step ahead forecast of a univariate time series, hence $H$ is the forecasting horizon. From $\bsY$ we observe the realized respectively materialized vector $\bsy = (y_1,\ldots, y_H)'$.
Let $\bsX = (X_1,\ldots, X_H)$ be the forecast random vector for $\bsY$. In practice we have to think about 
the reporting of these forecasts (see Section \ref{sec:reporting}). 
Here, we simply assume that we have the full multivariate (cumulative) distribution function $\bsF_{\bsX}$ as reported forecast available.

Given a forecast $\bsF_{\bsX}$ and a random variable $\bsY$ we can define a scoring rule $\text{S}(\bsF_{\bsX}, \bsY)$ that maps into $\R$ (or $\ov{\R}$).
They are designed in such a way that a good forecast yields small values (positively orientated), that is if $\bsF_{\bsY}$ is close (or closely related) to $\bsF_{\bsX}$.
Such a scoring rule is \emph{proper} if $\text{S}(\bsF_{\bsY}, \bsY) \leq \text{S}(\bsF_{\bsX}, \bsY)  $ holds for arbitrary random vectors $\bsX$, and 
\emph{strictly proper} if equality holds only if $\bsF_{\bsX} = \bsF_{\bsY}$. Hence the true distribution can be well separated.

In forecasting applications the random variable $\bsY$ is usually observed (or realized / materialized), therefore we often write directly $\text{S}(\bsF_{\bsX}, \bsy)$. 
Note in some literature, see e.g. \cite{gneiting2007strictly}, scoring rules are defined with inverse orientation. Obviously, all the theory holds as well. Here, we use the notation that is popular in applications,
that means the smaller the score the better the forecast.

As pointed out in the introduction, at the moment there is basically only a limited amount of scoring rules available. Here we recall the the major important ones, as
we will work with them in the latter part of this paper. We start with the most popular univariate scoring rule.

\subsection{Continuous ranked probability score}

The continuous ranked probability score (CRPS) for a random variable $X$ with distribution function $F_{X}$ and given forecast $y$ is defined by
\begin{align}
 \CRPS(F_{X}, y) = \int_{-\infty}^{\infty} (F_{X}(z) - \mathbbm{1}\{y < z\})^2 \,d z  =  \E | X - y | - \frac{1}{2} \E| X -\wtilde{X}|  
\label{eq_crps}
 \end{align}
where $\wtilde{X}$ is an iid copy of $X$. 
It is a special case of the energy score, which we will introduce in the next section, for one dimension and $\beta=1$. 
It is a strictly proper scoring rule with respect to the distribution of $Y$.
Another univariate score is for example the quantile loss or pinball score on a grid of quantiles ($0\leq\alpha_1<\alpha_2<\ldots <\alpha_N\leq 1$). Note that for an equidistant grid for $|\alpha_i - \alpha_{i+1}| \to 0$ the corresponding score converges to the CRPS, see e.g. \cite{nowotarski2017recent}.

If $Y$ respectively $X$ has a density, the logarithmic score is also a candidate for univariate scoring. However, in practical applications 
we have the problem that for sophisticated models we do not have an explicit formula for the forecasted density available.
Still, it may be approximated \cite{lerch2017forecaster}.
% , which makes it less favorable.

For more details on univariate scoring rules we recommend \cite{gneiting2007strictly}. 
In the subsequent sections, we will introduce some multivariate scores.

\subsection{Energy score}

For the $H$-dimensional random variable $\bsX$ and 
the observation vector $\bsy$ of the target distribution $\bsY$ the (Euclidean) \textbf{energy score} (see \cite{gneiting2007strictly}) is given by
\begin{equation}
\ES_\beta ( \bsF_{\bsX}, \bsy) = 
% \underbrace{
 \E\left( \left\| \bsX - \bsy \right\|_2^\beta \right) - %}_{\text{ED}_\beta:=}  - 
  %\underbrace{ 
  \frac{1}{2} \E\left( \| \bsX- \wtilde{\bsX} \|_2^\beta \right) %}_{\text{EI}_\beta:=} 
% = \text{ED}_\beta - \text{EI}_\beta
 \label{eq_energy_score}
\end{equation}
where $\wtilde{\bsX}$ is an i.i.d. copy of $\bsX$, so it is drawn independently from the same distribution 
$\bsF_{\bsX}$ as $\bsX$. Moreover, $\beta \in (0,2)$ and $\|\cdot\|_2$ is the the Euclidean norm.
As described in \cite{gneiting2007strictly}, $\text{ES}_\beta ( \bsF_{\bsX}, \bsy)$ is a strictly proper evaluation score for all $\beta$, 
but $\beta=1$ seems to be the standard choice in application. \cite{szekely2013energy} points out that for heavy tailed data 
the choice of small $\beta$ values should be favored to guarantee that the corresponding moments exists.

\subsection{Variogram score}

The second multivariate score of interest is the \textbf{variogram score} (see \cite{scheuerer2015variogram}) which is motivated by the variogram that is a popular tool in geostatistics, see \cite{cressie1985fitting}.
The variogram score is defined by 
\begin{equation}
 \VS_{\bsW,p}( \bsF_{\bsX} , \bsy )
= \sum_{i=1}^H \sum_{j=1}^H w_{i,j} ( |y_{i}-y_{j}|^p  - \E| X_{i} - X_{j}|^p      )^2
\label{eq_variogram_score}
\end{equation}
with $p>0$ and weight matrix $\bsW = (w_{i,j})_{i,j}$. 
In applications standard cases for $p$ are $p=0.5$ and $p=1$, the weight matrix is usually chosen with $w_{i,j}=1$. %; in this case WLOG $w_{i,j}=0$ for $j\leq i$ .
It is designed to capture differences in the correlation structure.

\subsection{Dawid-Sebastiani score}

Another alternative is motivated by the multivariate \textbf{logarithmic score} (or log-score) which is available if $\bsX$ has a density $\bsf_{\bsX}$. It is given by
$$ \text{LogS}( \bsF_{\bsX} , \bsy) = \log(\bsf_{\bsX}( \bsy )) .$$
If $\bsY$ has a density, this score is strictly proper as well. Moreover, it has locality properties which might be an advantage, depending on the application. 
As mentioned in the introduction, in many applications a density forecast is not available.
Therefore the multivariate log-score has a limited range of applications, even though the density might be approximated, see \cite{lerch2017forecaster}.

Under the normality assumption for $\bsX$ (up to some dropped constants) the log-score yields the \textbf{Dawid-Sebastiani score}, which is given by
\begin{equation}
 \DSS( \bsF_{\bsX} , \bsy) = 
 \log( \det( \bsSigma_{\bsX}) ) + (\bsy - \bsmu_{\bsX})'\bsSigma_{\bsX}^{-1}(\bsy - \bsmu_{\bsX}) 
\label{eq_dawid_sebastiani_score}
\end{equation}
 with $\bsmu_{\bsX}$ and $\bsSigma_{\bsX}$ as mean and covariance matrix of $\bsX$ and $\det$ as determinant.
It is clear this score is only proper but not strictly proper as it only matches the mean and the covariance matrix of the forecast with the observations.

\subsection{Marginal-copula score}

After we introduced the most well known and adapted scoring rules, we present a new scoring approach using copulas in the subsequent part. The idea is to describe a given multivariate distribution by its marginals and the copula for the dependency structure. Hence, we have to combine a scoring rule for the marginals with one for the copula. We start with the description of the marginal score.

\subsubsection{Marginal score}
Let $\MS_h$ be a scoring rule for the random variable $Y_h$, the \emph{$h$-th marginal score}. 
$\MS_h$ is a function depending on the forecast $X_h$ of $Y_h$ and the observation $y_h$,
so precisely $\MS_h = \MS_h(F_{X_h}, y_h)$ with $F_{X_h}$ as forecasted (cumulative) distribution function of $X_h$, which is the $h$-th marginal distribution of $\bsX$.
Following  \cite{gneiting2007strictly}, we could use every plausible scoring rule for univariate distributions, as the aforementioned CRPS.
% The likely most popular choice is the continuous ranked probability score.
We construct a marginal score as follows. Let $\text{\textbf{MS}} = (\MS_1, \ldots, \MS_H)'$ denote the vector of marginal scores for each dimension and $\bsa = (a_1,\ldots, a_H)'$ be a coefficient vector with $a_h>0$, then we define the (joint) 
\emph{marginal score} by  
$$\MS(\bsa) = \bsa' \text{\textbf{MS}} = \sum_{h=1}^H a_h \MS_h .$$
It is obvious, that if $\MS_h$ is a (strictly) proper scoring rule for $Y_h$ then the resulting marginal score $\MS(\bsa)$ is (strictly) proper for $\bsY$ for any choice of $\bsa$. 
However, the most intuitive choice is clearly $\bsa = \bsone_H$ which equally weights the marginals. This is usually a plausible choice as all the $i$-th step ahead forecasts live on the same scale.
In many applications $\bsa = \bsone_H /H$ or $\bsa = \bsone_H$ is used by default, e.g. in the Global Energy Forecasting Competition (GEFCom2014) (\cite{hong2016probabilistic}). Still, other weighting schemes for evaluation the marginals are plausible as well.
Any marginal score can correctly identify the true marginal distribution. 
However, the major drawback of those scores is that they are not feasible for identification of the correlation structure of the distribution. In order to account for that, we combine the marginal score with another score for the dependency structure. Here the copula comes into play.

\subsubsection{Copula score}

First, we briefly recall some copula theory.
The core of all copula theory is Sklar's theorem. It states that any random vector $\bsY = (Y_1,\ldots,Y_H)'$ can be represented
by its marginal distributions $F_{Y_1}, \ldots, F_{Y_H}$ and the copula function $\bsC_{\bsY}$ defined on a $H$-dimensional unit cube. 
Vice versa, given marginal distribution and a copula function, they describe the joint distribution function.
If the marginal distributions $F_{Y_1}, \ldots, F_{Y_H}$ are continuous this representation is unique.

As $F_{Y_1}, \ldots, F_{Y_H}$ characterize the marginal distributions, 
the copula $\bsC_{\bsY}$ represents the full dependency structure of $\bsY$. We want make use of this characteristic, as e.g. the energy score shows weaknesses in discriminating well between dependency structures. Further, remember that the copula $\bsC_{\bsY}$ is a (cumulative) distribution function - so we can apply all characteristics to them.

Let $\CS$ denote a \emph{copula score}, i.e.~an evaluation score for the copula $\bsC_{\bsY}$, which is a multivariate (cumulative) distribution function.
If the $\CS$ is a strictly proper scoring rule it can identify the true dependency structure of $\bsY$. 
In the forecasting evaluation literature there is no specific focus on the copulas, even though their are often used to link marginal distribution forecast with a copula forecast to represent the full multivariate distribution or adjust ensemble forecasts (e.g. \cite{moller2013multivariate}, \cite{madadgar2014towards}).
However, as every copula is just a special multivariate distribution we can use the forecasting evaluation and scoring rule literature on multivariate forecasting evaluation
to construct copula scores. Essentially we have three possible options, the energy score, the variogram score and the Dawid-Sebastiani score. Note that only the former one is a strictly proper scoring rule and thus ad-hoc preferable. Still, we will discuss all options within the paper.

\subsubsection{Combining marginal and copula score}

Considering that we have the marginal score $\MS$ and the copula score $\CS$ we have to link both to create a proper scoring rule.
Therefore, we need a transformation $g:\R\times\R \to \R$ that preserves the properties of $\MS$ and $\CS$.
It is easy to observe that the functions $g$ should satisfy some monotonicity conditions, e.g. $g$ is strictly monotone in both components.
% In more detail, if $\MS_h$ is positive and $g$ is strictly monotonic increasing in both components, then
% we can state that: if $\MS_h$ and $\CS$ are strictly proper scoring rules (for $F_{Y_h}$ and $\bsC_{\bsY}$) then $g(\MS,\CS)$ is strictly proper as well. 
In more detail, we need that $g$ is strictly isotonic (also known as $2$-monotonic) on the supports. Thus, we require $g$ such that 
$g(x_1,y_1) - g(x_1,y_2) - g(x_2,y_1) + g(x_2,y_2) > 0 $ holds for $x_1,x_2\in\text{supp}(\MS)$ and $y_1,y_2\in\text{supp}(\CS)$ with $x_1<x_2$ and $y_1<y_2$. %($>$ strictly)
% wobei $x_1<x_2$ aus dem support von MS und $y_1<y_2$ aus dem support von CS $\rightarrow$ 
The choice  $g(x,y) = x+y$ is strictly isotonic on $\R\times \R$, while $g(x,y)=xy$ works on $(0,\infty)\times(0,\infty)$.
Other options would be feasible as well, but for practical application a simple transformation $g$ is favored. 
%  $x_1 y_1 + x_2y_2 > x_1y_2+x_2y_1$ holds because of $x_1 y_1 + x_2y_2 + (y_2-y_1)(x_2-x_1)= x_1y_2+x_2y_1$, 
%  Every strictly increasing bivariate distribution function can act as $g$, this might 
% be used to map the score directly to $[0,1]$.
% 

The construction of a combined score is done as follows: Let $\MS(\bsa)$ be the marginal score and $\CS$ be the copula score, then we propose the \emph{marginal-copula score} by
\begin{align}
\MCS(\bsa) =  \MS(\bsa) \cdot \CS = \CS \sum_{h=1}^H a_h \MS_h \ \ .
\label{eq_msc}
\end{align}
%where $c\in \R$ sensitivity coefficient.
Hence, formally we choose $g(\MS(\bsa),\CS) =  \MS(\bsa) \cdot  \CS  $ which is a multiplicative structure. 
% For the plausible choice $c=0$ it is simply the product of the marginal score $\MS$ with the copula 
% score $\CS$. 

Note that the additive option $g(\MS, \CS) = \lambda \MS + (1-\lambda)\CS$ with $\lambda\in (0,1)$ looks appealing at the beginning as well. However, it turns out to be impractical 
in application as $\MS$ and $\CS$ live on different scales. 
Consider for example a forecast for $\bsY$ and $c \bsY$ with $c>0$. For most marginal scores
it follows that $\MS(F_{\bsY}, \bsY ; \bsa) \neq \MS(F_{c\bsY}, c\bsY ; \bsa)$. For the popular CRPS we even have $\MS(F_{\bsY}, \bsY ; \bsa) = \frac{1}{c} \MS(F_{c\bsY}, c\bsY ; \bsa)$.
On the other hand, for the copula it holds $\CS(\bsC_{\bsY}, \bsU_{\bsY} ) = \CS(\bsC_{c\bsY}, \bsU_{c\bsY} )$ with $\bsU_{\bsY} \sim \bsC_{\bsY}$ and 
$\bsU_{c\bsY} \sim \bsC_{c\bsY}$. So the marginal score changes with its scale but the copula 
does not change at all. Thus the additive approach does not seem to be practical for empirical application. One could solve this problem with an adequate scaling for the marginal score but since in general practice the real distribution is unknown it is difficult to find proper bounds.

Moreover, the multiplicative structure of \eqref{eq_msc} can be justified by looking at the 2-dimensional normal distribution as
in the energy score study in \cite{pinson2013discrimination}. For the bivariate normal distribution of $\bsY = (Y_1,Y_2)'$ only the correlation $\rho$ determines the copula. 
If $Y_1$ and $Y_2$ have a certain unit (e.g. $\$$, $kWh$ or $m^2/s$), then $\mu_1$, $\mu_2$, $\sigma_1$ and $\sigma_2$ have the same unit, but $\rho$
has no unit at all. Usually the marginal scores inherits the units (e.g. $\$$, $kWh$ or $m^2/s$) but the copula score can never have a unit. Therefore 
the definition $c \MS + (1-c)\CS$ provides interpretation troubles.

Here we want to summarize everything discussed above using the corollary that follows directly from the theorem of Sklar:
\begin{theorem}
Let $\MS_h$ be a strictly proper score for $Y_h$ and $\CS$ be a 
 strictly proper score for the copula $\bsC_{\bsY}$ of $\bsY = (Y_1,\ldots, Y_H)'$. Then
 the marginal-copula score
$$  \MCS(\bsF_{\bsX}, \bsy ; \bsa) 
=  \MS( (F_{X_1},\ldots,F_{X_H})', \bsy ; \bsa) \cdot \CS( \bsC_{\bsX}, \bsu_{\bsy} ) = \CS( \bsC_{\bsX}, \bsu_{\bsy} ) \sum_{h=1}^H a_h \MS_h(F_{X_h}, y_h)  $$
with $\bsF_{\bsX}$ being the cumulative distribution function, continuous marginals $F_{X_1},\ldots,F_{X_H}$, copula $\bsC_{\bsX}$ of  forecasts $\bsX = (X_1,\ldots, X_H)'$,
observation vector $\bsy$, copula observations 
$\bsu_{\bsY} = (u_{\bsY,1}, \ldots, u_{\bsY,H})' = (F_{Y_1}(y_1), \ldots, F_{Y_H}(y_H))'$ and weight vector
 $\bsa $ with $a_h>0$ % and $c\in \R$ 
is a strictly proper scoring rule.
\end{theorem}
% As pointed out above this could be generalised using the strictly monotonic function $g$.
% \FZ{TODO 
% We suggest the use of the CRPS for the marginal scores $\MS_h$ and the scaled energy score of the coupla $\bsC_{\bsY}$ of $\bsY$ TODOREF 
% The empricial results show that $c=?$ and $\bsa=\bsone$ yields promosing results in several simulation studies.}
Unfortunately, this holds only for continuous random variables. In case of non continuous marginals the theorem holds for normal (but not strict) propriety at least. Still, many applications are covered.

Another feature that we want to point out is that 
the copula is invariant against strictly monotonic transformations.
So if we have strictly monotonic transformations $g_h$ then the copula of $(g_1(Y_1), \ldots,g_H(Y_H))'$ 
has the same copula as $\bsY = (Y_1,\ldots, Y_H)'$. Thus, only the marginal score components $\MS_h$ get affected by 
strictly monotonic transformations; especially the scoring rule remains strictly proper after monotonic transformations of the 
random variables $\bsY$ of interest.

In the next part, we discuss in more details possible choices for the copula score $\CS$. 
% As mentioned above concerning the actual literature there are three plausible choices for 
% constructing the copula score $\CS$. 
We discuss the energy score, the variogram score and the Dawid-Sebastiani score applied to copulas. 
% However, only the first one will be strictly proper. Still, for many practical applications the variogram score based approach works well too.
Remember that latter score evaluates only the first two moments. 
As all copulas have uniform marginals, they have all the same mean  and variance structure, thus
 only the correlation 
structure would be evaluated. %\FZ{TODO: Should we add it, maybe someone is only interested in the correlation structure?}
The log-score is not suitable for application as we would require a copula density forecast which is rarely available in practice.

We are interested in evaluating the fit of the copula $\bsC_{\bsX}$ belonging to the forecast $\bsF_{\bsX}$ to the copula $\bsC_{\bsY}$ of $\bsY$. Usually we require for forecasting evaluation the observed value $\bsy=(y_1,\ldots, y_H)'$ of $\bsY$. 
In the copula case we apply the probability integral transformation of each component to receive
$$\bsU_{\bsY} = (U_{\bsY,1}, \ldots, U_{\bsY,H})' = (F_{Y_1}(Y_1), \ldots, F_{Y_H}(Y_H))' = \bsF_{\bsY} (\bsY) $$
which has uniform marginals 
and define the (pseudo) \emph{copula observations} as 
$$\bsu_{\bsY} = (u_{\bsY,1}, \ldots, u_{\bsY,H})' = (F_{Y_1}(y_1), \ldots, F_{Y_H}(y_H))' = \bsF_{\bsY} (\bsy) .$$
Note that in application we can never observe $\bsu_{\bsY}$ as the required marginal distributions $F_{Y_h}$ are unknown. We will further address this issue in section \ref{sec:copobs}.

\subsubsection{Copula energy score}

Now, we define the copula energy score ($\CES$) as the energy score of the copula scaled by $H^{-\frac12}$
\begin{align}
 \CES( \bsC_{\bsX}, \bsu_{\bsY} ) 
 &= \frac{1}{\sqrt{H}} \ES(\bsC_{\bsX}, \bsu_{\bsY}) \nonumber \\
 &=  
 \frac{1}{\sqrt{H}}\left(\E\left( \left\| \bsU_{\bsX} - \bsu_{\bsY} \right\|_2 \right) - %}_{\CES_{1}:=}  - 
  \frac{1}{2} \E\left( \| \bsU_{\bsX} - \wtilde{\bsU}_{\bsX} \|_2 \right)  - \text{lb}_{\text{CES}}\right)  %}_{\CES_{2}:=} 
%  &= \frac{1}{\sqrt{H}}\left(\CES_{1} - \frac12 \CES_{2} \right)
 \label{eq_copula_energy_score}
 \end{align}
where $\wtilde{\bsU}_{\bsX}$ is an iid copy of $\bsU_{\bsX}$, so $\bsU_{\bsX}, \wtilde{\bsU}_{\bsX} \stackrel{\text{iid}}{\sim} \bsC_{\bsX}$.
The additional $ \text{lb}_{\text{CES}}$ term is inserted because it corresponds to the lower bound where we choose $ \text{lb}_{\text{CES}} = \frac{1}{4}- \frac{1}{2 \sqrt{6}}$ and discussed below.
Note that in \cite{gneiting2007strictly} a more general version of the energy score is introduced which involves an additional parameter $\beta$ which 
is chosen here (but also in other studies like \cite{pinson2013discrimination}) to be $1$. 
Furthermore we consider the energy-score with negative
orientation, thus its optimum is a minimum (for more details see \cite{gneiting2007strictly}). This is in line with many applications.
% \FZ{The energy score can be seen as a multivariate version of Hoeffding’s $\Phi^2$ by ($L_2$ measure), see copula book 10.5.2. }

In  \eqref{eq_copula_energy_score} we defined the copula energy score scaled by $H^{-\frac12}$ to adjust for the dependency in the dimension $H$. This allows a better comparison of score values for different forecast horizons, for instance. 
Clearly, it would be nice to have a copula score which is scaled in such a way that it allows easy interpretation, 
like the correlation which is a bounded measure and takes values 
between -1 and 1 and allows for interpretation of the linear dependency structure. However, such a scaling is not feasible for the copula energy score 
as its outcome will always depend on the true distribution $\bsF_{\bsY}$, especially the lower and upper bounds will change.
Still, both the lower an upper bound grow in $H$ always with a rate of $\sqrt{H}$. Hence, we suggest to scale the outcome with $H^{-\frac12}$.

Since we use negatively orientated scoring rules, 
the lower bound is the score of the optimal forecast. 
We know that for a given copula $\bsC_{\bsX}$ the energy score $\CES( \bsC_{\bsX}, \bsu_{\bsY} )$ 
is minimized if the forecasted distribution $\bsX$ equals the true distribution $\bsY$ since the energy score is strictly proper.

We studied the lower bound of the energy score of the copula. However, the authors were not able to proof a strict lower bound. Additional research is ongoing.
% Still, based on extensive simulation studies it seems that the energy score of a copula is minimized by $\ES(\bsM_{H}, \frac{1}{2}\bsone)$
% with $\bsM_{H}$ the upper Fr\'{e}chet-Hoeffding bound (comonotonicity copula) of dimension $H$, see e.g. \cite{durante2010copula}. 
Instead, we consider the lower bound $\text{lb}_{\text{CES}} = \frac{1}{4}- \frac{1}{2}\frac{1}{ \sqrt{6}}$ where the $\frac{1}{4}$ corresponds to a lower bound of the first term
$\E\left( \| \bsU_{\bsX} - \bsu_{\bsY} \|_2 \right)$
in \ref{eq_copula_energy_score}, and the $\frac{1}{ \sqrt{6}}$ to an upper bound of second term $\E\left( \| \bsU_{\bsX} - \wtilde{\bsU}_{\bsX} \|_2 \right)$. 
In the Appendix we show upper and lower bounds for both terms in Lemma \ref{lemma_1_appendix} and Lemma \ref{lemma_2_appendix}.

%\E\left( \| \bsU_{\bsX} - \wtilde{\bsU}_{\bsX} \|_2 \right)

% Further, it holds $\ES(\bsM_{H}, \frac{1}{2}\bsone) = \frac{\sqrt{H}}{12}$ 
% which justifies the scaling constant in  \eqref{eq_copula_energy_score}, see Lemma \ref{lemma_3_appendix} in the Appendix for the calculation.
% Note that this result is not unique, e.g. $\bsC_{\bsX} = (\bsZ, 1-\bsZ_i)$ with $\bsZ \sim \bsM_{H-1}$ yields the same bound.

% These bounds hold true but they are very likely not sharp. 
% However, we have some assumptions on sharper (probably even strict) bounds 
% Based on extensive simulation studies, it seems to hold that 
% \begin{align}
%  \frac{\sqrt{H}}{4} - \frac{1}{2} \frac{\sqrt{H}}{3} = \sqrt{H}/12 \leq \ES( \bsC_{\bsX}, \bsu_{\bsY} )  \leq  C_H - \frac{1}{2} \frac{\sqrt{H}}{3} 
% \label{eq_bound_ces_strict}
%  \end{align}
% where $C_H$ given explained in  Lemma \ref{lemma_3_appendix}. 
% 
% (see Lemma \ref{lemma_3_appendix} and \ref{lemma_4_appendix} in the Appendix)
% 
% The first inequality in \eqref{eq_bound_ces_strict}
% results from the choice $\bsC_{\bsX} = \bsM_{H}$ and $\bsu_{y} = \frac{1}{2}\bsone$. 
% Note that this is not unique, e.g. $\bsC_{\bsX} = (\bsZ, 1-\bsZ_i)$ with $\bsZ \sim \bsM_{H-1}$ yields the same result.
% The second inequality in \eqref{eq_bound_ces_strict} can be derived by choosing so that $\bsU_{\bsX} \sim \bsC_{\bsX}$
% with $\bsU_{\bsX} = (\bsZ, 1-\bsZ_i)$ with $\bsZ \sim \bsM_{H-1}$ and $\bsy = \bsnull$. Also this choice is not unique. Any $\bsy\in \{0,1\}^H$ yields the same results.

\subsubsection{Copula variogram score}

As mentioned, we can apply the variogram score as a copula score  $\bsC_X$ as well.
We define the variogram copula score as a scaled variogram score of the copula:
\begin{align}
\text{CVS}_{p}( \bsC_{\bsX}, \bsu_{\bsY}; \bsW ) 
&= \frac{1}{\bsone' \bsW \bsone} \text{CS}_{p}( \bsC_{\bsX}, \bsu_{\bsY}; \bsW )  \\
&= \sum_{i=1}^H \sum_{j=1}^H w_{i,j} ( |u_{\bsY,i}-u_{\bsY,j}|^p  - \E| U_{\bsX,i} - U_{\bsX,j}|^p      )^2
\end{align}
with $\bsU_{\bsX} = (U_{\bsX,1},\ldots,U_{\bsX,H})' \sim \bsC_{\bsX}$,
$p>0$ and weight matrix $\bsW = (w_{i,j})_{i,j}$. 
The standard cases are $p=0.5$, $1$ and $2$. For the weight weight matrix we assume $w_{i,j}=1$ which gives
a scaling coefficient of $\bsone' \bsW \bsone= H^2$.
Similarly to the energy score the scaling is motivated by the fact to adjust the dependency on the dimension of the resulting score.
%; in this case WLOG $w_{i,j}=0$ for $j\leq i$ .
As for the energy score case we can derive an upper scaling bound
\begin{align}
\VS_{\bsW, p}( \bsC_{\bsX}, \bsu_{\bsY} ) 
&=
\sum_{i=1}^H \sum_{j=1}^H w_{i,j} ( |u_{\bsY,i}-u_{\bsY,j}|^p  - \E| U_{\bsX,i} - U_{\bsX,j}|^p      )^2 \\
&\leq
\sum_{i=1}^H \sum_{j=1}^H w_{i,j} ( 1^p  - 0       )^2 = \bsone' \bsW \bsone,
\end{align}
which justifies the scaling constant.
Obviously, the lower bound of $\text{CVS}_{p}$ is zero, as for is holds 
$ \VS_{\bsW,p}( \bsM_{H}, \bsU_{\bsY}) =0$ if  $\bsU_{\bsY} \sim \bsM_H $.

 As the variogram score is only a proper scoring rule, the resulting MS-$\text{CVS}_{p}$ score can not be strictly proper. 
 However, from practical perspective the use of MS-$\text{CVS}_{p}$ can be more appealing as the disadvantages occur mainly 
 in the marginal distributions. There is no very simple example where a MS-$\text{CVS}_{p}$ score is not strictly proper if MS is strictly proper.

 \subsubsection{Copula Dawid-Sebastiani score}

 Finally, we can also consider the Dawid-Sebastiani score (DSS). As mentioned above, the DSS can only evalutate the correlation structure in the forecast,
 so it is only appealing if we are interested in evaluating the linearly dependency structure in the copula of the forecasts, but not recommended in general.
 \begin{align}
 \text{CDSS}( \bsC_{\bsX} , \bsu_{\bsy}) 
 &= 
  \DSS( \bsC_{\bsX} , \bsu_{\bsy})  \nonumber \\
 &= 
 \log( \det( \bsSigma_{\bsU_{\bsX}} ) ) + (\bsu_{\bsy} - \bsmu_{\bsU_{\bsX}})'\bsSigma_{\bsU_{\bsX}}^{-1}(\bsu_{\bsy} - \bsmu_{\bsU_{\bsX}}) 
 \end{align}
 with $\bsmu_{\bsU_{\bsX}}$ and $\bsSigma_{\bsU_{\bsX}}$ as mean and covariance matrix of $\bsU_{\bsX}\sim \bsC_{\bsX}$.
From properties of the uniform distribution it follows that
 $\bsmu_{\bsU_{\bsX}} = \frac{1}{2}\bsone$ and $\bsSigma_{\bsU_{\bsX}} = \bsS \bsR_{\bsU_{\bsX}} \bsS$ where $\bsS = \frac{1}{\sqrt{12}} \bsI$ and correlation matrix $\bsR_{\bsU_{\bsX}}$. 
 So we receive
 \begin{align}
\text{CDSS}( \bsC_{\bsX} , \bsu_{\bsy})  = -H \log( 12 \det( \bsR_{\bsU_{\bsX}} ) ) + \frac{1}{12}\left(\bsu_{\bsy} - \frac{1}{2}\bsone\right)'\bsR_{\bsU_{\bsX}}^{-1}\left(\bsu_{\bsy} - \frac{1}{2}\bsone\right)    
\label{eq_CDSS_2}
 \end{align}
 as $ \det( \bsSigma_{\bsU_{\bsX}}) = 12^{-H} \det( \bsR_{\bsU_{\bsX}} ) $. Hence, due to \eqref{eq_CDSS_2} a scaling by $\frac{1}{H}$ might be considered 
for the Copula Dawid-Sebastiani score to adjust for impact of the dimension.

In contrast to the copula energy score and copula variogram score we do not apply any scaling constant. The reason is that DSS is an unbounded score.
For instance, when considering a copula with the the constant correlation matrix $\bsR_{\bsU_{\bsX}}(\delta) = (1-\delta) \bsI + \delta \bsone\bsone'$
then it holds for the limit %s $ \lim_{\delta \to 0} \det( \bsR_{\bsU_{\bsX}}(\delta)^{-1} ) = 1$ and 
$ \lim_{\delta \to 1} \det( \bsR_{\bsU_{\bsX}}(\delta) ) = 0$. Additionally, the second term is a quadratic form and minimal if $\bsu_{\bsy} - \frac{1}{2}\bsone$. 
Thus, the copula Dawid-Sebastiani score has no lower bound and can take negative values.

%\FZ{WE should scale it by $1/H$ oder?}

 %  We covered and propose here only the use of two copula scores. 
%  However, there are more potential candidates that might serve as.
%  
%  If $C_{\bsX}$ has a density that is reported in the forecasting scheme, that the log score is an alternative.
%  However, this restricts the case field of applications, as most the time only ensemble forecasts are available, see section \ref{sec_reporting}.
% 
%  We also have seen the Dawid-Sebastiani Score (DSS). Theoretically, we can use it as a Copula Score as well. 
%  However, DSS depends only on the mean and covariance matrix of the forecast.
%  As all copulas have uniform marginals, they have all the same mean of $\frac{1}{2}\bsone$ 
%  the covariance matrix is $\bsS \bsR \bsS$ where $\bsS = \text{diag}( \frac{1}{\sqrt{12}}\bsone )$ and an arbitrary correlation matrix $\bsR$. So 
%  in fact a DSS-based copula score would only explore differences in the correlation structure of the forecast, so only detect differences in the linear dependency 
%  structure.
% 

 \section{Reporting multivariate forecasts and evaluation design} \label{sec:reporting}

% Assuming we have a univariate time series $(Y_t)_{t\in \Z}$ that we are interested to forecast.
% Further, assume that at the moment when the forecast is going to be created we are at time $T$. Consequently 
% we want to compute a forecast of $Y_t$ for $T+1,\ldots, T+H$ where $H$ is the forecasting horizon.
% So we are interest in $\bsY = (Y_{T+1},\ldots, Y_{T+H})'$ given all available information at time $T$. 
% \FZ{Different notation $\bsY_1$ instead of $\bsY_T$? then we have later on $\bsY_1,\ldots, \bsY_N$ which seems to be more natural...}
% More precisely, we want to know the full multivariate distribution $\bsF_{\bsY_{1}}$ of $\bsY_{1}$, which characterize all
% properties of $\bsY_{1}$, like the mean, the variance, higher moments, the marginals, the tail behavior, the correlation structure or 
% the dependency structure in general. 
 
Remember, we want to forecast $\bsY$ by $\bsX$.
% . But from the 
% forecasting evaluation perspective it not relevant which information goes into the forecast, as long as the information is available at time $T$.
%  So here $\bsX_{1}$ is the forecast of $\bsY_{1}$. 
 In statistical communities, often $\what{\bsY}$ is usually used instead of $\bsX$ to emphasize relationship to the target $\bsY$.

% As mentioned we are interested in the distribution $\bsF_{\bsY_{1}}$ of $\bsY_{1}$. 
% We denote the corresponding forecasting of $\bsF_{\bsY_{T}}$ of $\bsY_{T}$ we denote by $\bsF_{\bsX_{T}}$. 

In practice, there are two options for describing the forecasting distribution $\bsF_{\bsX}$ of $\bsX$ or to report this forecast:
\begin{enumerate}
 \item \label{item_mdistr} Reporting the forecasting distribution $\bsF_{\bsX}$ explicitly 
 \item \label{item_simul} Providing a independent large sample of the (forecasting) model of $\bsX$ of the target distribution $\bsY$, 
 we refer to this large sample as \emph{ensemble} and the method as \emph{ensemble forecasting}.
\end{enumerate}

If we are going for option \ref{item_mdistr}, then we have to report the forecasted distribution $\bsF_{\bsY}$ explicitly or an equivalent representation. 
A popular way of providing the equivalent information is by reporting an estimate $\bsf_{\bsX}$
of the joint density $\bsf_{\bsY}$ which exists for (absolutely) continuous distribution.
Alternatively, $\bsF_{\bsX}$ can be described using copulas, in detail we have to report estimates 
$f_{X_{1}},\ldots, f_{X_{H}}$ of
the marginal distributions
$f_{Y_1},\ldots, f_{Y_H}$ of $\bsf_{\bsY}$ and a copula estimate $\bsC_{\bsX}$ for the copula $\bsC_{\bsY}$. For our copula-score based evaluation approach
this is a very appealing way of reporting, unfortunately not a very common one.
% the multivariate copula $\bsC$ itself.
The last option we want to mention is to report an estimator $\bsvarphi_{\bsX}$ of the characteristic function defined through
$\bsvarphi_{\bsY}(\bsz) = \E(\exp(i\bsz'\bsY))$ which determines uniquely $\bsF_{\bsY}$.

From the evaluation point of view we must be able to draw from the reported distribution on a computer; or solve some characteristics (esp. some expected values) 
of the reported distribution which is only feasible in limited cases. 
Unfortunately for sophisticated forecasting models, an explicit description of  $\bsF_{\bsX}$ (or an equivalent counterpart) 
is either very challenging or not feasible. Still, option \ref{item_simul} can be applied in such cases. Therefore option \ref{item_simul} turned into the standard alternative in application.

In the second reporting option \ref{item_simul} we provide $M$ independent simulations resp. draws from the underlying model of $\bsX$ 
which essentially describes $\bsF_{\bsX}$. We denote the paths/trajectories of the ensemble by $\XX = \bsX^{(1)},\ldots,\bsX^{(M)}$, so $\bsX^{(j)}$ is an independent and identically distributed (iid) copy of $\bsX$.
This methodology is basically a Monte-Carlo simulation, but it also known as ensemble forecasting, ensemble simulation, path simulation, trajectory simulation, scenario simulation or scenario generation (but depending on the community the meaning might be differ slightly as well).
Within this paper we assume that the sample size denoted by $M$ is large, so the true underlying distribution function is described well
by the given simulated sample. This approximation goes back to the multivariate Glivenko-Cantelli theorem which characterize that the multivariate 
empirical (cumulative) distribution function is converging almost surely to the drawn distribution.

For comparing forecasts in a scientific way it is not sufficient to do forecasting evaluation based on a single forecast 
 $\bsY$ of $\bsX$. We require multiple forecasts in a (pseudo-)out-of-sample study design to make significance statements later on.
Therefore we assume to have $N$ forecasts %$\bsX_1,\ldots, \bsX_N$ 
of $\bsY$ %$\bsY_1,\ldots, \bsY_N$
available. How these 
forecasted values are generated is theoretically not relevant as long only information is taken into account that was available at the time point of generating the forecast.

Now, we assume to have a univariate time series $(Y_t)_{t\in \Z}$ that we are interested to forecast.
We denote these multiple targets within the forecasting window by
$\bsY_{ i} = (Y_{T+s_i+1},\ldots, Y_{T+s_i+H})'$ with $s_i< s_{i+1}$ ($1\leq i<N$). 
So $\bsY_{1},\ldots, \bsY_{N}$ are ordered in time. The forecast of $\bsY_{1},\ldots, \bsY_{N}$ are 
denoted by $\bsX_{1},\ldots, \bsX_{N}$ in accordance with the notation of the previous section.
Note, that the forecasting periods of $\bsY_{i}$ and $\bsY_{i+1}$ my overlap, so if e.g. $s_1=0$, $s_2=1$ with $H=24$ we have 
forecasts $\bsX_{1} = (X_{T+1},\ldots, X_{T+24}) $ and $\bsX_{2} = (X_{T+2},\ldots, X_{T+25} )$ of 
$\bsY_{1} = (Y_{T+1},\ldots, Y_{T+24}) $ and $\bsY_{2} = (Y_{T+2},\ldots, Y_{T+25} )$. 
Obviously, this design should be tailored in accordance with the corresponding application. 
From the statistical point of view is useful to have nice properties of both the forecasts $\bsX_{i}$ and the true $\bsY_{i}$ in the evaluation window
like stationarity, ergodicity or finite variance. However, we can only influence the forecasting model, so $\bsX_{i}$.

For forecasting models which evaluate historic data we recommend the usage of a \emph{rolling window forecasting study} using
$N$ equidistant windows. It is the most suitable way to evaluate forecasts.  
In general there are several options for the design of a rolling window forecasting study. But they all share together that the in-sample data length is fixed and 
the in-sample window moves equally distant across the time range. So the window shift are $s_i = K(i-1)$. If $K<H$ then we have overlapping forecasting horizons.
If $K\geq H$ the forecasting horizons in the forecasting study do not overlap which is favorable from the statistical point of view - but does not necessarily meet practical applications. 
Here 
the special case $K=H$ is usually chosen.
A slightly different design of an rolling window forecasting study with 
$s_i = 0.25 H(i-1)$ and $H=12$ is visualized in Figure \ref{fig_sample_air} of the real data application section along with forecasting reporting option reporting option \ref{item_simul}.
In many practical applications such a design is clearly favorable.
Note that for forecasting evaluation expanding windows forecasting studies are not suitable. More details on the design of forecasting studies, can be found in \cite{diebold2015comparing}.

% \begin{figure}
%  
% %  \resizebox{.99\textwidth}{!}{
% \includegraphics[width=.99\textwidth]{fig/int_roll_1.pdf} 
% \includegraphics[width=.99\textwidth]{fig/int_roll_2.pdf} 
% \includegraphics[width=.99\textwidth]{fig/int_roll_3.pdf} 
% \caption{Illustration of a rolling window forecasting study with non-overlapping windows ($s_i=H(i-1)$) for $i = 1,\ldots, 3$ windows and $M=6$
% forecast samples $\bsx_{i}^{(1)},\ldots,\bsx_{i}^{(M)}$ for each window $i$. }
% \label{fig_rolling}
% \end{figure}

Finally, we require for the evaluation that we have observations of the full forecasting window of $\bsY_{1},\ldots, \bsY_{N}$
available. We denote these observations by $\bsy_{1}, \ldots, \bsy_{N}$. 
This is important for the evaluation method as well. So we do not want to compare the forecasted distributions 
of $\bsX_{1},\ldots, \bsX_{N}$
with the true underlying $\bsY_{1},\ldots, \bsY_{N}$ but with the observations $\bsy_{1}, \ldots, \bsy_{N}$. 
So from the statistical point of view we require a so called one sample testing framework for comparing distributions.

 \section{Estimation} \label{sec:estim}
  
 So far we had only a look at the theoretical properties score of a forecasting method. 
 However, as always in statistics, the theoretical scores involve objects that are unknown in practice.
 Therefore we assume that we are in a forecasting study framework, where we have the observations $\bsy_1,\ldots,\bsy_N$ of $\bsY_1,\ldots, \bsY_N$ such as the ensemble forecasts $\XX_i = (\bsX_{i}^{(1)},\ldots, \bsX_{i}^{(M)})'$ of the forecasting distribution $\bsX_i$ for $\bsY_i$ available.
 
 Subsequently, we give an overview on estimators for the standard evaluation methods, namely the energy score, the variogram score and the Dawid-Sebastiani score. Afterwards we deal with the problem of estimating copula observations and the marginal copula score.

\subsection{CRPS}

When estimating \eqref{eq_crps}, we have essentially two options, matching the two possible representations. 
% Either we utilise the first representati in 
The most standard form is to utilise the first one. It solves the corresponding integral numerically by replacing the unknown cdf by an ecdf.
$\what{\CRPS}_{i,h} =  \int_{-\infty}^{\infty} (\what{F}_{Y_{i,h}}(z) - \mathbbm{1}\{y_h < z\})^2 \,d z $
with standard estimator for $F_{Y_{i,h}}$ is the empirical distribution function (ecdf) of $h$-th coordinate of the ensemble $\XX_i$.
This is given by
$\what{F}_{Y_{i,h}}(z) = \what{F}_{Y_{i,h}}(z; \XX_i) = \frac{1}{M} \sum_{j=1}^M \mathbbm{1}\{X_{i,h}^{(j)} \leq z\}$.

Alternatively, we can utilize the second term in \eqref{eq_crps} and estimate the two expected values. This is e.g. done in 
\cite{taieb2016forecasting} in a load forecasting context. It additionally helps for interpretation. We discuss this resulting estimator as a special case 
of the estimation of the energy score in the subsequent subsection.

Given the scores $\what{\bsCRPS}_{i} = (\what{\CRPS}_{i,1},\ldots, \what{\CRPS}_{i,H})$ for each marginal distribution we can compute the corresponding marginal score
by $ \what{\CRPS}_i(\bsa) = \bsa'\what{\bsCRPS}_{i}$
with weight vector $\bsa = (a_1,\ldots, a_H)'$.
% \begin{equation}
%   \CRPS(\bsF_{X}, y)
%  \end{equation}

%  \CRPS(\bsF_{X}, y) = \int_{-\infty}^{\infty} (F_{X}(z) - \mathbbm{1}\{y < z\})^2 \,d z  =  \E | X - y | - \frac{1}{2} \E| X -\wtilde{X}|  

\subsection{Energy score}

% First, we consider an estimator for the energy score.
First, we can rewrite the energy score definition \eqref{eq_energy_score} as
\begin{align}
 \ES_{i,\beta} ( \bsF_{\bsX_i}, \bsy_i) %=&
 &= \text{ED}_{\beta,i}(\bsX_i, \bsy_i) - %}_{\text{ED}_\beta:=}  - 
  %\underbrace{ 
  \frac{1}{2} \text{EI}_{\beta,i}(\bsX_i, \bsy_i)
  %}_{\text{EI}_\beta:=} 
\end{align}

% Therefore we require samples of the underling distributions of $ \bsX_{i}$.
% If we are using the reporting option \ref{item_simul} we can directly use the $M$ samples drawn independently
% from the corresponding model for $\bsX_{i}$. We denote these $M$ iid copy draws by $\bsX_{i}^{(1)},\ldots,\bsX_{i}^{(M)}$.
% If reporting option \ref{item_mdistr} is used, then we have to draw from the given 
% multivariate distribution characterization. Without loss of generality, we can assume that we also draw $M$ samples from this distribution.

If reporting option \ref{item_mdistr} is used, we might be able to compute the two terms in \eqref{eq_energy_score} explicitly, 
e.g. under the normality assumption see \cite{pinson2013discrimination}. 
If we are able to do use explicit or numeric approximation formulas for the two terms in \eqref{eq_energy_score}, we should prefer them. 
They are usually more precise and much faster than the simulation approaches.

% In reporting option \ref{item_simul}
However, in the standard case of reporting option \ref{item_simul} we have to estimate the first term in $\text{ED}_{\beta,i}$ and the second term $\text{EI}_{\beta,i}$ in 
 \eqref{eq_energy_score} given the $M$ samples $\bsX_{i}^{(1)},\ldots,\bsX_{i}^{(M)}$ of the distribution $\bsF_{\bsX}$.
The estimation of the first term $\text{ED}_{\beta,i}$ is straight forward, we simply estimate the expectation by its sample mean.
So we have
\begin{equation}
 \what{\text{ED}}_{i,\beta} = 
\frac{1}{M} \sum_{j=1}^M  \left\| \bsX_{i}^{(j)} - \bsy_{i} \right\|_2^\beta .
\end{equation}

The second term $\text{EI}_{\beta,i}$ has multiple plausible options for the estimation.
The reason is that the definition in \eqref{eq_energy_score} of $\text{EI}_{\beta,i}$ suggests we require these independent copies $\wtilde{\bsX}_{i}$ 
of our forecasting distribution. As $\bsX_{i}^{(1)},\ldots,\bsX_{i}^{(M)}$ are independently drawn, 
we can simply use the first have of the data set as draws from $\bsX$ in \eqref{eq_energy_score} and the second half as draws from the iid copy $\wtilde{\bsX}$.
Hence we can simply use the estimator
\begin{equation}
\what{\text{EI}}^{\text{iid}}_{i,\beta} = \frac{1}{2M} \sum_{j=1}^{0.5M}  \left\| \bsX_{i}^{(j)} - \bsX_{i}^{(0.5M + j)} \right\|_2^\beta  
\end{equation}
where we assume without loss of generality that $M$ is even. 
Note that the sum in $\what{\text{EI}}^{\text{iid}}_{i,\beta}$ contains only $M/2$ summands. But these summands have nice statistical properties, as they are iid. 
Also \cite{moller2013multivariate} and \cite{junk2014comparison} mention the resulting estimator for the energy score as a plausible option.

Still, we can use an alternative estimator that uses more summands and provides in average higher accuracy.
Such an estimator is given by
\begin{equation}
 \what{\text{EI}}^{K\text{band}}_{i,\beta} = \frac{1}{M} \sum_{j=1}^{M} \sum_{k=j}^K  \left\| \bsX_{i}^{(j)} - \bsX_{i}^{(j+k)} \right\|_2^\beta 
\label{eq_est_energy_score_alt2_second_term}
\end{equation}
for an integer $K$ with $1\leq K\leq M$ where we set for notational purpose $\bsX_{i}^{(M+k)} = \bsX_{i}^{(k)}$.
 
 The expression $\what{\text{EI}}^{K\text{band}}_{i,\beta}$ has the advantage by using a larger 
amount of summands for approximating the sum.  This should increase the precision in general. However, as the elements of the sum become pairwise dependent, this improvement is weaker as if they would be all independent. %, i.e. 

For the important case $K=1$ we have $M$ summands across the summation diagonal, for $K=M$ have $M(M-1)/2$ different pairs $\bsX_{i}^{(j)} - \bsX_{i}^{(l)}$ that are summed up. 
The latter cases uses the maximal amount of information for estimating $\text{EI}_{\beta,i}$. %Like in $\what{\text{EI}}^{\text{lin}}_{i,\beta}$ the summands are depended. 
In the \texttt{R} packages \cite{Rpackage_scoringRules} and \cite{Rpackage_energy}, only this estimator is provided. It has the mentioned advantage of high accuracy but a high computation demand for large values of $M$
as the amount of summands is increasing quadratically in $M$. Thus, if $M$ is large (which is recommended to choose), we suggest to use a small $K$ to keep the computational complexity manageable. Then the energy score is estimated by
$$  \what{\ES}^{K\text{band}}_{i,\beta}  =  \what{\text{ED}}_{i,\beta} -  0.5 \what{\text{EI}}^{K\text{band}}_{i,\beta} .$$

\subsection{Variogram score}

The computation is similarly to the energy score. If we apply the reporting option \ref{item_mdistr}, we might be able to solve some special cases explicitly, like the case of multivariate normality.

If we consider the simulation reporting option \ref{item_simul}, we have can estimate the expected values in \eqref{eq_variogram_score} by the sample means.
Thus, we have
\begin{align}
 \what{\VS}_{i,\bsW,p} %( \bsX_i , \bsy_i ; \bsW )
=  \sum_{j=1}^H \sum_{k=1}^H w_{j,k} ( |y_{i,j}-y_{i,k}|^p  - \sum_{m=1}^M | X_{i, j}^{(m)} - X_{i,k}^{(m)}|^p      )^2. 
\label{eq_variogram_score_est}
\end{align}
for a given $\bsW$ and  $p$. Due to symmetries $y_{i,j}-y_{i,k} = y_{i,k}-y_{i,j}$ and $X_{i,j}-X_{i,k} = X_{i,k}-X_{i,j}$ the computational cost can be halved in 
the implementation. Thus it holds
\begin{align}
 \what{\VS}_{i,\bsW,p} %( \bsX_i , \bsy_i ; \bsW )
=  2\sum_{j=1}^H \sum_{k=j+1}^H w_{j,k} ( |y_{i,j}-y_{i,k}|^p  - \sum_{m=1}^M | X_{i, j}^{(m)} - X_{i,k}^{(m)}|^p      )^2. 
\label{eq_variogram_score_estalt}
\end{align}

\subsection{Dawid-Sebastiani score}

The Dawid-Sebastiani score depends only on the first two moments. Under reporting option \ref{item_mdistr}, we are usually able to compute the corresponding moments explicitly.
Under the reporting option \ref{item_simul}, we estimate the moments by their sample counterparts:

Then we have for the mean and covariance matrix estimates
\begin{align}
\what{\bsmu}_{i,\bsX} &= \frac{1}{M} \sum_{m=1}^M \bsX_{i}^{(m)}  \\
\what{\bsSigma}_{i,\bsX} &= \frac{1}{M-1} \sum_{m=1}^M (\bsX_{i}^{(m)} - \what{\bsmu}_{i,\bsX} ) (\bsX_{i}^{(m)} - \what{\bsmu}_{i,\bsX} )'
\end{align}
where the latter one is the unbiased sample covariance matrix.
The resulting plug-in estimator for the Dawid-Sebastiani score is 
\begin{equation}
 \what{\DSS}_i = 
 \log( \det( \what{\bsSigma}_{i,\bsX} ) ) + (\bsy_i - \what{\bsmu}_{i,\bsX})'\what{\bsSigma}_{i,\bsX}^{-1}(\bsy_i - \what{\bsmu}_{i,\bsX}) .
\label{eq_dawid_sebastiani_score_est}
\end{equation}
Note that we require $M>H$ to receive a positive semidefinite sample covariance matrix. 

\subsection{Copula observations}
\label{sec:copobs}
Remember that for all copula scores we are interested in evaluating 
$\bsU_{\bsY_i} = (U_{\bsY_i,1}, \ldots, U_{\bsY_i,H})' = (F_{Y_{i,1}}(Y_{i,1}), \ldots, F_{Y_{i,H}}(Y_{i,H}))' $.
There we defined the 
copula observations $\bsu_{\bsY_i} = (u_{\bsy_i,1}, \ldots, u_{\bsy_i,H})' = (F_{Y_{i,1}}(y_{i,1}), \ldots, F_{Y_{i,H}}(y_{i,H}))'$.
However, as mentioned already above 
a crucial problem here is that the definition of $\bsu_{\bsY_i}$ includes the marginals $F_{Y_{i,h}}$ of the true underlying distribution 
which is unknown in practice.
Thus, the only way around this to estimate $F_{Y_{i,h}}$ to get an estimator for $\bsu_{\bsY_i}$. 

Of course the standard estimator for $F_{Y_{i,h}}$ is the empirical distribution function (ecdf) of the ensemble $\XX_i$.
However, we can use the standard estimator
$\what{F}_{Y_{i,h}}(z) = \what{F}_{Y_{i,h}}(z; \XX_i) = \frac{1}{M} \sum_{j=1}^M \mathbbm{1}\{\bsx_i^{(j)} \leq z\}$.\footnote{
The mid-point rule 
$\what{F}_{Y_{i,h}}^{\text{mid}}(z) = \frac{1}{2M} \sum_{j=1}^M \mathbbm{1}\{\bsx_i^{(j)} \leq z\} + \mathbbm{1}\{\bsx_i^{(j)} < z\}$
might be even a better choice.}
Then we receive \emph{estimated copula observations}
$$\what{\bsu}_{\bsY_i} = \what{\bsu}_{\bsY_i}(\XX_i) = (\what{F}_{Y_{i,1}}(y_{i,1}; \XX_i), \ldots, \what{F}_{Y_{i,H}}(y_{i,H}; \XX_i))'.$$ 

Intuitively, the use of $\what{\bsu}_{\bsY_i}$ instead of $\bsu_{\bsY_i}$ should not be a problem if $\what{F}_{Y_{i,h}}$ is a good estimator for $F_{Y_{i,1}}$;
and here intuition is right.
Unfortunately it is not a good choice if this is not the case. 
This quickly leads to problems concerning the application of $\what{\bsu}_{\bsY_i}$ for the evaluation of copula based scores.
There are situations with a misspecified forecasting marginal distribution and the use of $\what{\bsu}_{\bsY_i}$ as copula observation estimator 
where the corresponding copula score of the misspecified forecast is smaller than the true forecasting distribution. 
If for example $X_{i,h}$ of the forecast $\bsX_i$ has a variance that is too large then $\what{u}_{\bsY_i,h}(\XX_i)$ takes too many values around the center (usually around 0.5) and too less values in
the tails around $0$ and $1$. 
This is a problem for practical application and  restricts the potential applications of $\what{\bsu}_{\bsY_i}$ to the comparison of forecasts with the same marginals which is far away from practical needs.

Fortunately, there is a relatively simple way to avoid the problem with the misspecified marginals. % in the forecast $\bsX_i$. 
% Remember, that $\bsu_{\bsY_i}$ is a copula observation, so the realization of random variable with uniform marginals.
% If $\bsX_i$ has misspecified marginals then $\what{\bsu}_{\bsY_i}(\XX_i)$ has misspecified marginals as well, so $\what{u}_{\bsY_i,h}(\XX_i)$ is not uniformly distributed on $(0,1)$. 
% This is somehow a contradiction of the evaluation idea that we what to compare the forecasted copula with the observed copula observations. But if the observed copula observations 
% are not a draws from a copula copula due to the estimation of $F_{Y_{i,h}}$ by $F_{X_{i,h}}$ we expect problems.
% If for example $X_{i,h}$ has a variance that is too large then $\what{u}_{\bsY_i,h}(\XX_i)$ takes too many values around the center (usually around 0.5) and too less values in
% the tails around $0$ and $1$. 
% To avoid the mentioned problem 
We can force the misspecified marginal distribution to be a (plausible) draw from a copula, so that the general monotonic ordering across is preserved,
but such that they have uniform marginals. %However, for such an adjustment we have to know about the degree of adjustment. 
To perform such an adjustment we have to learn about the distribution of the estimated
copula observations $\what{\bsu}_{\bsY_i}$. Given a single $\what{\bsu}_{\bsY_i}$ this is impossible. Therefore, we explore the distributional structure across 
the full out-of-sample period $\what{\bsu}_{\bsY_1},\ldots, \what{\bsu}_{\bsY_N}$. The most obvious adjustment is to adjust the 
copula observations on each marginal $\what{u}_{\bsY_1,h},\ldots, \what{u}_{\bsY_N,h}$ so that they follow a perfect uniform distribution on $(0,1)$.
This can be easily done by ranking the estimated observations $\what{u}_{\bsY_1,h},\ldots, \what{u}_{\bsY_N,h}$ for each $h$. 
Hence, denote $R_{i,h}$ the rank of $\what{u}_{\bsY_i,h}$ within $\what{u}_{\bsY_1,h},\ldots, \what{u}_{\bsY_N,h}$. 
Then we define the \emph{adjusted estimated copula observations} by
\begin{align}
\what{u}^{*}_{\bsY_i,h} = \frac{2R_{i,h} - 1}{2N} . 
\label{eq_est_copula_obs_adj} 
\end{align}
Note that for the ranks $R_{i,h}$ there can appear ties due to the stepwise structure of the ecdf.
To preserve the optimal distribution, we highly suggest break the ties in the ranks at random (uniform sampling). 
% , so we properly have $r_i \sim \UU( \{ r | \what{u}_{\bsY_1,h,(r)} = \what{u}_{\bsY_i,h} \})$.
This ranking procedure guarantees that $\what{u}^{*}_{\bsY_1,h},\ldots, \what{u}^{*}_{\bsY_N,h}$ have perfect uniform marginals, i.e. they take the values $\frac{1}{2N},\ldots, \frac{2N-1}{2N}$.
Perfect means here that for these values the corresponding ecdf minimizes the Kolmogorov-Smirnov distance (but also the L\'evy distance or L\'evy metric which characterizes convergence in distribution) 
with respect to the uniform distribution. This procedure assumes that across the sample the comonotonicity within the implied copula structure is preserved.

% \FZ{ not sure
% 
% }

For the forecasted copula $\bsC_{\bsX_i}$ of $\bsX_i$ we simply consider the empirical copula of the ensemble $\XX_i$ with elements $\bsx_i^{(j)} = (x_{i,1}^{(j)},\ldots, x_{i,H}^{(j)})'$
as a corresponding estimator.
Here we suggest to consider
\begin{equation}
\what{\bsC}_{\bsX_i} (u_1,\ldots, u_H) = \frac{1}{M} \sum_{j=1}^M \mathbbm{1}\{ \wtilde{R}_{i,j,1}/M \leq u_1, \ldots, \wtilde{R}_{i,j,H}/M \leq u_H \} 
\label{eq_est_copula}
\end{equation}
with the ranks
$$ \wtilde{R}_{i,j,h} = \frac{1}{2} \sum_{k=1}^M \mathbbm{1}\{ x_{i,h}^{(k)} \leq x_{i,h}^{(j)}  \} + \mathbbm{1}\{ x_{i,h}^{(k)} < x_{i,h}^{(j)}  \}$$
where we consider the mid-point rule for the ecdf. Similarly as above potential ties should be broken at random (uniform sampling).

% estimate, simple replacing by marginals - only ok if marginals correctly estimated. If not, force to be copula 
% depended on the full window(!). [Otherwise no chance, as we can not discriminate between misspecification and dependency evaluation with a single observation.]

\subsection{Marginal copula score}

Whenever, we want to estimate a marginal copula score \eqref{eq_msc} we apply the plug-in principle.
Still, remember that \eqref{eq_msc} has a multiplicative structure and it holds $\E(X)\E(Y) =\E( XY) - \cov(X,Y)$.
Thus, we get the plug-in estimator 
\begin{align}
\what{\MS(\bsa)\text{-}\CS}_i =  \what{\MS}_i(\bsa)  \what{\CS}_i - \what{\sigma}_{\MS,\CS} %= \CS \sum_{h=1}^H a_h \MS_h \ \ .
\label{eq_msc_est}
\end{align}
where $\what{\MS}_i(\bsa)$ and  $\what{\CS}_i$ are the sample means across the marginal and copula score. 
$\what{\sigma}_{\MS,\CS} $ is the covariance between $\MS$ and $\CS$ across $i$. Here we assume implicitly some covariance stationarity across the out-of-sample window study
between the marginal and copula score.

For applications the estimated marginal scores $\what{\MS}_i$ can be estimated by the CRPS and the copula score can be 
either the estimated energy score $\what{\ES}_{i,\beta}$, the estimated variogram score $\what{\VS}_{i,\bsW, p}$ or the estimated Dawid-Sebastiani score $\what{\DSS}_{i}$
of the copula observations and the copula of the forecast.
Here, we suggest the estimates $\what{u}^{*}_{\bsY_i,h}$ (eqn. \eqref{eq_est_copula_obs_adj}) and $\what{\bsC}_{\bsX_i}$ (eqn. \eqref{eq_est_copula}) for the copula observations and the copula as derived in the previous section.
As the Copula Dawid-Sebastiani score $\what{\DSS}_{i}$ is not bonded from below, the resulting scoring rule is not strictly proper. In fact, it results in a useless rule
for applications. 

 % The definition \eqref{eq_energy_score_i} of the energy score involves an expectation which is in practice unknown.

\section{Evaluation} \label{sec:eval}

Let $\SC_i$ be a score of the $i$-th forecasting experiment for $i=1,\ldots,N$ of the considered forecasting model. Further let $\SC^*$ be the score of the true model, so $\bsY \sim \bsF_{\bsX}$
with draws $\bsy$.

All covered (strictly) proper scoring rules are negatively oriented. Thus, it holds: the smaller the score the better the forecasting performance.
Obviously, for practical application the corresponding sample mean
\begin{equation}
\ov{\SC} = \frac{1}{N} \sum_{i=1}^N \SC_i . 
\label{eq_score_average} 
\end{equation}
will be the most relevant criterion when comparing the predictive performance of two forecasts.

In this section we shortly introduce two further evaluation methods, based on the scores. 
First, we consider the relative change in scores as used for deriving conclusions in \cite{pinson2013discrimination}.
Secondly, we consider the  Diebold-Mariano test, which allows for significance statements.

\subsection{Relative change in scores}

In \cite{pinson2013discrimination} the relative change in the score with respect to the best forecast is considered:
\begin{equation}
 \RelCh(\SC) = \frac{\ov{\SC} - \ov{\SC}^*}{\ov{\SC}^*} 
 \label{eq_relch}
\end{equation}
where $\ov{\SC}^* = 1/N \sum_{i=1}^N \SC^*_i$.
Obviously, it holds $\RelCh(\SC^*) = 0$.
The idea is to measure the sensitivity in the scores with respect to some biased non-optimal forecast in a relative manner.

\subsection{Significance evaluation: the Diebold-Mariano test}

The Diebold-Mariano (DM) test was designed for the point forecast evaluation in \cite{diebold1995comparing}. 
However, as pointed out in e.g. \cite{diebold2015comparing} the test design is very general and allows for several generalizations.
For instance, in \cite{ziel2018day}  and\cite{uniejewski2018variance} it was applied in multivariate settings. 
In \cite{moller2015spatially} it is considered in an application for the energy score.

The DM-test requires a (pseudo-)out-of-sample forecasting study which aims to forecast $\bsY_{1},\ldots,\bsY_{N}$  by  $\bsX_{1},\ldots,\bsX_{N}$
as described in section \ref{sec:reporting}. Its target is to compare the forecasting accuracy of two forecasts $\A$ and $\B$ on the same (pseudo-)out-of-sample environment
based on a scoring rule $\SC$.

The DM-test checks whether the scores $\ov{\SC}^{\A}$ is significantly different from $\ov{\SC}^{\B}$, see \eqref{eq_score_average}.
Therefore, score differences are required.
Given the losses $\SC_1^{\A}, \ldots, \SC_N^{\A}$ and $\SC_1^{\B}, \ldots, \SC_N^{\B}$, we define the loss differences by 
$$\Delta_i^{\A,\B} = \SC_i^{\A} - \SC_i^{\B}.$$
% For the energy score, this is
% $$\Delta^{\A,\B}_i = \text{ES}^{\A}_{i,\beta} - \text{ES}^{\B}_{i,\beta}$$
% where $\text{ES}^{\A}_{i,\beta}$ and $\text{ES}^{\B}_{i,\beta}$ are the losses of the energy score of forecast $A$ and $B$. 
Now, the key idea of the DM-test is to check if the mean loss (or score) difference 
$$\ov{\Delta}^{\A,\B} 
= \frac{1}{N} \sum_{i=1}^N \Delta^{\A,\B}_i$$ is significantly 
different from zero or not.

 Remember, scoring rules $\SC_i$ have a certain direction, so that either a positive loss or a negative loss represents a 
 better forecasting accuracy.  As we consider only negatively oriented scores, 
 a $\ov{\Delta}^{\A,\B}$ which is significantly smaller than zero lead to the conclusion that $\A$ is significantly better than $\B$ with respect to the considered scoring rule.

% This direction is important for the DM-test design. For the energy score as defined in \eqref{eq_energy_score}
% a smaller score indicates better forecasting performance, as well as the mentioned measure MAE, MSE, or the quantile/pinball loss.\footnote{Note that if the energy score is used as introduces by \cite{gneiting2007strictly} then the design is going to be inverted.}

% Remember, we are interested in showing that $\ov{\Delta}^{\A,\B}$ is significantly different from zero.
% The problem that arises right now is that strictly speaking this only possible in limited settings where the $\Delta_i^{\A,\B}$ share specific properties.
%$\Delta_i^{\A,\B}$ is the loss difference for the target $\bsX_{i}$. 

In general, the distribution of $\bsX_{i}$ is different from $\bsX_{j}$ for $i\neq j$, 
in the consequence of the distribution
% in  which consequence the distribution 
of $\Delta_i^{\A,\B}$ is usually different from $\Delta_j^{\A,\B}$ as well. 
So $\ov{\Delta}^{\A,\B}$ is the sum of $N$ different distributions. Moreover, 
 $\Delta_i^{\A,\B}$ and $\Delta_j^{\A,\B}$ exhibit a specific dependency structure.
  Usually they are not even independent or uncorrelated. 
Hence, from the statistical point of view
it is not possible to make further statements about $\ov{\Delta}^{\A,\B}$ unless we make some assumptions for 
the sequence $(\Delta_i^{\A,\B})_{i\in \Z}$ (or $\Delta_1^{\A,\B}, \ldots, \Delta_N^{\A,\B}$).
The standard assumptions in the DM-test are:
\begin{enumerate}
 \item $\E( \Delta_i^{\A,\B} ) = \mu_{\A,\B} $ for all $i$
 \item $\cov( \Delta_i^{\A,\B}, \Delta_{i-k}^{\A,\B} ) = \gamma_{\A,\B}(k) $ for all $i$ and $k\geq 0$
%  \item $\var( \Delta_i^{\A,\B} ) = \sigma_{\A,\B}^2 $ for all $i$ 
\end{enumerate}
Thus, $(\Delta_i^{\A,\B})_{i\in \Z}$ is a weakly stationary (or covariance stationary) process. 
These assumptions are very restrictive, and can be relaxed substantially.
Still, under these assumption given above it possible to show that 
$\ov{\Delta}^{\A,\B}$ is asymptotic normal \cite{diebold2015comparing}, i.e. it holds for $N\to \infty$ that

\begin{equation}
 \frac{\ov{\Delta}^{\A,\B} - \mu_{\A,\B} }{ \sigma(\ov{\Delta}^{\A,\B}) } \to N(0,1) 
\label{eq_null} 
\end{equation}
in distribution for where $\sigma(\ov{\Delta}^{\A,\B}) = \sqrt{\gamma_{\A,\B}(0) }$ is the standard deviation of $\ov{\Delta}^{\A,\B}$. 
This allows the creation of asymptotic tests, namely the Diebold-Mariano test with the null hypothesis $\text{H}_0: \mu_{\A,\B} = 0$. When replacing $\sigma(\ov{\Delta}^{\A,\B})$
with a suitable estimator, e.g. the sample standard deviation $\what{\sigma}(\ov{\Delta}^{\A,\B})$ the same distributional limit is attained.

Note that the DM-test is a \emph{asymptotic} test, as a consequence the out-of-sample window length $N$ should be as large as possible. 
% Otherwise the test is not applicable.
In the next section we also study in more detail the impact of the simulation sample size $M$ of the proposed evaluation procedure.

 \section{Simulation Studies} \label{sec:studies}

In the subsequent subsections we will carry out some simulation studies to compare the forecasting measure in various forecasting situations.
These experiment are designed with $N$ replications (or instances) indexed by $i$ as described in the reporting section.
For evaluation we will consider always 9 different multivariate measures:
\begin{enumerate}
 \item Energy score: $\ES_{i,1}$, estimated by $ \what{\ES}^{K\text{band}}_{i,1} $ with $K=1$
\item Variogram score: $\VS_{i,\bsone \bsone', 1}$, estimated by $\what{\VS}_{i,\bsone \bsone', 1}$
\item Dawid-Sebastiani score: $\DSS_{i}$, estimated by $\what{\DSS}_i$
\item CRPS-copula energy score: ${(\CRPS(\frac{1}{H}\bsone)\text{-}\CES_{1})}_i$, estimated by \eqref{eq_msc_est} with $\what{\CRPS}_{i}(\frac{1}{H}\bsone)$ and $\what{\CES}^{K\text{band}}_{i,1}$ with $K=1$
\item CRPS-copula variogram score: $\CRPS(\frac{1}{H}\bsone)\text{-}\CVS_{i,\bsone \bsone', 1}$, estimated by \eqref{eq_msc_est} with $\what{\CRPS}_{i}(\frac{1}{H}\bsone)$ and $\what{\CVS}_{i,\bsone \bsone', 1}$
% \item CRPS-copula Dawid-Sebastiani score: $\CRPS(\frac{1}{H}\bsone)\text{-}\CDSS_{i}$, estimated by \eqref{eq_msc_est} with $\what{\CRPS}_{i}(\frac{1}{H}\bsone)$ and $\what{\CDSS}_i$
\item Continuous ranked probability score: $\CRPS_i$, estimated by $\what{\CRPS}_{i}(\frac{1}{H}\bsone)$.
\item Copula energy score: $\CES_{i,1}$, estimated by $\what{\CES}^{K\text{band}}_{i,1}$ with $K=1$
\item Copula variogram score: $\CVS_{i,\bsone \bsone', 1}$, estimated by $\what{\CVS}_{i,\bsone \bsone,1}$
\item Copula Dawid-Sebastiani score: $\CDSS_{i}$, estimated by $\what{\CDSS}_{i}$
\end{enumerate}
The first three are the established multivariate forecasting criterion. The next two are marginal-copula scores with the CRPS as marginal score. 
The sixth score is the CRPS itself, which evaluated only the marginals. The latter three ones only evaluate the dependency structure using the copulas.

% $ \what{\ES}^{K\text{band}}_{i,\beta} $
 
\subsection{Sensitivity study I}

Here we replicate the study of \cite{pinson2013discrimination} and enlarge the study design to some extend. This study was the reason for \cite{pinson2013discrimination}
to conclude that the energy score has bad discrimination abilities with respect to the dependency behavior.
We consider a bivariate normal distribution with zero mean and variance of $1$ in each component.
So the true model is $\bsY \sim \NN_2( \bsmu, \bsSigma(\rho) )$ with $\bsmu= (0,0)'$ and $\bsSigma(\rho) = \left( \begin{array}{cc}
                                                                                                 1 & \rho \\ \rho & 1
                                                                                                \end{array} \right)$ with correlation $\rho$.
In the simulation setup we choose the true $\rho$ on a grid $\{-1,-0.8, \ldots, 0.8, 1\}$.
 For the forecasting model we consider $\bsY \sim \NN_2( \bsmu, \bsSigma(\varrho) )$. Here we choose the 
 forecasted correlation $\varrho$ out of a slightly denser grid $\{-1,-0.9, -0.8 \ldots, 1\}$.

\begin{figure}[h!]
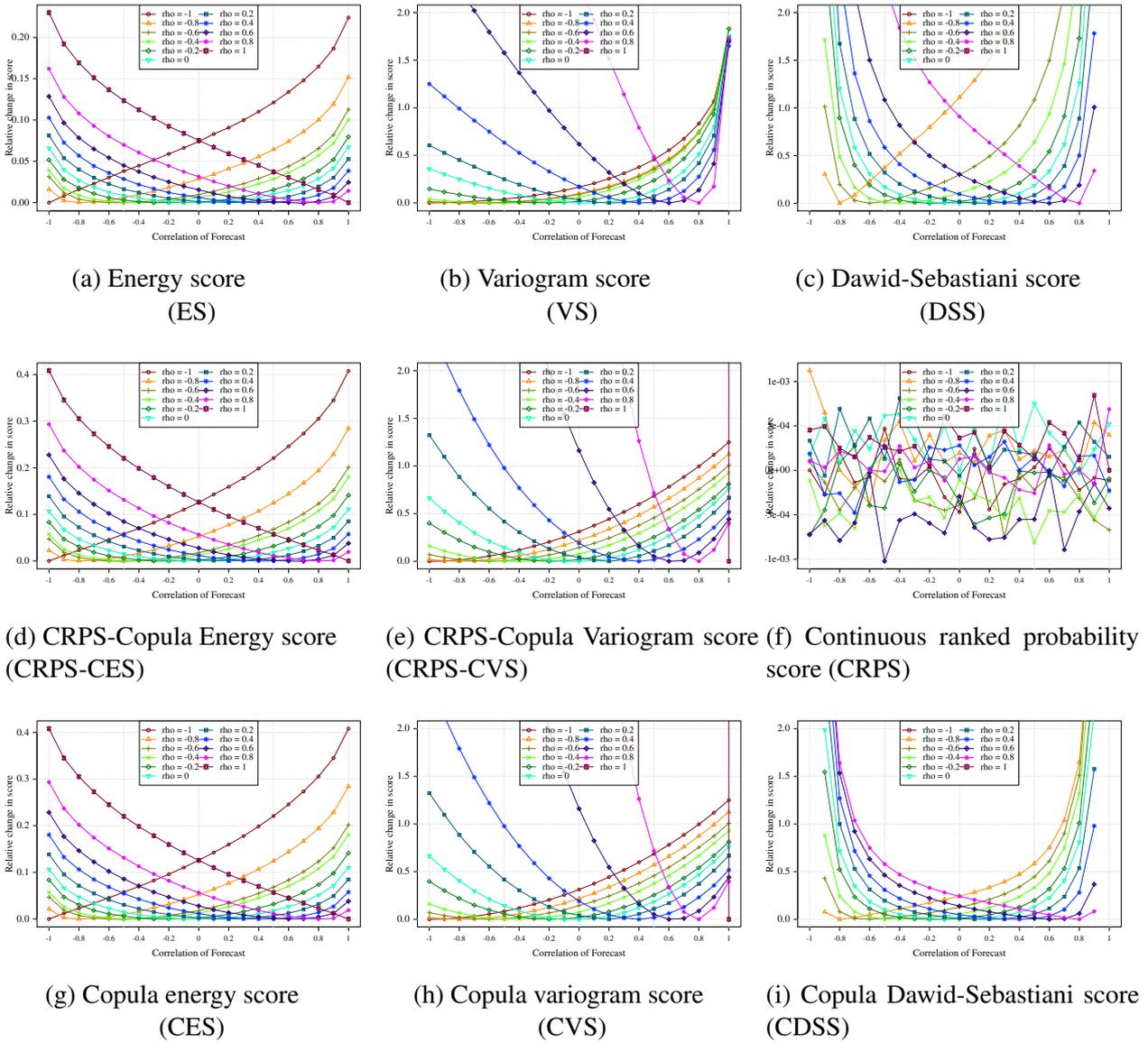

\begin{subfigure}[b]{0.32\textwidth}
 \resizebox{1.00\textwidth}{!}{\input{fig/pinsonstudy/rel_nY=512_nX=16384_Score=1.tex}}
        \caption{Energy score \newline (ES)}
\label{fig_sim_study_pinson_relch1}
    \end{subfigure}
\begin{subfigure}[b]{0.32\textwidth}
\resizebox{1.00\textwidth}{!}{\input{fig/pinsonstudy/rel_nY=512_nX=16384_Score=2.tex}}
        \caption{Variogram score \newline(VS)}
\label{fig_sim_study_pinson_relch2}
    \end{subfigure}
\begin{subfigure}[b]{0.32\textwidth}
\resizebox{1.00\textwidth}{!}{\input{fig/pinsonstudy/rel_nY=512_nX=16384_Score=3.tex}}
        \caption{Dawid-Sebastiani score \newline (DSS)}
\label{fig_sim_study_pinson_relch3}
    \end{subfigure}

    % \resizebox{.32\textwidth}{!}{\input{fig/pinsonstudy/rel_nY=512_nX=16384_Score=1.tex}}
\begin{subfigure}[b]{0.32\textwidth}
\resizebox{1.00\textwidth}{!}{\input{fig/pinsonstudy/rel_nY=512_nX=16384_Score=4.tex}}
        \caption{CRPS-Copula Energy score \newline (CRPS-CES)}
\label{fig_sim_study_pinson_relch4}
    \end{subfigure}
\begin{subfigure}[b]{0.32\textwidth}
\resizebox{1.00\textwidth}{!}{\input{fig/pinsonstudy/rel_nY=512_nX=16384_Score=5.tex}}
\caption{CRPS-Copula Variogram score (CRPS-CVS)}
\label{fig_sim_study_pinson_relch5}
\end{subfigure}
\begin{subfigure}[b]{0.32\textwidth}
\resizebox{1.00\textwidth}{!}{\input{fig/pinsonstudy/rel_nY=512_nX=16384_Score=7.tex}}
        \caption{Continuous ranked probability score (CRPS)}
\label{fig_sim_study_pinson_relch7}
    \end{subfigure}

\begin{subfigure}[b]{0.32\textwidth}
\resizebox{1.00\textwidth}{!}{\input{fig/pinsonstudy/rel_nY=512_nX=16384_Score=8.tex}}
        \caption{Copula energy score \newline (CES)}
\label{fig_sim_study_pinson_relch8}
    \end{subfigure}
\begin{subfigure}[b]{0.32\textwidth}
\resizebox{1.00\textwidth}{!}{\input{fig/pinsonstudy/rel_nY=512_nX=16384_Score=9.tex}}
        \caption{Copula variogram score \newline (CVS)}
\label{fig_sim_study_pinson_relch9}
    \end{subfigure}
\begin{subfigure}[b]{0.32\textwidth}
\resizebox{1.00\textwidth}{!}{\input{fig/pinsonstudy/rel_nY=512_nX=16384_Score=10.tex}}
        \caption{Copula Dawid-Sebastiani score  (CDSS)}
\label{fig_sim_study_pinson_relch10}
    \end{subfigure}
\caption{Relative changes $\RelCh$ (see \eqref{eq_relch})
for all considered scoring rules for the extended simulation study of \cite{pinson2013discrimination}
with an ensemble sample size of $M=2^{14}=16384$ and a rolling window length of $N=2^9=512$.}
\label{fig_sim_study_pinson_relch}
\end{figure}

\begin{figure}[h!]
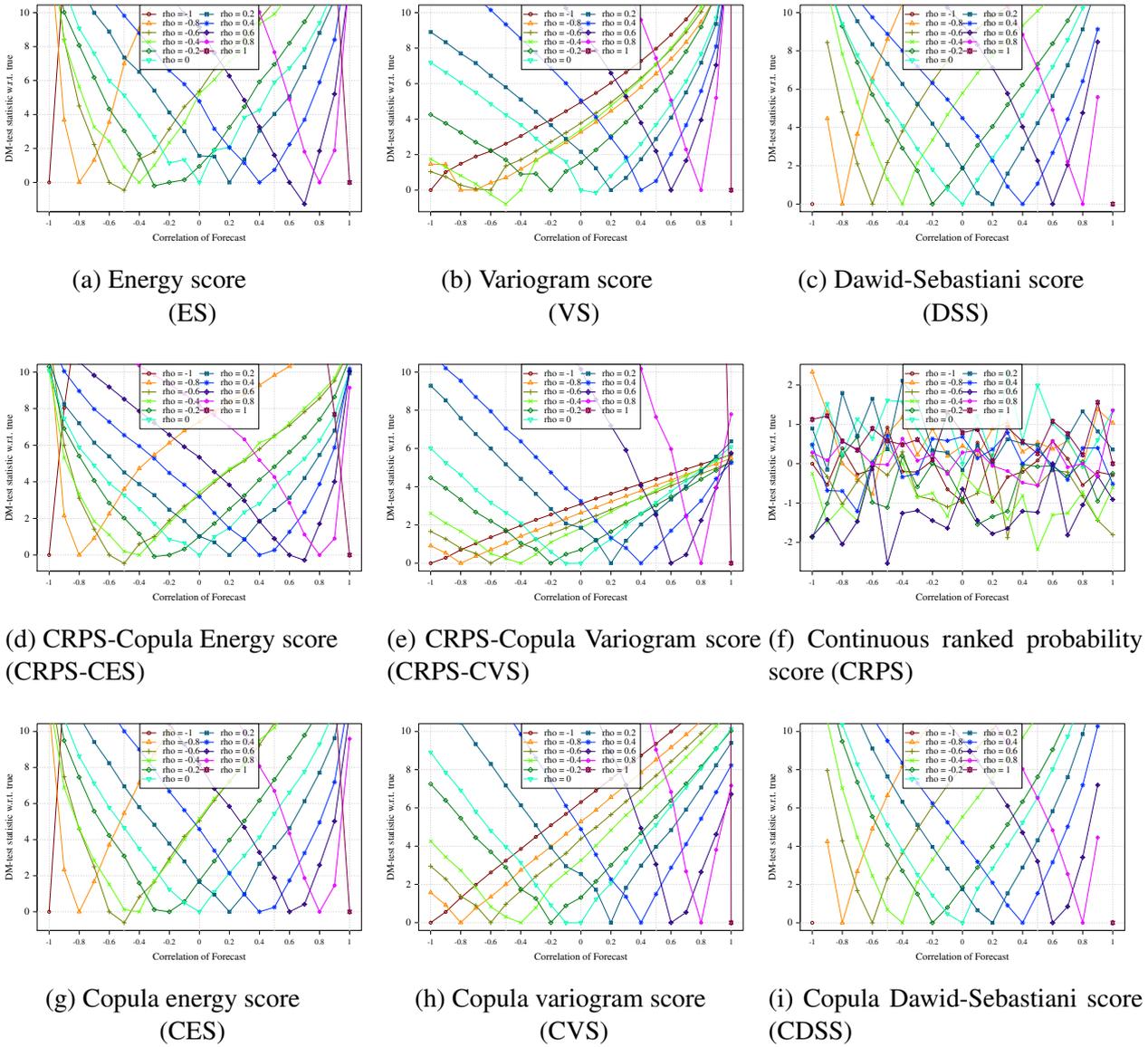

\begin{subfigure}[b]{0.32\textwidth}
 \resizebox{1.00\textwidth}{!}{\input{fig/pinsonstudy/dm_nY=512_nX=16384_Score=1.tex}}
        \caption{Energy score \newline (ES)}
\label{fig_sim_study_pinson_dm1}
    \end{subfigure}
\begin{subfigure}[b]{0.32\textwidth}
\resizebox{1.00\textwidth}{!}{\input{fig/pinsonstudy/dm_nY=512_nX=16384_Score=2.tex}}
        \caption{Variogram score \newline(VS)}
\label{fig_sim_study_pinson_dm2}
    \end{subfigure}
\begin{subfigure}[b]{0.32\textwidth}
\resizebox{1.00\textwidth}{!}{\input{fig/pinsonstudy/dm_nY=512_nX=16384_Score=3.tex}}
        \caption{Dawid-Sebastiani score \newline (DSS)}
\label{fig_sim_study_pinson_dm3}
    \end{subfigure}

    % \resizebox{.32\textwidth}{!}{\input{fig/pinsonstudy/dm_nY=512_nX=16384_Score=1.tex}}
\begin{subfigure}[b]{0.32\textwidth}
\resizebox{1.00\textwidth}{!}{\input{fig/pinsonstudy/dm_nY=512_nX=16384_Score=4.tex}}
        \caption{CRPS-Copula Energy score \newline (CRPS-CES)}
\label{fig_sim_study_pinson_dm4}
    \end{subfigure}
\begin{subfigure}[b]{0.32\textwidth}
\resizebox{1.00\textwidth}{!}{\input{fig/pinsonstudy/dm_nY=512_nX=16384_Score=5.tex}}
\caption{CRPS-Copula Variogram score (CRPS-CVS)}
\label{fig_sim_study_pinson_dm5}
\end{subfigure}
\begin{subfigure}[b]{0.32\textwidth}
\resizebox{1.00\textwidth}{!}{\input{fig/pinsonstudy/dm_nY=512_nX=16384_Score=7.tex}}
        \caption{Continuous ranked probability score (CRPS)}
\label{fig_sim_study_pinson_dm7}
    \end{subfigure}

\begin{subfigure}[b]{0.32\textwidth}
\resizebox{1.00\textwidth}{!}{\input{fig/pinsonstudy/dm_nY=512_nX=16384_Score=8.tex}}
        \caption{Copula energy score \newline (CES)}
\label{fig_sim_study_pinson_dm8}
    \end{subfigure}
\begin{subfigure}[b]{0.32\textwidth}
\resizebox{1.00\textwidth}{!}{\input{fig/pinsonstudy/dm_nY=512_nX=16384_Score=9.tex}}
        \caption{Copula variogram score \newline (CVS)}
\label{fig_sim_study_pinson_dm9}
    \end{subfigure}
\begin{subfigure}[b]{0.32\textwidth}
\resizebox{1.00\textwidth}{!}{\input{fig/pinsonstudy/dm_nY=512_nX=16384_Score=10.tex}}
        \caption{Copula Dawid-Sebastiani score  (CDSS)}
\label{fig_sim_study_pinson_dm10}
    \end{subfigure}

\caption{DM-test statistics with corresponding p-value given in squared brackets with respect to the true model for all considered scoring rules for the extended simulation study of \cite{pinson2013discrimination}
with an ensemble sample size of $M=2^{14}=16384$ and a rolling window length of $N=2^9=512$.}
\label{fig_sim_study_pinson_dm}
\end{figure}

 In the experiment we draw $M=2^{14}=16384$ times for a single bivariate forecast. The window length is $N=2^9=512$.
Then we compute the relative change in the score and the DM-test statistic with respect to the true model. Note 
that \cite{pinson2013discrimination} evaluated only the relative change.
The results are given in Figures \ref{fig_sim_study_pinson_relch} and \ref{fig_sim_study_pinson_dm}.

In Figures \ref{fig_sim_study_pinson_relch} we observe the same pattern for the energy score as reported in \cite{pinson2013discrimination}.
The relative change in the score is quite small. For instance, if the true correlation is $-0.8$ and the forecasted one is $0.3$, then 
the relative change is only 5\% (Fig. \ref{fig_sim_study_pinson_relch1}). In this situation, the variogram score yields a relative change of about 25\% (Fig. \ref{fig_sim_study_pinson_relch2}) and the Dawid-Sebastiani score
even a relative change of approximately 175\% (Fig. \ref{fig_sim_study_pinson_relch3}). However, the latter score is designed for evaluating multivariate normal distributions. Thus, the high sensitivity is not surprising. 
Further, the energy score and the Dawid-Sebastiani score exhibit symmetric evaluation behavior for positive and negative correlations. In contrast, the variogram score 
reacts more sensitive for positive correlations than for negative ones.
The CRPS-copula energy score (CRPS-CES) was designed to overcome this issue of the energy score. Indeed, the relative changes are substantially larger than 
the energy score values. In the mentioned example the relative change is about 12\% (Fig. \ref{fig_sim_study_pinson_relch4}), thus more than twice as sensitive in the relative change. 
Similarly, we observe an increase of the relative change of the CRPS-copula variogram score (CRPS-CVS) score when comparing with the variogram score (Fig. \ref{fig_sim_study_pinson_relch5}). However, we want to highlight that the relative change is not a suitable measure to evaluate if a scoring rule can discriminate well a suboptimal forecast and the optimal one. Hence, e.g. the CRPS-CES is not necessarily better than the energy score itself in doing so.
Obviously, the CRPS can not evaluate any change in the correlation structure and has tiny relative changes which are not significantly different from zero.
This is also a reason why, the curve pattern for the energy score look very similar to the CRPS-copula energy score and copula energy score
(Fig. \ref{fig_sim_study_pinson_relch1}, \ref{fig_sim_study_pinson_relch4}, \ref{fig_sim_study_pinson_relch8}), and analogously for the variogram score.
For the copula Dawid-Sebastiani score we see smaller relative changes than for the Dawid-Sebastiani score. This matches again the fact that 
Dawid-Sebastiani score should be optimal in the considered situation.

Let us turn to Figure \ref{fig_sim_study_pinson_dm} and DM-test statistic results. We capped the statistics at 10 to improve the interpretability of the graphs.
First, we observe that all scoring rules except the CRPS (Fig. \ref{fig_sim_study_pinson_dm7}) can identify the correct model.
Due to the normal distribution setting it is not surprisingly that the Dawid-Sebastiani score (Fig. \ref{fig_sim_study_pinson_dm3}) yields the 
largest test statistics, among all measures. Still, the related copula Dawid-Sebastiani score (Fig. \ref{fig_sim_study_pinson_dm10}) has only slightly 
smaller DM-test statistics, despite the observation on the relative changes of Figure \ref{fig_sim_study_pinson_relch}.
When comparing the energy score, the copula energy score and the CRPS-copula energy score (Fig. \ref{fig_sim_study_pinson_dm1}, Fig. \ref{fig_sim_study_pinson_dm4}, Fig. \ref{fig_sim_study_pinson_dm8}) we observe that the largest test statistics for the energy score, followed by the copula energy score, followed by the CRPS-copula 
energy score. Thus, among these three scoring rules the energy score seems to be the most sensitive with respect to the DM-test.
For the variogram score this corresponding ordering slightly changes (Fig. \ref{fig_sim_study_pinson_dm2}, Fig. \ref{fig_sim_study_pinson_dm5}, Fig. \ref{fig_sim_study_pinson_dm9}). So the copula variogram score is more sensitive than the variogram score 
which is more sensitive than the CRPS-copula variogram score.
Hence, we focus on the comparison between the energy score, the variogram score and the Dawid-Sebastiani score (Fig. \ref{fig_sim_study_pinson_dm1}, Fig. \ref{fig_sim_study_pinson_dm2}, Fig. \ref{fig_sim_study_pinson_dm3}).
Let us consider a true correlation of 0. In this simulation results, the energy score gives DM-statistics approximately 4 or greater
if for the forecasted correlation $\rho$ it holds $\rho \leq -0.5$ or $\rho \geq 0.4$. In contrast, the more sensitive Dawid-Sebastiani score
requires $\rho \leq -0.4$ or $\rho \geq 0.4$ for a DM-statistic above 4. This somewhat surprising result shows, that the energy score is almost as 
powerful as the Dawid-Sebastiani score in discriminating the forecasts. 
For the variogram score we receive a similar picture to the energy score, we need $\rho \leq -0.5$ or $\rho \geq 0.6$ for a DM-statistic above 4.
Still, the variogram score has a weaker performance than the energy score.
Basically the same results hold for other correlations as well.
However, as the variogram score is more sensitive to larger correlations than to smaller ones, the results improve for the variogram score when 
larger correlations are considered.

\subsection{Sensitivity study II}

This study is based on the study from \cite{scheuerer2015variogram} but it is also to some extend adjusted. % but to some extend adjusted.
Again, we consider the multivariate normal distribution case, more precisely the bivariate normal distribution with zero mean, variance equal to one and 
a correlation of $\rho = \sqrt{2}/2 \approx 0.707$. 
Now, we bias this underlying distribution and analyze the change in the score. The major difference to the previous study is that we analyze different changes and control the magnitude of the change, whereas in \cite{scheuerer2015variogram} these changes were chosen ad hoc. 
With $\bsmu= (0,0)'$ and $\bsSigma(\rho) = \left( \begin{array}{cc}
                                                                                                 1 & \rho \\ \rho & 1
                                                                                                \end{array} \right)$ we define the following prediction models:
\begin{enumerate}
 \item (true setting): $\bsX \sim \NN_2( \bsmu, \bsSigma)$ with $\rho = \sqrt{2}/2$
 \item (symmetric mean bias): $\bsX \sim \NN_2( \bsmu + a_1\bsone, \bsSigma(\rho))$  with $\rho = \sqrt{2}/2$
 \item (asymmetric mean bias): $\bsX \sim \NN_2( \bsmu + (a_2,-a_2)', \bsSigma(\rho))$   with $\rho = \sqrt{2}/2$
 \item (smaller variance): $\bsX \sim \NN_2( \bsmu, a_3\bsSigma(\rho))$ with $a_3<1$ and $\rho = \sqrt{2}/2$
 \item (larger variance): $\bsX \sim \NN_2( \bsmu, a_4\bsSigma(\rho))$ with $a_4>1$ and  $\rho = \sqrt{2}/2$
 \item (smaller correlation): $\bsX \sim \NN_2( \bsmu, \bsSigma(a_5))$ with $a_5<\rho$ 
 \item (larger correlation): $\bsX \sim \NN_2( \bsmu, \bsSigma(a_6))$ with $a_6>\rho$  %"Mean sym.", "Mean asym.", "Var sm.", "Var gr.", "Cor sm.", "Cor gr.)
\end{enumerate}
We determine $a_i$ so that the likelihoods of all settings except the true coincide. As we have one degree of freedom we choose $a_5=0$. This corresponds to a likelihood reduction with respect
to the true model of $\delta = \frac{1}{2}\log(2)$.
Then, the computation of the remaining $a_i$'s is simple, for the first two ones we even have explicit solutions. 
They are given by $a_1 = \sqrt{ \frac{\delta}{2-\sqrt{2}} }$,  $a_2 = \sqrt{ \frac{\delta}{2+\sqrt{2}} }$, $a_3\approx 0.48124$, $a_4\approx 2.62729$ and $a_6 \approx 0.89032$.

\begin{figure}[h!]
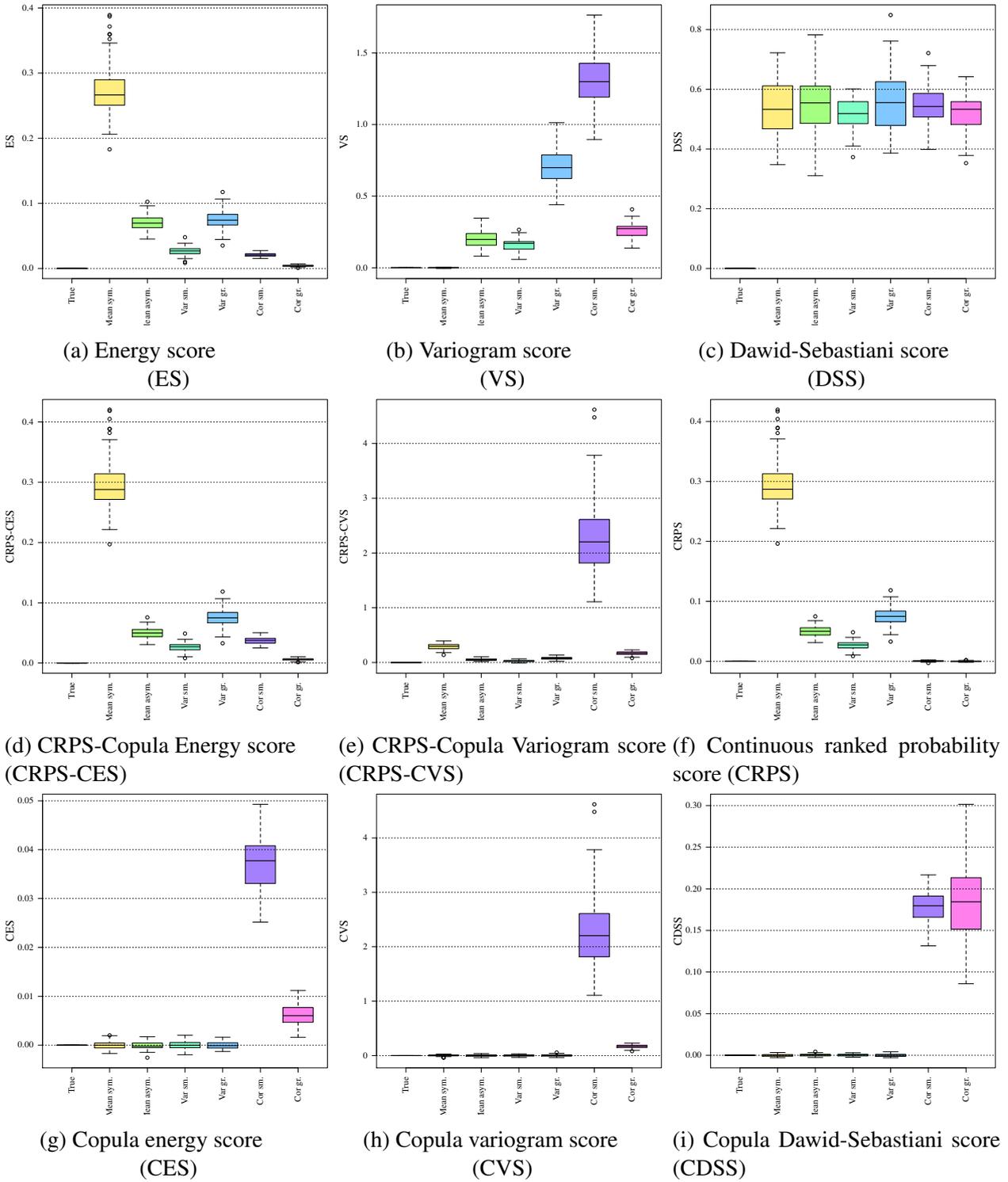

\begin{subfigure}[b]{0.32\textwidth}
\resizebox{1.00\textwidth}{!}{\input{fig/variostudy/rel_nEns=64nY=256nFC=8192_s=1.tex}}
        \caption{Energy score \newline (ES)}
\label{fig_sim_study_sens_relch1}
    \end{subfigure}
\begin{subfigure}[b]{0.32\textwidth}
\resizebox{1.00\textwidth}{!}{\input{fig/variostudy/rel_nEns=64nY=256nFC=8192_s=2.tex}}
        \caption{Variogram score \newline (VS)}
\label{fig_sim_study_sens_relch2}
    \end{subfigure}
\begin{subfigure}[b]{0.32\textwidth}
\resizebox{1.00\textwidth}{!}{\input{fig/variostudy/rel_nEns=64nY=256nFC=8192_s=3.tex}}
        \caption{Dawid-Sebastiani score \newline (DSS)}
\label{fig_sim_study_sens_relch3}
    \end{subfigure}

    \begin{subfigure}[b]{0.32\textwidth}
\resizebox{1.00\textwidth}{!}{\input{fig/variostudy/rel_nEns=64nY=256nFC=8192_s=4.tex}}
        \caption{CRPS-Copula Energy score \newline (CRPS-CES)}
\label{fig_sim_study_sens_relch4}
    \end{subfigure}
\begin{subfigure}[b]{0.32\textwidth}
\resizebox{1.00\textwidth}{!}{\input{fig/variostudy/rel_nEns=64nY=256nFC=8192_s=5.tex}}
\caption{CRPS-Copula Variogram score (CRPS-CVS)}
\label{fig_sim_study_sens_relch5}
    \end{subfigure}
\begin{subfigure}[b]{0.32\textwidth}
\resizebox{1.00\textwidth}{!}{\input{fig/variostudy/rel_nEns=64nY=256nFC=8192_s=7.tex}}
        \caption{Continuous ranked probability score (CRPS)}
\label{fig_sim_study_sens_relch7}
    \end{subfigure}

    \begin{subfigure}[b]{0.32\textwidth}
\resizebox{1.00\textwidth}{!}{\input{fig/variostudy/rel_nEns=64nY=256nFC=8192_s=8.tex}}
        \caption{Copula energy score \newline (CES)}
\label{fig_sim_study_sens_relch8}
    \end{subfigure}
\begin{subfigure}[b]{0.32\textwidth}
\resizebox{1.00\textwidth}{!}{\input{fig/variostudy/rel_nEns=64nY=256nFC=8192_s=9.tex}}
        \caption{Copula variogram score \newline (CVS)}
\label{fig_sim_study_sens_relch9}
    \end{subfigure}
\begin{subfigure}[b]{0.32\textwidth}
\resizebox{1.00\textwidth}{!}{\input{fig/variostudy/rel_nEns=64nY=256nFC=8192_s=10.tex}}
        \caption{Copula Dawid-Sebastiani score  (CDSS)}
\label{fig_sim_study_sens_relch10}
    \end{subfigure}

\caption{Box plots of relative change in score $\RelCh$ (see \eqref{eq_relch}) for all considered scoring rules among $L=64$ replications
with an ensemble sample size of $M=2^{13}=8192$ and a rolling window length of $N=2^8=265$.}
\label{fig_sim_study_sens_relch}
\end{figure}

\begin{figure}[h!]
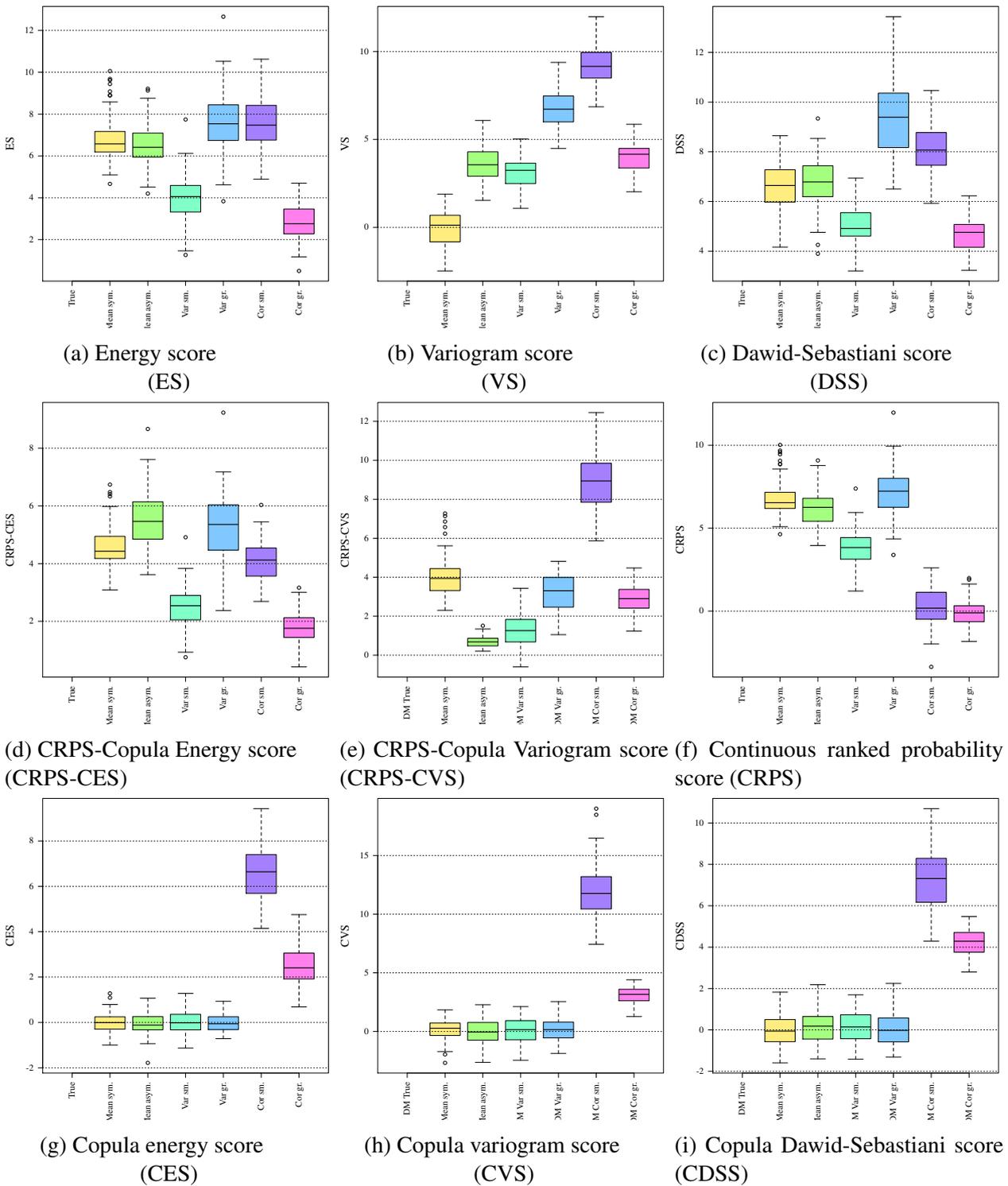

\begin{subfigure}[b]{0.32\textwidth}
 \resizebox{1.00\textwidth}{!}{\input{fig/variostudy/dm_nEns=64nY=256nFC=8192_s=1.tex}}
        \caption{Energy score \newline (ES)}
\label{fig_sim_study_sens_dm1}
    \end{subfigure}
\begin{subfigure}[b]{0.32\textwidth}
\resizebox{1.00\textwidth}{!}{% Created by tikzDevice version 0.11 on 2018-08-28 00:28:05
% !TEX encoding = UTF-8 Unicode
\begin{tikzpicture}[x=1pt,y=1pt]
\definecolor{fillColor}{RGB}{255,255,255}
\path[use as bounding box,fill=fillColor,fill opacity=0.00] (0,0) rectangle (433.62,433.62);
\begin{scope}
\path[clip] ( 50.40, 62.40) rectangle (430.02,430.02);
\definecolor{fillColor}{RGB}{255,237,128}

\path[fill=fillColor] (119.70,115.04) --
	(159.87,115.04) --
	(159.87,150.54) --
	(119.70,150.54) --
	cycle;
\definecolor{drawColor}{RGB}{0,0,0}

\path[draw=drawColor,line width= 1.2pt,line join=round] (119.70,137.25) -- (159.87,137.25);

\path[draw=drawColor,line width= 0.4pt,dash pattern=on 4pt off 4pt ,line join=round,line cap=round] (139.78, 76.02) -- (139.78,115.04);

\path[draw=drawColor,line width= 0.4pt,dash pattern=on 4pt off 4pt ,line join=round,line cap=round] (139.78,178.85) -- (139.78,150.54);

\path[draw=drawColor,line width= 0.4pt,line join=round,line cap=round] (129.74, 76.02) -- (149.82, 76.02);

\path[draw=drawColor,line width= 0.4pt,line join=round,line cap=round] (129.74,178.85) -- (149.82,178.85);

\path[draw=drawColor,line width= 0.4pt,line join=round,line cap=round] (119.70,115.04) --
	(159.87,115.04) --
	(159.87,150.54) --
	(119.70,150.54) --
	(119.70,115.04);
\definecolor{fillColor}{RGB}{164,255,128}

\path[fill=fillColor] (169.91,202.81) --
	(210.08,202.81) --
	(210.08,235.49) --
	(169.91,235.49) --
	cycle;

\path[draw=drawColor,line width= 1.2pt,line join=round] (169.91,218.16) -- (210.08,218.16);

\path[draw=drawColor,line width= 0.4pt,dash pattern=on 4pt off 4pt ,line join=round,line cap=round] (190.00,170.54) -- (190.00,202.81);

\path[draw=drawColor,line width= 0.4pt,dash pattern=on 4pt off 4pt ,line join=round,line cap=round] (190.00,277.53) -- (190.00,235.49);

\path[draw=drawColor,line width= 0.4pt,line join=round,line cap=round] (179.95,170.54) -- (200.04,170.54);

\path[draw=drawColor,line width= 0.4pt,line join=round,line cap=round] (179.95,277.53) -- (200.04,277.53);

\path[draw=drawColor,line width= 0.4pt,line join=round,line cap=round] (169.91,202.81) --
	(210.08,202.81) --
	(210.08,235.49) --
	(169.91,235.49) --
	(169.91,202.81);
\definecolor{fillColor}{RGB}{128,255,200}

\path[fill=fillColor] (220.12,192.97) --
	(260.30,192.97) --
	(260.30,220.27) --
	(220.12,220.27) --
	cycle;

\path[draw=drawColor,line width= 1.2pt,line join=round] (220.12,210.72) -- (260.30,210.72);

\path[draw=drawColor,line width= 0.4pt,dash pattern=on 4pt off 4pt ,line join=round,line cap=round] (240.21,159.89) -- (240.21,192.97);

\path[draw=drawColor,line width= 0.4pt,dash pattern=on 4pt off 4pt ,line join=round,line cap=round] (240.21,252.60) -- (240.21,220.27);

\path[draw=drawColor,line width= 0.4pt,line join=round,line cap=round] (230.17,159.89) -- (250.25,159.89);

\path[draw=drawColor,line width= 0.4pt,line join=round,line cap=round] (230.17,252.60) -- (250.25,252.60);

\path[draw=drawColor,line width= 0.4pt,line join=round,line cap=round] (220.12,192.97) --
	(260.30,192.97) --
	(260.30,220.27) --
	(220.12,220.27) --
	(220.12,192.97);
\definecolor{fillColor}{RGB}{128,200,255}

\path[fill=fillColor] (270.34,275.56) --
	(310.51,275.56) --
	(310.51,310.29) --
	(270.34,310.29) --
	cycle;

\path[draw=drawColor,line width= 1.2pt,line join=round] (270.34,292.55) -- (310.51,292.55);

\path[draw=drawColor,line width= 0.4pt,dash pattern=on 4pt off 4pt ,line join=round,line cap=round] (290.42,239.92) -- (290.42,275.56);

\path[draw=drawColor,line width= 0.4pt,dash pattern=on 4pt off 4pt ,line join=round,line cap=round] (290.42,355.21) -- (290.42,310.29);

\path[draw=drawColor,line width= 0.4pt,line join=round,line cap=round] (280.38,239.92) -- (300.47,239.92);

\path[draw=drawColor,line width= 0.4pt,line join=round,line cap=round] (280.38,355.21) -- (300.47,355.21);

\path[draw=drawColor,line width= 0.4pt,line join=round,line cap=round] (270.34,275.56) --
	(310.51,275.56) --
	(310.51,310.29) --
	(270.34,310.29) --
	(270.34,275.56);
\definecolor{fillColor}{RGB}{164,128,255}

\path[fill=fillColor] (320.55,334.20) --
	(360.72,334.20) --
	(360.72,368.45) --
	(320.55,368.45) --
	cycle;

\path[draw=drawColor,line width= 1.2pt,line join=round] (320.55,349.85) -- (360.72,349.85);

\path[draw=drawColor,line width= 0.4pt,dash pattern=on 4pt off 4pt ,line join=round,line cap=round] (340.64,295.78) -- (340.64,334.20);

\path[draw=drawColor,line width= 0.4pt,dash pattern=on 4pt off 4pt ,line join=round,line cap=round] (340.64,416.40) -- (340.64,368.45);

\path[draw=drawColor,line width= 0.4pt,line join=round,line cap=round] (330.60,295.78) -- (350.68,295.78);

\path[draw=drawColor,line width= 0.4pt,line join=round,line cap=round] (330.60,416.40) -- (350.68,416.40);

\path[draw=drawColor,line width= 0.4pt,line join=round,line cap=round] (320.55,334.20) --
	(360.72,334.20) --
	(360.72,368.45) --
	(320.55,368.45) --
	(320.55,334.20);
\definecolor{fillColor}{RGB}{255,128,237}

\path[fill=fillColor] (370.77,213.85) --
	(410.94,213.85) --
	(410.94,240.00) --
	(370.77,240.00) --
	cycle;

\path[draw=drawColor,line width= 1.2pt,line join=round] (370.77,232.24) -- (410.94,232.24);

\path[draw=drawColor,line width= 0.4pt,dash pattern=on 4pt off 4pt ,line join=round,line cap=round] (390.85,182.07) -- (390.85,213.85);

\path[draw=drawColor,line width= 0.4pt,dash pattern=on 4pt off 4pt ,line join=round,line cap=round] (390.85,272.39) -- (390.85,240.00);

\path[draw=drawColor,line width= 0.4pt,line join=round,line cap=round] (380.81,182.07) -- (400.90,182.07);

\path[draw=drawColor,line width= 0.4pt,line join=round,line cap=round] (380.81,272.39) -- (400.90,272.39);

\path[draw=drawColor,line width= 0.4pt,line join=round,line cap=round] (370.77,213.85) --
	(410.94,213.85) --
	(410.94,240.00) --
	(370.77,240.00) --
	(370.77,213.85);
\end{scope}
\begin{scope}
\path[clip] (  0.00,  0.00) rectangle (433.62,433.62);
\definecolor{drawColor}{RGB}{0,0,0}

\path[draw=drawColor,line width= 0.4pt,line join=round,line cap=round] ( 89.57, 62.40) -- (390.85, 62.40);

\path[draw=drawColor,line width= 0.4pt,line join=round,line cap=round] ( 89.57, 62.40) -- ( 89.57, 56.40);

\path[draw=drawColor,line width= 0.4pt,line join=round,line cap=round] (139.78, 62.40) -- (139.78, 56.40);

\path[draw=drawColor,line width= 0.4pt,line join=round,line cap=round] (190.00, 62.40) -- (190.00, 56.40);

\path[draw=drawColor,line width= 0.4pt,line join=round,line cap=round] (240.21, 62.40) -- (240.21, 56.40);

\path[draw=drawColor,line width= 0.4pt,line join=round,line cap=round] (290.42, 62.40) -- (290.42, 56.40);

\path[draw=drawColor,line width= 0.4pt,line join=round,line cap=round] (340.64, 62.40) -- (340.64, 56.40);

\path[draw=drawColor,line width= 0.4pt,line join=round,line cap=round] (390.85, 62.40) -- (390.85, 56.40);

\node[text=drawColor,rotate= 90.00,anchor=base east,inner sep=0pt, outer sep=0pt, scale=  1.00] at ( 93.01, 50.40) {True};

\node[text=drawColor,rotate= 90.00,anchor=base east,inner sep=0pt, outer sep=0pt, scale=  1.00] at (143.23, 50.40) {Mean sym.};

\node[text=drawColor,rotate= 90.00,anchor=base east,inner sep=0pt, outer sep=0pt, scale=  1.00] at (193.44, 50.40) {Mean asym.};

\node[text=drawColor,rotate= 90.00,anchor=base east,inner sep=0pt, outer sep=0pt, scale=  1.00] at (243.65, 50.40) {Var sm.};

\node[text=drawColor,rotate= 90.00,anchor=base east,inner sep=0pt, outer sep=0pt, scale=  1.00] at (293.87, 50.40) {Var gr.};

\node[text=drawColor,rotate= 90.00,anchor=base east,inner sep=0pt, outer sep=0pt, scale=  1.00] at (344.08, 50.40) {Cor sm.};

\node[text=drawColor,rotate= 90.00,anchor=base east,inner sep=0pt, outer sep=0pt, scale=  1.00] at (394.30, 50.40) {Cor gr.};

\path[draw=drawColor,line width= 0.4pt,line join=round,line cap=round] ( 50.40,134.39) -- ( 50.40,369.86);

\path[draw=drawColor,line width= 0.4pt,line join=round,line cap=round] ( 50.40,134.39) -- ( 44.40,134.39);

\path[draw=drawColor,line width= 0.4pt,line join=round,line cap=round] ( 50.40,252.12) -- ( 44.40,252.12);

\path[draw=drawColor,line width= 0.4pt,line join=round,line cap=round] ( 50.40,369.86) -- ( 44.40,369.86);

\node[text=drawColor,anchor=base east,inner sep=0pt, outer sep=0pt, scale=  1.00] at ( 38.40,130.94) {0};

\node[text=drawColor,anchor=base east,inner sep=0pt, outer sep=0pt, scale=  1.00] at ( 38.40,248.68) {5};

\node[text=drawColor,anchor=base east,inner sep=0pt, outer sep=0pt, scale=  1.00] at ( 38.40,366.41) {10};
\end{scope}
\begin{scope}
\path[clip] (  0.00,  0.00) rectangle (433.62,433.62);
\definecolor{drawColor}{RGB}{0,0,0}

\node[text=drawColor,rotate= 90.00,anchor=base,inner sep=0pt, outer sep=0pt, scale=  1.30] at ( 12.00,246.21) {VS};
\end{scope}
\begin{scope}
\path[clip] (  0.00,  0.00) rectangle (433.62,433.62);
\definecolor{drawColor}{RGB}{0,0,0}

\path[draw=drawColor,line width= 0.4pt,line join=round,line cap=round] ( 50.40, 62.40) --
	(430.02, 62.40) --
	(430.02,430.02) --
	( 50.40,430.02) --
	( 50.40, 62.40);
\end{scope}
\begin{scope}
\path[clip] ( 50.40, 62.40) rectangle (430.02,430.02);
\definecolor{drawColor}{RGB}{0,0,0}

\path[draw=drawColor,line width= 0.4pt,dash pattern=on 1pt off 3pt ,line join=round,line cap=round] ( 50.40,134.39) -- (430.02,134.39);

\path[draw=drawColor,line width= 0.4pt,dash pattern=on 1pt off 3pt ,line join=round,line cap=round] ( 50.40,252.12) -- (430.02,252.12);

\path[draw=drawColor,line width= 0.4pt,dash pattern=on 1pt off 3pt ,line join=round,line cap=round] ( 50.40,369.86) -- (430.02,369.86);
\end{scope}
\end{tikzpicture}}
        \caption{Variogram score \newline (VS)}
\label{fig_sim_study_sens_dm2}
    \end{subfigure}
\begin{subfigure}[b]{0.32\textwidth}
\resizebox{1.00\textwidth}{!}{\input{fig/variostudy/dm_nEns=64nY=256nFC=8192_s=3.tex}}
        \caption{Dawid-Sebastiani score \newline (DSS)}
\label{fig_sim_study_sens_dm3}
    \end{subfigure}

    \begin{subfigure}[b]{0.32\textwidth}
\resizebox{1.00\textwidth}{!}{\input{fig/variostudy/dm_nEns=64nY=256nFC=8192_s=4.tex}}
        \caption{CRPS-Copula Energy score \newline (CRPS-CES)}
\label{fig_sim_study_sens_dm4}
    \end{subfigure}
\begin{subfigure}[b]{0.32\textwidth}
\resizebox{1.00\textwidth}{!}{\input{fig/variostudy/dm_nEns=64nY=256nFC=8192_s=5.tex}}
\caption{CRPS-Copula Variogram score (CRPS-CVS)}
\label{fig_sim_study_sens_dm5}
    \end{subfigure}
\begin{subfigure}[b]{0.32\textwidth}
\resizebox{1.00\textwidth}{!}{\input{fig/variostudy/dm_nEns=64nY=256nFC=8192_s=7.tex}}
        \caption{Continuous ranked probability score (CRPS)}
\label{fig_sim_study_sens_dm7}
    \end{subfigure}

    \begin{subfigure}[b]{0.32\textwidth}
\resizebox{1.00\textwidth}{!}{\input{fig/variostudy/dm_nEns=64nY=256nFC=8192_s=8.tex}}
        \caption{Copula energy score \newline (CES)}
\label{fig_sim_study_sens_dm8}
    \end{subfigure}
\begin{subfigure}[b]{0.32\textwidth}
\resizebox{1.00\textwidth}{!}{\input{fig/variostudy/dm_nEns=64nY=256nFC=8192_s=9.tex}}
        \caption{Copula variogram score \newline (CVS)}
\label{fig_sim_study_sens_dm9}
    \end{subfigure}
\begin{subfigure}[b]{0.32\textwidth}
\resizebox{1.00\textwidth}{!}{\input{fig/variostudy/dm_nEns=64nY=256nFC=8192_s=10.tex}}
        \caption{Copula Dawid-Sebastiani score  (CDSS)}
\label{fig_sim_study_sens_dm10}
    \end{subfigure}
\caption{Box plots of DM-test statistics with respect to the true model for all considered scoring rules among $L=64$ replications
with an ensemble sample size of $M=2^{13}=8192$ and a rolling window length of $N=2^8=265$.}
\label{fig_sim_study_sens_dm}
\end{figure}

For the simulation study, we consider in total $M=2^{13}=8192$ paths in the simulated ensemble, a rolling window length of $N=2^{8}=256$ and consider 
$L=2^6= 64$ replications of the experiment.
Again, we evaluate the relative change in the scores such as the DM-test test statistic.
The results are visualized in Figures \ref{fig_sim_study_sens_relch} and \ref{fig_sim_study_sens_dm}.

First, we focus on Figures \ref{fig_sim_study_sens_relch} with the relative changes with respect to the true model.
We observe in Figure \ref{fig_sim_study_sens_relch3} that the constructions seems to be correct. So the Dawid-Sebastiani score reacts with 
a relative change sensitivity for all distorted setting of about 0.5.
Moreover, we see a similar picture as reported in \cite{scheuerer2015variogram} for the energy score (Fig. \ref{fig_sim_study_sens_relch1}).
For changes in the mean structure we have a relative high relative changes compared to changes in the variance structure which is still
more distinct than changes in the correlation structure. Especially for large correlations, the relative changes are very small, and appears to be 
hardly significantly different from zero. The CRPS-copula energy score (CRPS-CES) shows a very similar picture (Fig. \ref{fig_sim_study_sens_relch4}).
For the variogram score (Fig. \ref{fig_sim_study_sens_relch2}) we observe a better behavior in the relative change concerning the correlations. Still,
we see as well that the variogram score can not identify a shift in the mean. The CRPS-copula variogram score (CRPS-CVS) shows a completely different pattern (Fig. \ref{fig_sim_study_sens_relch5}), most 
importantly it can identify all settings correctly. Still, the CRPS-CVS is not strictly proper.
Finally, we observe that the CRPS (Fig. \ref{fig_sim_study_sens_relch7}) can not detect changes in the correlation structure. In contrast, all copula scores (Fig. \ref{fig_sim_study_sens_relch8}, \ref{fig_sim_study_sens_relch9}, \ref{fig_sim_study_sens_relch10}) can not identify changes in the structure of
the marginal distribution.

Now, let us turn to the interesting results of Figures \ref{fig_sim_study_sens_dm}, the DM-statistics.
For the Dawid-Sebastiani score (Fig. \ref{fig_sim_study_sens_dm3}), the average DM-test statistics is for all settings between about 5 and 9, where the smallest DM-statistic appear for the \emph{greater correlation} case.
The energy score (Fig. \ref{fig_sim_study_sens_dm1}) has DM-statistics averages are between about 3 and 8 for all settings. Thus, the energy score can identify those models correctly even though the power is noticeable smaller than for the Dawid-Sebastiani score. 
For the variogram score (Fig. \ref{fig_sim_study_sens_dm2}) we observe that for changes in mean a DM-statistic that is not significantly different from zero. 
Still, for the remaining models we receive DM-statistics between 3 and 9. Thus, similarly to the energy score the remaining settings can be identified.
The CRPS-copula energy score (CRPS-CES) shows similar pattern to the energy score (Fig. \ref{fig_sim_study_sens_dm4}), but all with smaller average DM-statistics in all settings, they vary only between about 2 and 5. Thus, it is clearly less powerful than the energy score in this simulation study setting.
Also the CRPS-copula variogram score (CRPS-CVS) does not perform great (Fig. \ref{fig_sim_study_sens_dm4}). 
In two setting (asymmetric mean shift, smaller variance) average DM-statistics are only about 1. Thus, for standard significance levels (e.g. $5\%$) we could not 
reject the null hypothesis.
For the marginal settings, the CRPS (Fig. \ref{fig_sim_study_sens_dm7}) has average DM-statistics between about 4 and 7 which seems to be a similar level 
as the energy score. Interestingly, the copula energy score (CES) shows similar 
average DM-statistics in the correlation settings to the energy score (Fig. \ref{fig_sim_study_sens_dm8}). It seems that 
the energy score combines to some extent the marginal discrimination ability of the CRPS with the dependency discrimination ability of the copula energy score.
However, among the three  copula scores the copula energy score performs worst with respect to the discrepancy measurement in the correlation structure.

\subsection{Random peak study}

 In this section we discuss in detail the application of the energy score and the Diebold-Mariano test to a toy example.
This example is motivated by \cite{haben2014new} with application to smart-meter electricity load data.
It illustrates nicely proper multivariate forecasting evaluation as it is an example where simple measures usually fail.

We assume that $\bsX_{i}$ is an $H$-dimensional process. It is defined by:
$\bsX_{i} = \bsY_{i} + Q \bsZ_{i}$ with $\bsY_{i} \sim \NN_H(\bsnull, \bsI)$ and 
$\bsZ_{i} \sim \UU( \{\bse_1,\ldots,\bse_H\})$ with $\NN_H$ as $H$-dimensional normal distribution, $\UU(\AA)$ as uniform distribution 
on $\AA$, a constant $Q$ such as unit vectors $\bse_1=(1, 0, 0, \ldots)'$, $\ldots$ and $\bse_H=(0, \ldots, 0, 0 , 1)'$ and identity matrix $\bsI$. Each
$\bsY_{i}$ and $\bsZ_{i}$ are independent from each other.
So $\bsX_{i}$ is a $H$-dimensional standard normal distribution where at one uniformly chosen coordinate we add an additional 
constant $Q=5$. This can be interpreted as a peak that occurs always at one of the $H$ dimensions. 
An interpretation of this experiment based on \cite{haben2014new} could be that we know that a person will have a shower every morning, say either at 
6:00, 7:00 or 8:00. In this $H=3$-dimensional example we expect a peak at exactly one of these hours but non at the others.
Obviously, with standard point forecasting measures MAE (mean absolute error) or RMSE (root mean square error) 
the \emph{flat forecast} (having never a shower) outperforms 
the forecast \emph{fixed peak forecast} which stats that a person has a shower always at e.g. 7:00. This seems to be counterintuitive as the \emph{fixed peak forecast} seems 
to be a better forecast than the \emph{flat forecast} as at least the peak is detected, even though not the correct one.

Back to the simulation study: Next to the perfect model %that is identical to $\bsX_{i}$ and denoted as $\bsX_{i}^{[k]}$
we will consider seven other models that could serve as (more or less) plausible forecast model.
They are defined by:
\begin{enumerate}
 \item (true)
 $\bsX_{i} = \bsY_{i} + Q \bsZ_{i}$ with $\bsY_{i} \stackrel{\text{iid}}{\sim} \NN_H(\bsnull, \bsI)$ and 
$\bsZ_{i} \stackrel{\text{iid}}{\sim} \UU( \{\bse_1, \ldots,\bse_H\})$ 
\item (average mean) 
$\bsX_{i} \stackrel{\text{iid}}{\sim} \NN_H(\mu \bsone, \bsI)$ with $\mu = \frac{Q}{H} $
\item (zero mean) 
$\bsX_{i} \stackrel{\text{iid}}{\sim} \NN_H(\bsnull, \bsI)$ 
\item (fixed peak)
$\bsX_{i} \stackrel{\text{iid}}{\sim} \NN_H(\bsmu, \bsI)$ with $\bsmu= (Q,0, \ldots ,0)'$
\item (rolling peak)
$\bsX_{i} \stackrel{\text{iid}}{\sim} \NN_H(\bsmu_i, \bsI)$ with 
                                       $\bsmu_i = \bse_{1 + (i-1) \text{mod} H }$ with as unit vector for the $i$'th coordinate.
                                                                          
\item (mixture normal with same marginals)
$\bsX_{i} = (X_{i,1}, \ldots ,X_{i,H})'$ with \newline
 $X_{i,j}\stackrel{\text{iid}}{\sim}  \begin{cases}
\NN_1(0,1)&  U_j \leq \frac{H-1}{H} \\
\NN_1(Q,1) &  U_j > \frac{H-1}{H}
 \end{cases}$
where $U_j \stackrel{\text{iid}}{\sim}  \UU( [0,1] )$.
\item (shifted mean)
 $\bsX_{i} = \bsY_{i} + Q \bsZ_{i}$ with $\bsY_{i} \stackrel{\text{iid}}{\sim} \NN_H(Q/H \bsone, \bsI)$ and 
$\bsZ_{i} \stackrel{\text{iid}}{\sim} \UU( \{\bse_1,\ldots,\bse_H\})$ 
\item (normal with true mean and covariance)
$\bsX_{i} \stackrel{\text{iid}}{\sim} \NN_H(\bsmu , \bsSigma)$ with $\bsmu = \frac{Q}{H} \bsone$ and $\bsSigma = 
(H+ Q^2)/H \bsI  -Q^2/H^2 \bsone \bsone' $
%  \bsY_{i} + Q \bsZ_{i}$ with $\bsY_{i} \stackrel{\text{iid}}{\sim} \NN_3(\bsnull, \bsI)$ and 
% $\bsZ_{T,i} \stackrel{\text{iid}}{\sim} \UU( \{\bse_1,\bse_2,\bse_3\})$ 
\end{enumerate}
The second and third model are two multivariate normal models with a constant mean, around the overall mean $\mu$ for model 2 and around $0$ for model 3. 
In the contest of peak or spike forecasting model $3$ is the above mentioned \emph{flat forecast}. 
But both model 2 and 3 miss completely the occurring peak. 

Model 4 instead, uses a multivariate normal distribution as well and  predicts a peak of size $Q$. But this peak is predicted to occur always
in the first coordinate and corresponds to the \emph{fixed peak forecast} example.
Similarly does model 5, it always predicts a peak of size $Q$. 
But the time of occurrence is moves with $i$. For $i=1$ it is in the first dimension, with $i=2$ in the second, up to $i=H$ in the $H$'s component 
and with $i=H+1$ in the first again. So the peak is predicted to move cyclic (esp. deterministic) within the $H$ dimensions.

The next considered model 6 is a mixture normal model. Every single random variable $X_{i,j}$ 
is with probability $1/H$ a normal distribution with mean $Q$ (which corresponds to the peak) and with probability $(H-1)/H$ it has a zero mean (no peak).
% is normally distributed with the same variance. But the
% mean is $0$ with a probability of $2/3$ and $Q$ with a probability of $1/3$. 
As a consequence the model can have peaks as well. But the number of peaks in $\bsX_{t,i}$ can vary between $0$ and $H$. 
If $H=3$, there are cases
with no peak (about $30\%$ of all cases), and cases with three peaks (about $4\%$ of all cases). Still, the peaks are randomly spread around the $H$ dimensions.
The seventh model corresponds to the true one with a mean shift. And the latter one is a normal distribution, with exactly the same first and second moment as 
the true distribution. Hence, we expect the Dawid-Sebastiani score to fail here.

\begin{figure}[h!]
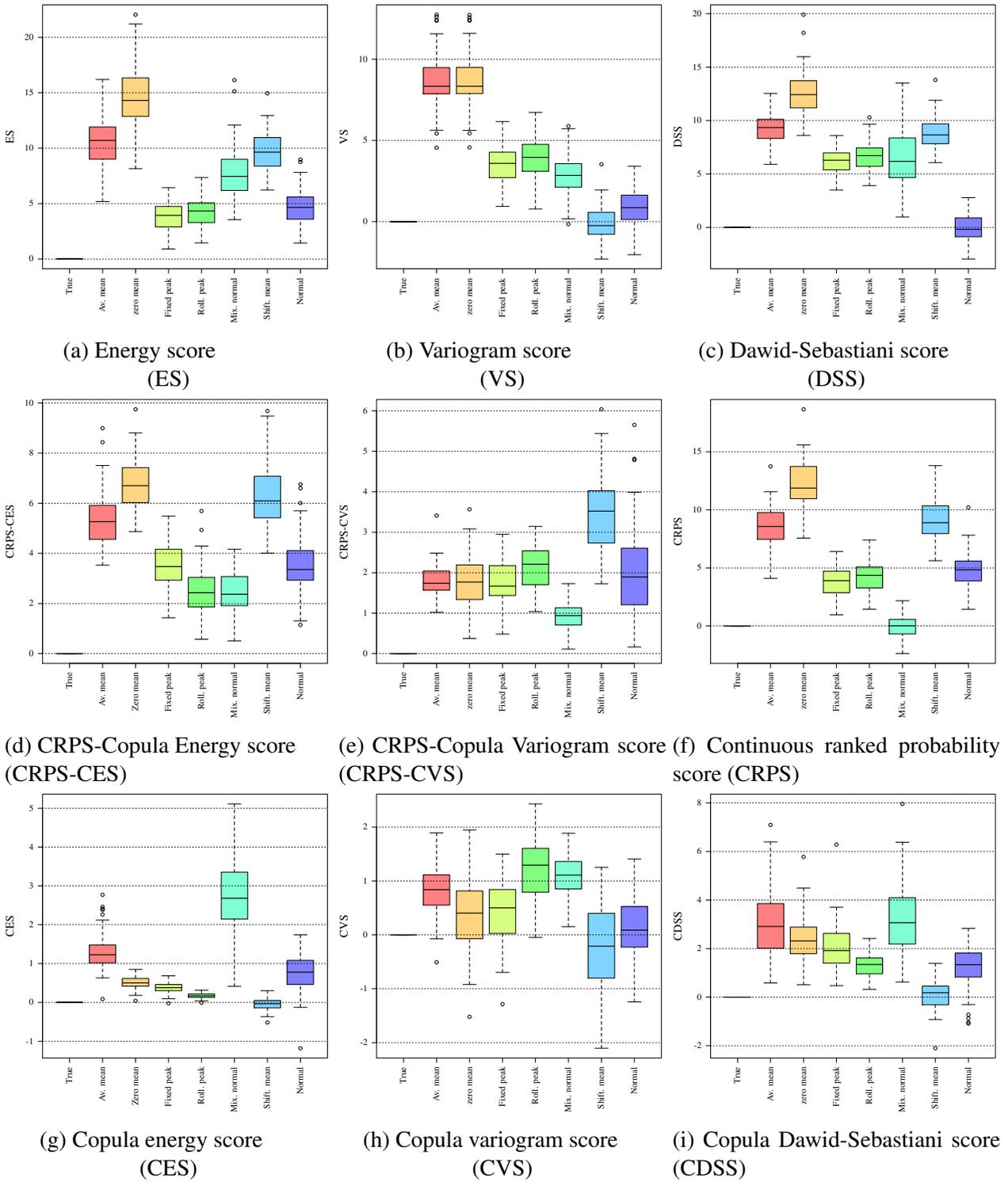

\begin{subfigure}[b]{0.32\textwidth}
\resizebox{1.00\textwidth}{!}{\input{fig/peakstudy/dm_H=3_Q=5_nEns=64nY=32nFC=16384_s=1.tex}}
        \caption{Energy score \newline (ES)}
\label{fig_sim_study_peak_dm1}
    \end{subfigure}
\begin{subfigure}[b]{0.32\textwidth}
\resizebox{1.00\textwidth}{!}{\input{fig/peakstudy/dm_H=3_Q=5_nEns=64nY=32nFC=16384_s=2.tex}}
        \caption{Variogram score \newline (VS)}
\label{fig_sim_study_peak_dm2}
    \end{subfigure}
\begin{subfigure}[b]{0.32\textwidth}
\resizebox{1.00\textwidth}{!}{\input{fig/peakstudy/dm_H=3_Q=5_nEns=64nY=32nFC=16384_s=3.tex}}
        \caption{Dawid-Sebastiani score \newline (DSS)}
\label{fig_sim_study_peak_dm3}
    \end{subfigure}

    \begin{subfigure}[b]{0.32\textwidth}
\resizebox{1.00\textwidth}{!}{\input{fig/peakstudy/dm_H=3_Q=5_nEns=64nY=32nFC=16384_s=4.tex}}
        \caption{CRPS-Copula Energy score \newline (CRPS-CES)}
\label{fig_sim_study_peak_dm4}
    \end{subfigure}
\begin{subfigure}[b]{0.32\textwidth}
\resizebox{1.00\textwidth}{!}{\input{fig/peakstudy/dm_H=3_Q=5_nEns=64nY=32nFC=16384_s=5.tex}}
\caption{CRPS-Copula Variogram score (CRPS-CVS)}
\label{fig_sim_study_peak_dm5}
    \end{subfigure}
\begin{subfigure}[b]{0.32\textwidth}
\resizebox{1.00\textwidth}{!}{\input{fig/peakstudy/dm_H=3_Q=5_nEns=64nY=32nFC=16384_s=7.tex}}
        \caption{Continuous ranked probability score (CRPS)}
\label{fig_sim_study_peak_dm7}
    \end{subfigure}

    \begin{subfigure}[b]{0.32\textwidth}
\resizebox{1.00\textwidth}{!}{\input{fig/peakstudy/dm_H=3_Q=5_nEns=64nY=32nFC=16384_s=8.tex}}
        \caption{Copula energy score \newline (CES)}
\label{fig_sim_study_peak_dm8}
    \end{subfigure}
\begin{subfigure}[b]{0.32\textwidth}
\resizebox{1.00\textwidth}{!}{\input{fig/peakstudy/dm_H=3_Q=5_nEns=64nY=32nFC=16384_s=9.tex}}
        \caption{Copula variogram score \newline (CVS)}
\label{fig_sim_study_peak_dm9}
    \end{subfigure}
\begin{subfigure}[b]{0.32\textwidth}
\resizebox{1.00\textwidth}{!}{\input{fig/peakstudy/dm_H=3_Q=5_nEns=64nY=32nFC=16384_s=10.tex}}
        \caption{Copula Dawid-Sebastiani score  (CDSS)}
\label{fig_sim_study_peak_dm10}
    \end{subfigure}
\caption{Box plots of DM-test statistics with respect to the true model for all considered scoring rules for the random peak study 
with dimension $H=3$ and peak size $Q=5$ among $L=2^{6}=64$ replications
with an ensemble sample size of $M=2^{14}=16384$ and a rolling window length of $N=2^5=32$.}
\label{fig_sim_study_peak_dm}
\end{figure}

We conduct the random peak study for the $H=3$-dimensional case with a peak size of $Q=5$. 
We carry out $L=2^{6}=64$ experiment replications with an ensemble sample size of $M=2^{14}=16384$ and a rolling window length of $N=2^5=32$.
As we pointed out, that the relative change is not suitable for evaluation, we only report the statistics of the Diebold-Mariano, see Figure \ref{fig_sim_study_peak_dm}.

We observe that the variogram score and Dawid-Sebastiani score (Fig. \ref{fig_sim_study_peak_dm2}, \ref{fig_sim_study_peak_dm3}) 
can not identify the true model in all settings. This, is not surprising, as they are not strictly proper.
In contrast, the energy score (Fig. \ref{fig_sim_study_peak_dm1}) has an average DM-statistic in all settings between about 4 and 14. Thus, again it can clearly identify the true model.
Also the strictly proper CRPS-Copula energy score (CRPS-CES) can identify all models (Fig. \ref{fig_sim_study_peak_dm4}), even though the average DM-statistics 
vary in the settings between 2 and 6. Similarly, the CRPS-Copula variogram score (CRPS-CVS) can identify all settings, even though it is not strictly proper.
Obviously, the CRPS (Fig. \ref{fig_sim_study_peak_dm7}) fails in separating the mixture normal model from the true one, as they have the same marginal distribution.
Interestingly, the copula energy score (Fig. \ref{fig_sim_study_peak_dm8}) is very sensitive to exactly this normal mixture setting. This is at least an indication why 
it might be a good idea to combine the copula energy score with the CRPS. Still, overall copula scores show rather small DM-statistics.

\subsection{The role of the ensemble size and forecasting horizon.}

In this simulation study we want to highlight the importance of the ensemble sample size $M$ and the forecasting horizon $H$ when reporting multivariate forecasts. First we focus on the ensemble sample size $M$.
It is clear that for $M\to \infty$ the forecasting distribution is arbitrarily well covered. However, in real applications we face the problem that 
can only simulate finite amount of time, and hope that for large $M$ the underlying distribution is approximated sufficiently well. Now, we want to study the impact 
of the sample size.

\begin{figure}[h!]
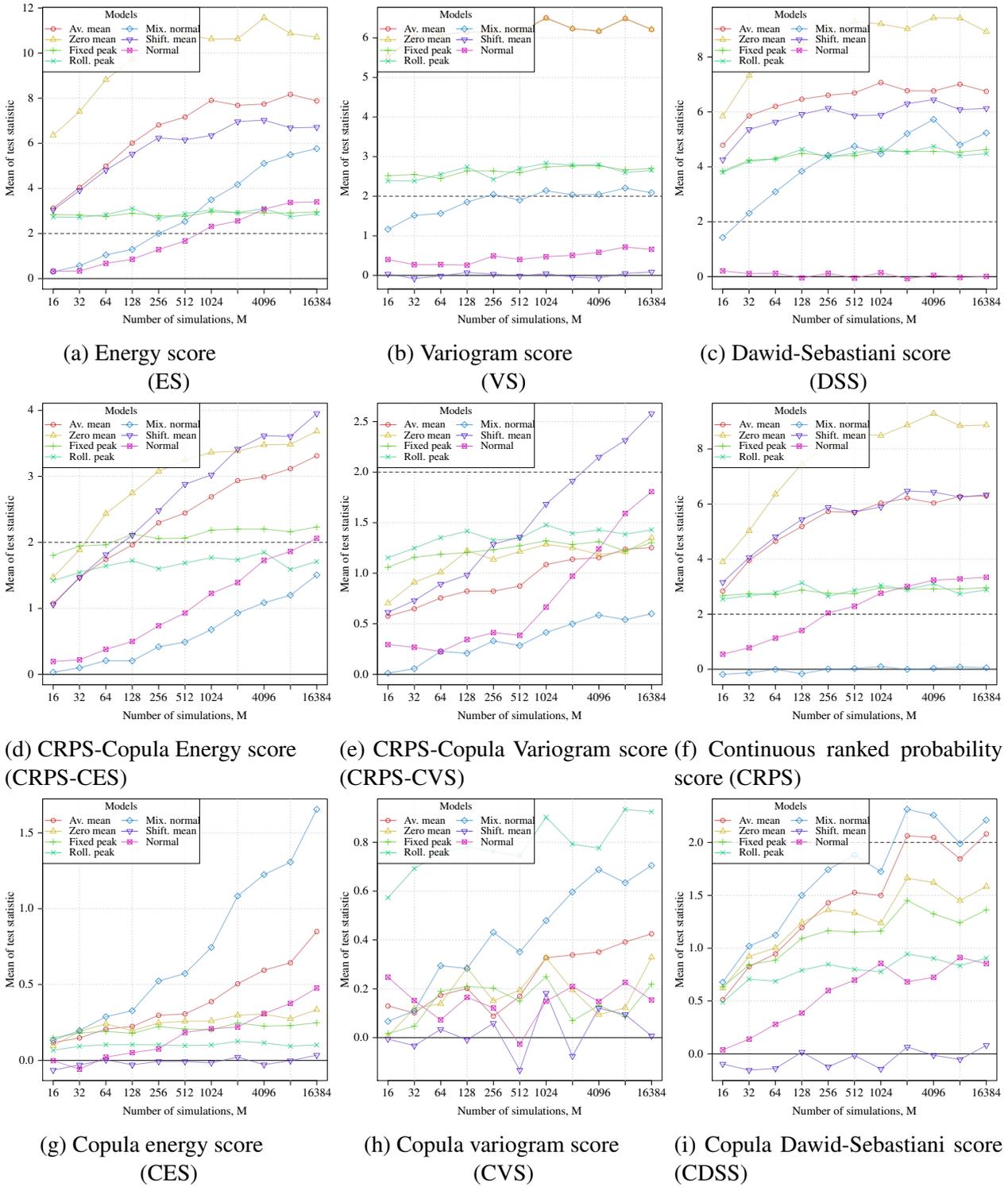

\begin{subfigure}[b]{0.32\textwidth}
\resizebox{1.00\textwidth}{!}{\input{fig/peakstudy/depM_av_H=3_Q=5_nEns=128nY=16nFC=16384_s=1_k=8.tex}}
        \caption{Energy score \newline (ES)}
\label{fig_sim_study_peak_M1_dm1}
    \end{subfigure}
\begin{subfigure}[b]{0.32\textwidth}
\resizebox{1.00\textwidth}{!}{\input{fig/peakstudy/depM_av_H=3_Q=5_nEns=128nY=16nFC=16384_s=2_k=8.tex}}
        \caption{Variogram score \newline (VS)}
\label{fig_sim_study_peak_M1_dm2}
    \end{subfigure}
\begin{subfigure}[b]{0.32\textwidth}
\resizebox{1.00\textwidth}{!}{\input{fig/peakstudy/depM_av_H=3_Q=5_nEns=128nY=16nFC=16384_s=3_k=8.tex}}
        \caption{Dawid-Sebastiani score \newline (DSS)}
\label{fig_sim_study_peak_M1_dm3}
    \end{subfigure}

    \begin{subfigure}[b]{0.32\textwidth}
\resizebox{1.00\textwidth}{!}{\input{fig/peakstudy/depM_av_H=3_Q=5_nEns=128nY=16nFC=16384_s=4_k=8.tex}}
        \caption{CRPS-Copula Energy score \newline (CRPS-CES)}
\label{fig_sim_study_peak_M1_dm4}
    \end{subfigure}
\begin{subfigure}[b]{0.32\textwidth}
\resizebox{1.00\textwidth}{!}{\input{fig/peakstudy/depM_av_H=3_Q=5_nEns=128nY=16nFC=16384_s=5_k=8.tex}}
\caption{CRPS-Copula Variogram score (CRPS-CVS)}
\label{fig_sim_study_peak_M1_dm5}
    \end{subfigure}
\begin{subfigure}[b]{0.32\textwidth}
\resizebox{1.00\textwidth}{!}{\input{fig/peakstudy/depM_av_H=3_Q=5_nEns=128nY=16nFC=16384_s=7_k=8.tex}}
        \caption{Continuous ranked probability score (CRPS)}
\label{fig_sim_study_peak_M1_dm7}
    \end{subfigure}

    \begin{subfigure}[b]{0.32\textwidth}
\resizebox{1.00\textwidth}{!}{\input{fig/peakstudy/depM_av_H=3_Q=5_nEns=128nY=16nFC=16384_s=8_k=8.tex}}
        \caption{Copula energy score \newline (CES)}
\label{fig_sim_study_peak_M1_dm8}
    \end{subfigure}
\begin{subfigure}[b]{0.32\textwidth}
\resizebox{1.00\textwidth}{!}{\input{fig/peakstudy/depM_av_H=3_Q=5_nEns=128nY=16nFC=16384_s=9_k=8.tex}}
        \caption{Copula variogram score \newline (CVS)}
\label{fig_sim_study_peak_M1_dm9}
    \end{subfigure}
\begin{subfigure}[b]{0.32\textwidth}
\resizebox{1.00\textwidth}{!}{\input{fig/peakstudy/depM_av_H=3_Q=5_nEns=128nY=16nFC=16384_s=10_k=8.tex}}
        \caption{Copula Dawid-Sebastiani score  (CDSS)}
\label{fig_sim_study_peak_M1_dm10}
    \end{subfigure}
\caption{Average of DM-test statistics across $L=2^8=128$ experiment replications with respect to the true model for all considered scoring rules for the random peak study 
with dimension $H=3$, peak size $Q=5$ and a rolling window length of $N=2^4=16$ on a ensemble size $M$ on the grid $\MM= \{16,32,\ldots, 16384\}$.}
\label{fig_sim_study_peak_M1}
\end{figure}

Therefore, we consider again the framework of the peak study of the previous subsection but vary across the ensemble sample size $M$.
This grid is chosen to be $\MM=\{2^i| i \in \{4,\ldots, 14\}\} = \{2^4, 2^5, \ldots, 2^{14}\}= \{16,32,\ldots, 16384\}$.
We conduct experiments with $L=2^7=128$ replication and a rolling window length of $N=2^4=16$. 
As in the previous section we chose the dimension $H=3$ and a peak size of $Q=5$.
The results of the average DM-statistics across the $L$ replications are given in Figure \ref{fig_sim_study_peak_M1}.

There, we observe that small $M$ values are not sufficient to report a stable forecast. When focusing on the energy score (Fig. \ref{fig_sim_study_peak_M1_dm1}) we notice,
that for most settings (all except \emph{fixed peak} and \emph{rolling peak}) an ensemble size of $M=16$ is by far not sufficient to 
receive approximately maximal DM-statistics. For three settings (\emph{average mean}, \emph{zero mean} and \emph{shifted mean}) we require 
about $2^{10}=1024$ paths in the ensemble to reach stability in the DM-statistic. For the remaining tricky situation s
(\emph{mixture normal} and \emph{normal}) we observe that $M=2^{14}=163846$ might be even not sufficient to reach the theoretical maximum.
% For all the other scoring rules we observes similar pattern.
For the variogram score (Fig. \ref{fig_sim_study_peak_M1_dm2}) the results seem to be quite robust, even for very small ensemble sizes, like $M=16$ or $M=32$, they work relatively well.
The Dawid-Sebastiani score (Fig. \ref{fig_sim_study_peak_M1_dm3}) requires about $M=2^9=512$ paths in the ensemble seem to be sufficiently close to the optimum 
in the given setting. Also for the other scores we observe that larger sample sizes $M$ help  to increase the DM-statistics, even for the CRPS which is only based on univariate scoring rules.
However, the copula based scoring rules have relatively small average test statistics for all cases.

In literature the ensemble sample size $M$ is sometimes not reported.
If so, the considered ensemble size is usually surprisingly small, e.g. $M=8$ and $M=19$ in \cite{gneiting2008assessing}, $M=51$ in \cite{pinson2012adaptive} and \cite{junk2014comparison}. 
The most common reason is that for multivariate forecasts in the meteorologic applications, only a few meteorologic ensemble forecasts based on a physical/meteorologic weather model are considered. 
% Their number is
% usually relative small.

%  \FZ{TODO HERE}

% depM_av_H=9_Q=5_nEns=128nY=16nFC=16384_s=10_k=8

\begin{figure}[h!]
\begin{subfigure}[b]{0.32\textwidth}
\resizebox{1.00\textwidth}{!}{\input{fig/peakstudy/depM_av_H=9_Q=5_nEns=128nY=16nFC=16384_s=1_k=8.tex}}
        \caption{Energy score \newline (ES)}
\label{fig_sim_study_peak_M2_dm1}
    \end{subfigure}
\begin{subfigure}[b]{0.32\textwidth}
\resizebox{1.00\textwidth}{!}{\input{fig/peakstudy/depM_av_H=9_Q=5_nEns=128nY=16nFC=16384_s=2_k=8.tex}}
        \caption{Variogram score \newline (VS)}
\label{fig_sim_study_peak_M2_dm2}
    \end{subfigure}
\begin{subfigure}[b]{0.32\textwidth}
\resizebox{1.00\textwidth}{!}{\input{fig/peakstudy/depM_av_H=9_Q=5_nEns=128nY=16nFC=16384_s=3_k=8.tex}}
        \caption{Dawid-Sebastiani score \newline (DSS)}
\label{fig_sim_study_peak_M2_dm3}
    \end{subfigure}

    \begin{subfigure}[b]{0.32\textwidth}
\resizebox{1.00\textwidth}{!}{\input{fig/peakstudy/depM_av_H=9_Q=5_nEns=128nY=16nFC=16384_s=4_k=8.tex}}
        \caption{CRPS-Copula Energy score \newline (CRPS-CES)}
\label{fig_sim_study_peak_M2_dm4}
    \end{subfigure}
\begin{subfigure}[b]{0.32\textwidth}
\resizebox{1.00\textwidth}{!}{\input{fig/peakstudy/depM_av_H=9_Q=5_nEns=128nY=16nFC=16384_s=5_k=8.tex}}
\caption{CRPS-Copula Variogram score (CRPS-CVS)}
\label{fig_sim_study_peak_M2_dm5}
    \end{subfigure}
\begin{subfigure}[b]{0.32\textwidth}
\resizebox{1.00\textwidth}{!}{\input{fig/peakstudy/depM_av_H=9_Q=5_nEns=128nY=16nFC=16384_s=7_k=8.tex}}
        \caption{Continuous ranked probability score (CRPS)}
\label{fig_sim_study_peak_M2_dm7}
    \end{subfigure}

    \begin{subfigure}[b]{0.32\textwidth}
\resizebox{1.00\textwidth}{!}{\input{fig/peakstudy/depM_av_H=9_Q=5_nEns=128nY=16nFC=16384_s=8_k=8.tex}}
        \caption{Copula energy score \newline (CES)}
\label{fig_sim_study_peak_M2_dm8}
    \end{subfigure}
\begin{subfigure}[b]{0.32\textwidth}
\resizebox{1.00\textwidth}{!}{\input{fig/peakstudy/depM_av_H=9_Q=5_nEns=128nY=16nFC=16384_s=9_k=8.tex}}
        \caption{Copula variogram score \newline (CVS)}
\label{fig_sim_study_peak_M2_dm9}
    \end{subfigure}
\begin{subfigure}[b]{0.32\textwidth}
\resizebox{1.00\textwidth}{!}{\input{fig/peakstudy/depM_av_H=9_Q=5_nEns=128nY=16nFC=16384_s=10_k=8.tex}}
        \caption{Copula Dawid-Sebastiani score  (CDSS)}
\label{fig_sim_study_peak_M2_dm10}
    \end{subfigure}
\caption{Average of DM-test statistics across $L=2^8=256$ experiment replications with respect to the true model for all considered scoring rules for the random peak study 
with dimension $H=9$, peak size $Q=5$ and a rolling window length of $N=2^4=16$ on a ensemble size grid $\MM= \{16,32,\ldots, 4096\}$.}
\label{fig_sim_study_peak_M2}
\end{figure}

Finally, we want to vary the forecasting horizon. Instead of an $H=3$-dimensional we consider now a problem 
of the square size. Thus, we assume $H=9$. The average DM-statistics across the $L$ experiments are given in Figure \ref{fig_sim_study_peak_M2}.
There we see that for almost all scores the DM-statistic descrease compared to the 3-dimensional case \ref{fig_sim_study_peak_M2}. The CRPS for instance is an 
exception, here the statistics increase in some cases (Fig. \ref{fig_sim_study_peak_M2_dm7}). Likely, because there are more marginals that can be evaluated.
For larger sample sizes, only the energy score (Fig. \ref{fig_sim_study_peak_M2_dm1}) can identify the true model significantly. 
In constrast, the strictly proper CRPS-CES (Fig. \ref{fig_sim_study_peak_M2_dm4}) fails to seperate the true and the mixture normal model, as the copula energy score (CES) does as well.
The reason is likely that the ensemble size is too small such the underlying $9$-dimensional space is explored well. More on this is explained in the next subsection.

\subsection{Discussion}

Let us shortly discuss the simulation results.
As pointed out before, the relative change in a score with respect to the true (or optimal) model is not a good criterion for measuring 
the sensitivity of a score. In contrast, the DM-test is a suitable tool for evaluating the sensitivity, as it makes a statement 
if a non-optimal model can be identified as non-optimal with respect to the true one. 
With respect to the DM-test have seen that in all simulation studies (with multivariate normal design and the peak study) the energy score performs clearly best. It can identify the true model in all studies of article with relatively strong power. 

When the authors received this results they somehow wondered as a couple of researchers 
(e.g. \cite{scheuerer2015variogram, spath2015time, gamakumara2018probabilisitic}) mentioning to weak discrimination ability of the energy score. But, they are all referring the working paper simulation study of \cite{pinson2013discrimination} which uses the relative change in the score to conclude.
Therefore, it is worth studying again a bit deeper the properties of the energy score.

Let $\bsY$ be an $H$-dimensional random variable. The energy score results as the one-sided version of the 
energy distance $d_{E}$ (see \cite{szekely2013energy}). The energy distance is defined as
\begin{equation}
d_{E}(\bsX,\bsY) = %\left( 
\E \|\bsX-\bsY\|_2^{\beta} - \frac{1}{2} \E \|\bsX-\wtilde{\bsX}\|_2^{\beta} -\frac{1}{2} \E \|\bsY-\wtilde{\bsY}\|_2^{\beta} %\right)^{\frac{1}{2}}  
\label{eq_energy_dist}
\end{equation}
where $\wtilde{\bsX}$ and $\wtilde{\bsY}$ are iid-copies of $\bsX$ and $\bsY$, and $\beta \in (0,1)$. Again, $\beta=1$ gives the standard case which results 
in the 1-dimension the Cramer-distance $d_{C}$:
\begin{equation}
d_{C}(X,Y) = 
\int_\R | F_{X}(z)-F_{Y}(z) |^2 d \, z  = %\left( 
\E|X-Y| - \frac{1}{2}\E|X-\wtilde{X}| -\frac{1}{2} \E|Y-\wtilde{Y}| %\right)^{\frac{1}{2}} 
\label{eq_cramer_repr}
\end{equation}
If we replace $Y$ and $\wtilde{Y}$ by a constant $y$ we receive the popular CRPS. %Note that energy score results only as a generalization
%of the latter equation. 

The energy distance allows for many statistical applications, \cite{szekely2013energy, rizzo2016energy}. It allows to construct a
general (non-parametric) test for equality of two multivariate distributions. When replacing one distribution by a certain family of distributions
it allows the derivation of test for various parametric distributions. For instance, the resulting test 
for multivariate normality has a better empirical power than standard alternatives \cite{szekely2005new}. 
However, another important application of the energy distance is the construction of so called \emph{distance correlations}.
This is a general concept to characterize multivariate dependence, but not just linear dependency as measured by standard Pearson correlation.
Consequently it allows to construct the so called \emph{energy test of independence} which tests for multivariate independence.

To understand the energy distance and its power even better it is useful  to highlight a representation of the energy distance in \eqref{eq_energy_dist} using characteristic function.
The energy distance $d_{E}$ can be represented as a weighted $L^2$-distance of characteristic the characteristic functions $\bsvarphi_{\bsX}$ and $\bsvarphi_{\bsY}$ of $\bsX$ and $\bsY$
with $\bsvarphi_{\bsX}(\bsz) = \E( e^{i\bsz'\bsX})$ and $\bsvarphi_{\bsY}(\bsz) = \E( e^{i\bsz'\bsY})$:
\begin{equation}
d_{E}(\bsX,\bsY) = \frac{\pi^{\frac{H+1}{2}}}{\Gamma(\frac{H+1}{2})} \int_{\R^H } \frac{|\bsvarphi_{\bsX}(\bsz) - \bsvarphi_{\bsY}(\bsz)|^2 }{\|\bsz\|_2^{H+\beta}} \, d \bsz
% \ \ \text{ with } \ \ C_{d} = \frac{\pi^{\frac{d+1}{2}}}{\Gamma(\frac{d+1}{2})}
\label{eq_energy_dist_characteristic}
\end{equation}

 The weight function $\|\bsz\|_2^{-(H+\beta)}$ is crucial. 
 It is the only choice of any continuous function $g$ such that
 the weighted $L^2$-distance 
$ C \int_{\R^H } g(\bsz) |\bsvarphi_{\bsX}(\bsz) - \bsvarphi_{\bsY}(\bsz)|^2 \, d \bsz $
 between the characteristic functions $\bsvarphi_{\bsX}$ and $\bsvarphi_{\bsX}$ is scale equivariant and rotational invariant. Thus, it is suitable for 
 the tests described above.

 Of course, this does not directly explain the high empirical power of the energy score, but it gives an intuition why it works well.
 Moreover, we know that for any dimension $H$ $\bsX$ is uniquely determined by its complex valued 
 characteristic function $\bsvarphi_{\bsX}$. Thus, in \eqref{eq_energy_dist_characteristic} we are measuring distances in the complex plane which 
 is isomorphic to the 2-dimensional space $\R^2$. Hence, if $H$ is greater than $2$, the energy distance measures de facto the discrepancy in a  in a lower dimensional space. Consequently, we should prefer the energy distance especially for long forecasting horizons.
 
 Note that the energy score should not necessarily be the only score to be considered. Of course, any proper scoring rule can be applied. A perfect forecasting model performs well with respect to every proper scoring rule. As any measure is designed for a certain purpose this might help to detect problems in the forecasting model. For instance, if a model scores well in the energy score but the opposite in the Dawid-Sebastiani score we have a
 strong indication that either the first or second moment of our forecast is not well captured by the corresponding forecasting model.
 
Here, it also worth mentioning, that the forecaster should not restrict to $H$-dimensional scores. Instead the forecaster should evaluate scores on lower 
marginal distributions as well. The an important special case, this is the evaluation on the $H$ marginal distributions, e.g. by the 
CRPS or log-score. This gives an indication in which forecasting horizon which model suffers problems concerning the marginal distribution. This, seems to be quite common practice in some fields 
of application, see e.g. \cite{hong2016probabilistic}.
Similarly, we can evaluate any score on low dimensional marginals, especially bivariate distributions. For instance, we can look at the bivariate
copula energy score for the dimension 1 and 2, then 2 and 3, etc.. By doing that, it gives us information about the capturing of the pairwise 
dependency structure along the forecasting path. Thus, we can study how the predictive performance concerning the pairwise
dependency structure changes across the forecasting horizon.
The described procedure is also applied in the real data example of the next section.
 
%  The main reason for the special properties of the energy distance (or energy statistic) are due to the rotational invariant characterization as a
% weighted $L^2$-distance of the characteristic functions. Similarly to the cumulative distribution function the characteristic function $\bsvarphi_{\bsX}$
% characterizes $\bsX$ uniquely and vice versa. However, for any dimension $d$, $\bsvarphi_{\bsX}$ is only a complex valued function. Thus, we are measuring 
% discrepancies in the complex plane which is isomorphic to $\R\times \R = \R^2$. In turn, especially in high dimensions $d$, the energy score should be favored
% against measures where integrals in higher dimensional spaces must be computed. This suggests that the energy distance is also suitable for scenario reduction where
% we usually face higher dimensional problems.

% However, the key properties. \eqref{eq_energy_dist}
%We observe, that the energy distance results 

% 
% 
% - again shortly discuss with pinson. study 
% 
% - mention relationship to energy distance  [how powerful it is - allowing arbitrary two sample test, or test for multivariate independence] 
% - and characteristic function characterization. As charact. is complex valued and this is isomorph to $\R^2$ we are measuring effectively distances 
% in a 2-dimensional space even for higher dimensional settings.

\section{Illustrative real data example} \label{sec:realdata}

\begin{figure}
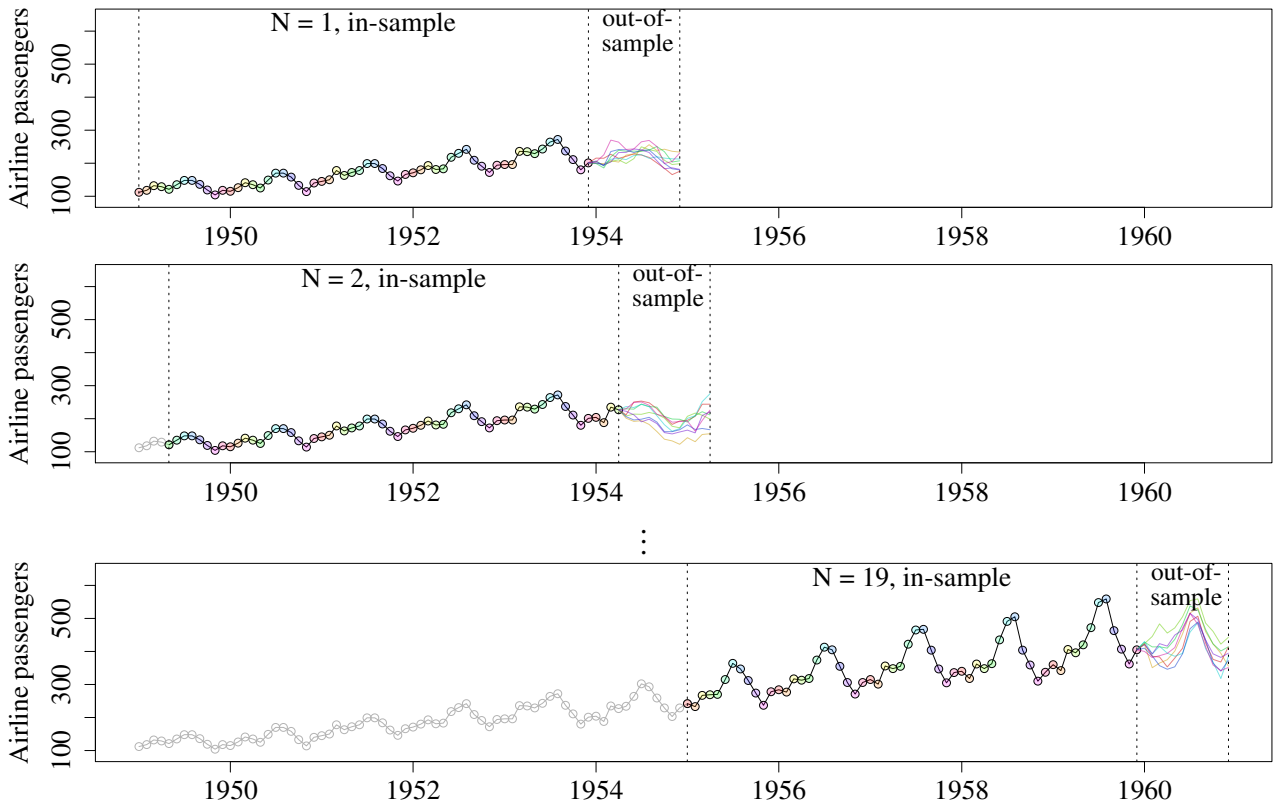

\resizebox{.99\textwidth}{!}{\input{fig/realdata/sample_iN1_H=12nY=19nFC=65536_s=3.tex}}
\resizebox{.99\textwidth}{!}{\input{fig/realdata/sample_iN2_H=12nY=19nFC=65536_s=3.tex}}
\centering{ $\vdots$ }
\resizebox{.99\textwidth}{!}{\input{fig/realdata/sample_iN19_H=12nY=19nFC=65536_s=3.tex}}
\caption{Illustration of the rolling window forecasting, for rolling window $1$, $2$ and $N=19$ with a small illustrative ensemble forecast 
with forecasting horizon $H=12$ and ensemble sample size $M=8$ of the AR(13) model \eqref{item_ar13} shown below.}
\label{fig_sample_air}
\end{figure}

For illustration purpose we consider the popular  classic Box and Jenkins airline data, which is denoted \texttt{AirPassengers} in \texttt{R}.
It contains monthly totals of international airline passengers from 1949 to 1960. Thus, there are in total 144 observations.
We apply a rolling window forecasting study with $T=5\times12 = 70$ as in-sample data length. We choose a forecasting horizon on $H=12$ which corresponds to a year
and shift the rolling window always by $4$ which corresponds to a quarter. Hence, we have in total $N=19$ forecasting windows. 
The forecasting study design is illustrated in Figure \ref{fig_sample_air}.

% /home/florian/Dropbox/Marginal Copula Score/forcast_eval/paper/fig/realdata/sample_mod_19_H=12nY=19nFC=65536_s=8.tex
% /home/florian/Dropbox/Marginal Copula Score/forcast_eval/paper/fig/realdata/sample_mod_19_H=12nY=19nFC=65536_s=5.tex
% /home/florian/Dropbox/Marginal Copula Score/forcast_eval/paper/fig/realdata/sample_mod_19_H=12nY=19nFC=65536_s=2.tex

% 
% 
% There hourly observations from 1st January 2009 0:00 to 31st December 2015 23:00 
% measured in Bremen, Germany. The data is measured and provided by the DWD.
% Note that the considered models are not sophisticated. They are only chosen to illustrate the methodology 
% if the true data generating model is unknown.
% 

\begin{figure}
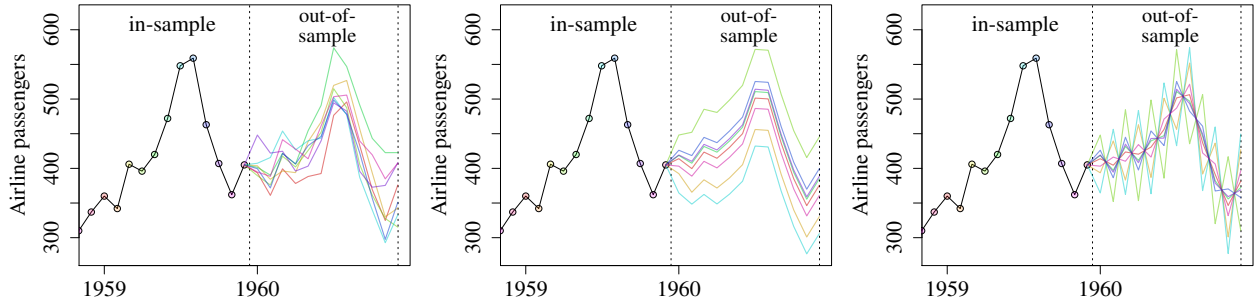

\resizebox{.32\textwidth}{!}{\input{fig/realdata/sample_mod_19_H=12nY=19nFC=65536_s=2.tex}}
\resizebox{.32\textwidth}{!}{\input{fig/realdata/sample_mod_19_H=12nY=19nFC=65536_s=5.tex}}
\resizebox{.32\textwidth}{!}{\input{fig/realdata/sample_mod_19_H=12nY=19nFC=65536_s=8.tex}}
\caption{Illustration of standard (left), comonotone (center) and countermonotone (right) model simulations for the AR(13) with $M=8$ paths
 for the last experiment ($N=19$).}
\label{fig_sample_monotone}
\end{figure}

We consider 9 very simple forecasting models, based on 3 AR models with 3 different error terms.
 \begin{enumerate}
% \item AR(1): $Y_t= \phi_0 + \phi_1 Y_{t-1} + \eps_t$.
\item AR(12): $Y_t= \phi_0 + \sum_{k=1}^{12} \phi_k Y_{t-k} + \eps_t$ with $\eps_t$ iid and $\E(\eps_t)=0$. \label{item_ar12}
\item AR(13): $Y_t= \phi_0 + \sum_{k=1}^{13} \phi_k Y_{t-k} + \eps_t$ with $\eps_t$ iid and $\E(\eps_t)=0$. \label{item_ar13}
\item AR($p$): $Y_t= \phi_0 + \sum_{k=1}^p \phi_k Y_{t-k} + \eps_t$ with $\eps_t$ iid, $\E(\eps_t)=0$ and $p\in \{1,\ldots, T/2\}$ such that the corresponding Akaike information criterion (AIC) is minimized.  \label{item_arp}

\item AR(12) as in \ref{item_ar12}. but with comonotone residuals (i.e. $(\what{\eps}_t,\what{\eps}_{t+1})$ have the copula $\bsM_2$)
\item AR(13) as in \ref{item_ar13}. but with comonotone residuals (i.e. $(\what{\eps}_t,\what{\eps}_{t+1})$ have the copula $\bsM_2$)
\item AR($p$) as in \ref{item_arp}. but with comonotone residuals (i.e. $(\what{\eps}_t,\what{\eps}_{t+1})$ have the copula $\bsM_2$)

\item AR(12) as in \ref{item_ar12}. but with countermonotone residuals (i.e. $(\what{\eps}_t,\what{\eps}_{t+1})$ have the copula $\bsW_2$)
\item AR(13)  as in \ref{item_ar13}. but with countermonotone residuals  (i.e. $(\what{\eps}_t,\what{\eps}_{t+1})$ have the copula $\bsW_2$)
\item AR($p$)  as in \ref{item_arp}. but with countermonotone residuals  (i.e. $(\what{\eps}_t,\what{\eps}_{t+1})$ have the copula $\bsW_2$)
\end{enumerate}
All models are estimated by solving the Yule-Walker equations which yields stationary estimated process. The ensemble simulations are generated using residual based bootstrap. For the reporting of the forecasts we consider $M=2^{16}=65536$ simulated paths.
Note that the comonotone and countermonotone models have exactly the same predicted marginal distribution, but a 
crucially different dependency structure. A small example of simulated paths is given in Figure \ref{fig_sample_monotone} simulated for 
the AR(13) in \ref{item_ar13}.
We evaluate all nine evaluation measures as in the simulation studies. 

\begin{figure}
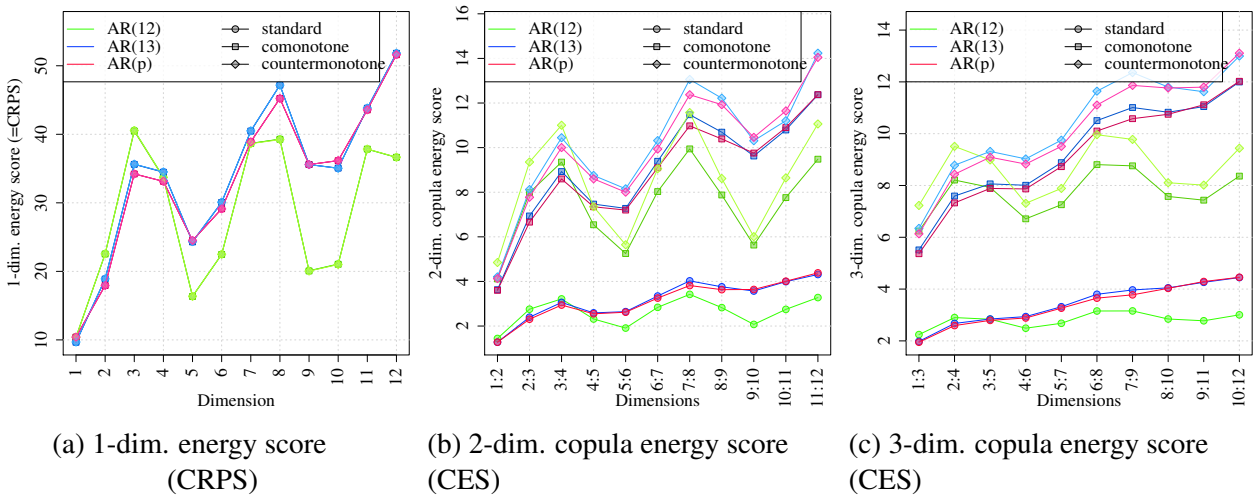


\begin{subfigure}[b]{0.32\textwidth}
\resizebox{1\textwidth}{!}{\input{fig/realdata/DM_E1_N=19M=65536.tex}}
\caption{1-dim. energy score \newline (CRPS)}
\label{fig_sample_marg_dim1}
    \end{subfigure}
% \begin{subfigure}[b]{0.32\textwidth}
% \resizebox{1\textwidth}{!}{\input{fig/realdata/DM_E2_N=19M=65536.tex}}
% \caption{2-dim. energy score (ES)}
% \label{fig_sample_marg_dim2}
%     \end{subfigure}
% \begin{subfigure}[b]{0.32\textwidth}
% \resizebox{1\textwidth}{!}{\input{fig/realdata/DM_E3_N=19M=65536.tex}}
% \caption{3-dim. energy score (ES)}
% \label{fig_sample_marg_dim3}
%     \end{subfigure}
    \begin{subfigure}[b]{0.32\textwidth}
\resizebox{1\textwidth}{!}{\input{fig/realdata/DM_CE2_N=19M=65536.tex}}
\caption{2-dim. copula energy score \newline (CES)}
\label{fig_sample_marg_dim2}
    \end{subfigure}
\begin{subfigure}[b]{0.32\textwidth}
\resizebox{1\textwidth}{!}{\input{fig/realdata/DM_CE3_N=19M=65536.tex}}
\caption{3-dim. copula energy score \newline (CES)}
\label{fig_sample_marg_dim3}
    \end{subfigure}

% \resizebox{.32\textwidth}{!}{\input{fig/realdata/DM_E1_N=19M=65536.tex}}
% \resizebox{.32\textwidth}{!}{\input{fig/realdata/DM_E2_N=19M=65536.tex}}
% \resizebox{.32\textwidth}{!}{\input{fig/realdata/DM_E3_N=19M=65536.tex}}

\caption{Low-dimensional scores across the forecasting horizon for all considered models.}
\label{fig_sample_marg}
\end{figure}

In Table \ref{tab_air_scores} the sample means (see eqn. \eqref{eq_score_average}) 
of all nine scores and nine models across all $N=19$ paths are given\footnote{Note that due to collinearity the copula-Dawid-Sebastiani score (CDSS) 
can not be computed for the comonotone and countermonotone AR models.}. 
DM-statistics and the corresponding p-values for the energy score is given in  are Table \ref{tab_DM_ES}. The corresponding tables for 
the remaining eight scores are provided in the Appendix. Additionally, to the considered scores for $H$-dimensional forecasts we compute 
the selected low-dimensional scores along the forecasting horizon are given in Figure \ref{fig_sample_marg}.

\setlength{\tabcolsep}{2pt}

% latex table generated in R 3.4.3 by xtable 1.8-2 package
% Mon Sep 10 15:40:15 2018
\begin{table}[ht]
\centering
\resizebox{.99\textwidth}{!}{
\begin{tabular}{rccccccccc}
  \hline
 \footnotesize Score$\backslash$Model & \footnotesize AR(12) & \footnotesize AR(13) & \footnotesize AR(p) & \footnotesize AR(12)-M & \footnotesize AR(13)-M & \footnotesize AR(p)-M & \footnotesize AR(12)-W & \footnotesize AR(13)-W & \footnotesize AR(p)-W \\ 
  \hline
ES & \cellcolor[rgb]{0.6,1,0.6} 120.8 & \cellcolor[rgb]{1,0.807,0.6} 137.5 & \cellcolor[rgb]{1,0.845,0.6} 134.9 & \cellcolor[rgb]{1,0.522,0.522} 197.3 & \cellcolor[rgb]{1,0.524,0.524} 196.2 & \cellcolor[rgb]{1,0.537,0.537} 190.9 & \cellcolor[rgb]{1,0.508,0.508} 203.2 & \cellcolor[rgb]{1,0.5,0.5} 206.6 & \cellcolor[rgb]{1,0.513,0.513} 201.0 \\ 
  VS & \cellcolor[rgb]{1,0.605,0.6} 158011 & \cellcolor[rgb]{1,0.932,0.6} 120657 & \cellcolor[rgb]{0.6,1,0.6} 111823 & \cellcolor[rgb]{1,0.5,0.5} 205513 & \cellcolor[rgb]{1,0.776,0.6} 133903 & \cellcolor[rgb]{1,0.906,0.6} 122607 & \cellcolor[rgb]{1,0.565,0.565} 175112 & \cellcolor[rgb]{1,0.892,0.6} 123631 & \cellcolor[rgb]{0.87,1,0.6} 114357 \\ 
  DSS & \cellcolor[rgb]{0.6,1,0.6}     95.92 & \cellcolor[rgb]{0.6,1,0.6}     90.99 & \cellcolor[rgb]{0.6,1,0.6}     91.78 & \cellcolor[rgb]{1,0.563,0.563} 683635 & \cellcolor[rgb]{1,0.684,0.6} 372563 & \cellcolor[rgb]{1,0.689,0.6} 364726 & \cellcolor[rgb]{1,0.5,0.5} 995252 & \cellcolor[rgb]{1,0.532,0.532} 838370 & \cellcolor[rgb]{1,0.547,0.547} 762531 \\ 
  CRPS-CES & \cellcolor[rgb]{0.6,1,0.6}  3.228 & \cellcolor[rgb]{1,0.937,0.6}  3.887 & \cellcolor[rgb]{1,0.947,0.6}  3.828 & \cellcolor[rgb]{1,0.553,0.553}  8.617 & \cellcolor[rgb]{1,0.508,0.508} 10.240 & \cellcolor[rgb]{1,0.513,0.513} 10.045 & \cellcolor[rgb]{1,0.547,0.547}  8.818 & \cellcolor[rgb]{1,0.5,0.5} 10.537 & \cellcolor[rgb]{1,0.505,0.505} 10.365 \\ 
  CRPS-CVS & \cellcolor[rgb]{0.6,1,0.6} 0.3346 & \cellcolor[rgb]{1,0.806,0.6} 0.4638 & \cellcolor[rgb]{1,0.895,0.6} 0.4166 & \cellcolor[rgb]{1,0.63,0.6} 0.6357 & \cellcolor[rgb]{1,0.568,0.568} 0.7712 & \cellcolor[rgb]{1,0.582,0.582} 0.7248 & \cellcolor[rgb]{1,0.559,0.559} 0.7989 & \cellcolor[rgb]{1,0.5,0.5} 0.9954 & \cellcolor[rgb]{1,0.515,0.515} 0.9464 \\ 
  CRPS & \cellcolor[rgb]{0.6,1,0.6} 28.28 & \cellcolor[rgb]{1,0.5,0.5} 33.94 & \cellcolor[rgb]{1,0.521,0.521} 33.36 & \cellcolor[rgb]{0.67,1,0.6} 28.32 & \cellcolor[rgb]{1,0.501,0.501} 33.90 & \cellcolor[rgb]{1,0.521,0.521} 33.36 & \cellcolor[rgb]{0.623,1,0.6} 28.29 & \cellcolor[rgb]{1,0.501,0.501} 33.92 & \cellcolor[rgb]{1,0.521,0.521} 33.35 \\ 
  CES & \cellcolor[rgb]{0.6,1,0.6} 0.1142 & \cellcolor[rgb]{0.613,1,0.6} 0.1144 & \cellcolor[rgb]{0.622,1,0.6} 0.1146 & \cellcolor[rgb]{1,0.507,0.507} 0.3044 & \cellcolor[rgb]{1,0.51,0.51} 0.3020 & \cellcolor[rgb]{1,0.511,0.511} 0.3010 & \cellcolor[rgb]{1,0.5,0.5} 0.3118 & \cellcolor[rgb]{1,0.501,0.501} 0.3105 & \cellcolor[rgb]{1,0.501,0.501} 0.3106 \\ 
  CVS & \cellcolor[rgb]{0.6,1,0.6} 0.01184 & \cellcolor[rgb]{1,0.919,0.6} 0.01367 & \cellcolor[rgb]{0.96,1,0.6} 0.01247 & \cellcolor[rgb]{1,0.579,0.579} 0.02247 & \cellcolor[rgb]{1,0.575,0.575} 0.02276 & \cellcolor[rgb]{1,0.587,0.587} 0.02171 & \cellcolor[rgb]{1,0.512,0.512} 0.02826 & \cellcolor[rgb]{1,0.5,0.5} 0.02935 & \cellcolor[rgb]{1,0.511,0.511} 0.02836 \\ 
  CDSS & \cellcolor[rgb]{0.6,1,0.6} -26.06 & \cellcolor[rgb]{1,0.567,0.567} -24.54 & \cellcolor[rgb]{1,0.5,0.5} -23.78 & \cellcolor[rgb]{.9,.9,.9}  -       & \cellcolor[rgb]{.9,.9,.9}  -    & \cellcolor[rgb]{.9,.9,.9}  -      & \cellcolor[rgb]{.9,.9,.9}  -       & \cellcolor[rgb]{.9,.9,.9}  -       & \cellcolor[rgb]{.9,.9,.9}  -       \\ 
   \hline
\end{tabular}}
\caption{Score averages $\ov{\text{SC}}$ across the $N=19$ out-of-sample windows for the considered scores and models.
-M indecates models with comonotone residuals, -W for countermonotone residuals.} 
\label{tab_air_scores}
\end{table}

% 
% /home/florian/Dropbox/Marginal Copula Score/forcast_eval/paper/fig/realdata/DM_E3_N=19M=1024.tex
% /home/florian/Dropbox/Marginal Copula Score/forcast_eval/paper/fig/realdata/DM_E2_N=19M=1024.tex
% /home/florian/Dropbox/Marginal Copula Score/forcast_eval/paper/fig/realdata/DM_E1_N=19M=1024.tex

% In Tab. \ref{tab_DM_ES} we see the results for the energy score. For the remaining scores they are given in the appendix.

% latex table generated in R 3.4.3 by xtable 1.8-2 package
% Fri Sep  7 13:28:49 2018
% latex table generated in R 3.4.3 by xtable 1.8-2 package
% Mon Sep 10 15:25:32 2018
\begin{table}[ht]
\centering
\begin{tabular}{rlllllllll}
  \hline
 & \footnotesize AR(12) & \footnotesize AR(13) & \footnotesize AR(p) & \footnotesize AR(12)-M & \footnotesize AR(13)-M & \footnotesize AR(p)-M & \footnotesize AR(12)-W & \footnotesize AR(13)-W & \footnotesize AR(p)-W \\ 
  \hline
AR(12) & \cellcolor[rgb]{0.6,0.6,0.6} & \cellcolor[rgb]{0.601,1,0.6} $\underset{[<0.001]}{  -4.04}$ & \cellcolor[rgb]{0.73,1,0.6} $\underset{[0.005]}{  -2.57}$ & \cellcolor[rgb]{0.6,1,0.6} $\underset{[<0.001]}{ -26.53}$ & \cellcolor[rgb]{0.6,1,0.6} $\underset{[<0.001]}{ -19.57}$ & \cellcolor[rgb]{0.6,1,0.6} $\underset{[<0.001]}{ -13.39}$ & \cellcolor[rgb]{0.6,1,0.6} $\underset{[<0.001]}{ -30.02}$ & \cellcolor[rgb]{0.6,1,0.6} $\underset{[<0.001]}{ -17.90}$ & \cellcolor[rgb]{0.6,1,0.6} $\underset{[<0.001]}{ -13.02}$ \\ 
  AR(13) & \cellcolor[rgb]{1,0.601,0.6} $\underset{[>0.999]}{   4.04}$ & \cellcolor[rgb]{0.6,0.6,0.6} & \cellcolor[rgb]{1,0.923,0.6} $\underset{[0.766]}{   0.73}$ & \cellcolor[rgb]{0.6,1,0.6} $\underset{[<0.001]}{ -11.15}$ & \cellcolor[rgb]{0.6,1,0.6} $\underset{[<0.001]}{ -31.58}$ & \cellcolor[rgb]{0.6,1,0.6} $\underset{[<0.001]}{ -13.67}$ & \cellcolor[rgb]{0.6,1,0.6} $\underset{[<0.001]}{ -13.57}$ & \cellcolor[rgb]{0.6,1,0.6} $\underset{[<0.001]}{ -35.14}$ & \cellcolor[rgb]{0.6,1,0.6} $\underset{[<0.001]}{ -14.51}$ \\ 
  AR(p) & \cellcolor[rgb]{1,0.73,0.6} $\underset{[0.995]}{   2.57}$ & \cellcolor[rgb]{0.923,1,0.6} $\underset{[0.234]}{  -0.73}$ & \cellcolor[rgb]{0.6,0.6,0.6} & \cellcolor[rgb]{0.6,1,0.6} $\underset{[<0.001]}{  -9.12}$ & \cellcolor[rgb]{0.6,1,0.6} $\underset{[<0.001]}{ -14.06}$ & \cellcolor[rgb]{0.6,1,0.6} $\underset{[<0.001]}{ -29.95}$ & \cellcolor[rgb]{0.6,1,0.6} $\underset{[<0.001]}{ -10.74}$ & \cellcolor[rgb]{0.6,1,0.6} $\underset{[<0.001]}{ -16.87}$ & \cellcolor[rgb]{0.6,1,0.6} $\underset{[<0.001]}{ -31.22}$ \\ 
  AR(12)-M & \cellcolor[rgb]{1,0.6,0.6} $\underset{[>0.999]}{  26.53}$ & \cellcolor[rgb]{1,0.6,0.6} $\underset{[>0.999]}{  11.15}$ & \cellcolor[rgb]{1,0.6,0.6} $\underset{[>0.999]}{   9.12}$ & \cellcolor[rgb]{0.6,0.6,0.6} & \cellcolor[rgb]{1,0.971,0.6} $\underset{[0.602]}{   0.26}$ & \cellcolor[rgb]{1,0.896,0.6} $\underset{[0.860]}{   1.08}$ & \cellcolor[rgb]{0.6,1,0.6} $\underset{[<0.001]}{  -7.18}$ & \cellcolor[rgb]{0.812,1,0.6} $\underset{[0.031]}{  -1.86}$ & \cellcolor[rgb]{0.94,1,0.6} $\underset{[0.292]}{  -0.55}$ \\ 
  AR(13)-M & \cellcolor[rgb]{1,0.6,0.6} $\underset{[>0.999]}{  19.57}$ & \cellcolor[rgb]{1,0.6,0.6} $\underset{[>0.999]}{  31.58}$ & \cellcolor[rgb]{1,0.6,0.6} $\underset{[>0.999]}{  14.06}$ & \cellcolor[rgb]{0.971,1,0.6} $\underset{[0.398]}{  -0.26}$ & \cellcolor[rgb]{0.6,0.6,0.6} & \cellcolor[rgb]{1,0.879,0.6} $\underset{[0.918]}{   1.39}$ & \cellcolor[rgb]{0.788,1,0.6} $\underset{[0.024]}{  -1.98}$ & \cellcolor[rgb]{0.6,1,0.6} $\underset{[<0.001]}{  -9.32}$ & \cellcolor[rgb]{0.896,1,0.6} $\underset{[0.138]}{  -1.09}$ \\ 
  AR(p)-M & \cellcolor[rgb]{1,0.6,0.6} $\underset{[>0.999]}{  13.39}$ & \cellcolor[rgb]{1,0.6,0.6} $\underset{[>0.999]}{  13.67}$ & \cellcolor[rgb]{1,0.6,0.6} $\underset{[>0.999]}{  29.95}$ & \cellcolor[rgb]{0.896,1,0.6} $\underset{[0.140]}{  -1.08}$ & \cellcolor[rgb]{0.879,1,0.6} $\underset{[0.082]}{  -1.39}$ & \cellcolor[rgb]{0.6,0.6,0.6} & \cellcolor[rgb]{0.749,1,0.6} $\underset{[0.011]}{  -2.28}$ & \cellcolor[rgb]{0.6,1,0.6} $\underset{[<0.001]}{  -4.18}$ & \cellcolor[rgb]{0.6,1,0.6} $\underset{[<0.001]}{  -9.51}$ \\ 
  AR(12)-W & \cellcolor[rgb]{1,0.6,0.6} $\underset{[>0.999]}{  30.02}$ & \cellcolor[rgb]{1,0.6,0.6} $\underset{[>0.999]}{  13.57}$ & \cellcolor[rgb]{1,0.6,0.6} $\underset{[>0.999]}{  10.74}$ & \cellcolor[rgb]{1,0.6,0.6} $\underset{[>0.999]}{   7.18}$ & \cellcolor[rgb]{1,0.788,0.6} $\underset{[0.976]}{   1.98}$ & \cellcolor[rgb]{1,0.749,0.6} $\underset{[0.989]}{   2.28}$ & \cellcolor[rgb]{0.6,0.6,0.6} & \cellcolor[rgb]{0.919,1,0.6} $\underset{[0.219]}{  -0.78}$ & \cellcolor[rgb]{1,0.959,0.6} $\underset{[0.642]}{   0.36}$ \\ 
  AR(13)-W & \cellcolor[rgb]{1,0.6,0.6} $\underset{[>0.999]}{  17.90}$ & \cellcolor[rgb]{1,0.6,0.6} $\underset{[>0.999]}{  35.14}$ & \cellcolor[rgb]{1,0.6,0.6} $\underset{[>0.999]}{  16.87}$ & \cellcolor[rgb]{1,0.812,0.6} $\underset{[0.969]}{   1.86}$ & \cellcolor[rgb]{1,0.6,0.6} $\underset{[>0.999]}{   9.32}$ & \cellcolor[rgb]{1,0.6,0.6} $\underset{[>0.999]}{   4.18}$ & \cellcolor[rgb]{1,0.919,0.6} $\underset{[0.781]}{   0.78}$ & \cellcolor[rgb]{0.6,0.6,0.6} & \cellcolor[rgb]{1,0.879,0.6} $\underset{[0.918]}{   1.39}$ \\ 
  AR(p)-W & \cellcolor[rgb]{1,0.6,0.6} $\underset{[>0.999]}{  13.02}$ & \cellcolor[rgb]{1,0.6,0.6} $\underset{[>0.999]}{  14.51}$ & \cellcolor[rgb]{1,0.6,0.6} $\underset{[>0.999]}{  31.22}$ & \cellcolor[rgb]{1,0.94,0.6} $\underset{[0.708]}{   0.55}$ & \cellcolor[rgb]{1,0.896,0.6} $\underset{[0.862]}{   1.09}$ & \cellcolor[rgb]{1,0.6,0.6} $\underset{[>0.999]}{   9.51}$ & \cellcolor[rgb]{0.959,1,0.6} $\underset{[0.358]}{  -0.36}$ & \cellcolor[rgb]{0.879,1,0.6} $\underset{[0.082]}{  -1.39}$ & \cellcolor[rgb]{0.6,0.6,0.6} \\ 
   \hline
\end{tabular}
\caption{DM-test statistics with corresponding p-value given in squared brackets for the energy score (ES).} 
\label{tab_DM_ES}
\end{table}

In Table \ref{tab_air_scores} we observe that the energy score is the smallest for the AR(12). 
Even though for the AR(13) and AR(p) the energy score has a magnitude their performance is significantly worse than the AR(12) with 
respect to the energy score, see Table \ref{tab_DM_VS}.
Moreover, we see that the three AR models with comonotone residuals have similar scores, as well as the three models with countermonotone residuals but with larger scores. It shows us that the energy score takes the dependency structure into account and favors a comonotone
to a countermonotone. This makes intuitively sense as the time series data tends to be positively correlated.

However, when looking at the variogram and Dawid-Sebastiani score we get a similar picture.
As both scores are proper and
the corresponding  DM-statistics (Tab.
\ref{tab_DM_VS}
 and \ref{tab_DM_DSS}) are clearly greater than 2 this indicated that the AR(12) model can not be optimal. It is likely that 
 the correlation structure of the AR(13) and AR(p) is better captured than for the AR(12). But the AR(12) seems to have a better predictive 
 performance concerning the marginal distributions as CRPS is significantly smaller (Tab. \ref{tab_air_scores} and \ref{tab_DM_CRPS}).
 This illustrative study, also shows that any proper scoring rule should never be used alone without considering a strictly proper one.
 For instance, variogram score (Tab. \ref{tab_DM_VS}) states that the 
 co- and countermonotone AR(p) provide significantly better forecasts than the AR(12), which is hard do agree if you just eyeball the corresponding forecasts.

 Given the simulation studies conducted before, it is not surprising that 
 the CRPS-copula energy score (Tab. \ref{tab_DM_CRPS-CES}) gives similar results to the energy score. In some cases it even more sensitive than 
 the energy score, e.g. the AR(12)-M has significantly worse forecasting accuracy than the AR(p), but the DM-statistic is about 9
 for the energy score but about 15 for the CRPS-CES. The reason is the strong sensitivity of the copula energy score (CES) (Tab. \ref{tab_DM_CES}). It detects easily that the comonotone and countermonotone AR processes do not match the data behaviour. All 
 test statistics are extremely large, from at least 28 to about 158 which is amazing having in mind that we have only $N=19$ rolling window experiments.
 Indeed, the copula energy score is the most sensitive score concerning the detection on mis-predicted dependency structures.
 However, it is followed by energy score which is still quite sensitive for dependency misspecifications.

 Now, we take a closer look at the lower dimensional marginal evaluation (Fig. \ref{fig_sample_marg}).
 Obviously, we observe that the 1-dimensional energy score is essentially the same for the standard, co- and countermonotone AR models of the same type
 (Fig. \eqref{fig_sample_marg_dim1}). But, we see that AR(12) has a larger CRPS than 
 the AR(13) and the similar performing AR(p) for the first three time steps. For larger forecasting horizons the 
 AR(12) is clearly best. Thus, it might be worth to combine both model aspects into a better one.
 When looking at the pairwise evaluation based on the copula energy score (Fig. \eqref{fig_sample_marg_dim2}) we observe that the co- and countermonotone 
 models can be well separated from the more plausible AR models. Interestingly the 3-dimensional copula energy score (Fig. \eqref{fig_sample_marg_dim3}) gives roughly the same picture as 
 the bivariate one. Which allows the conclusion that the main differences in the dependency structure is likely due to pairwise dependencies.

 %  .  TODO significanty diff?

% that were used in simulation studies before.

\section{Summary and Conclusion} \label{sec:conclusion}

We analyzed in detail established and newly introduced scoring rules for multivariate forecasting evaluation. 
% The focus in the evaluation is the dependency structure of the forecast, 
The introduced \textit{marginal-copula scoring rules} can excellently evaluate the dependency structure in forecasts, this 
holds especially for the \textit{copula energy score}. Additionally, we also discuss reporting and forecasting study design. 
Thus, our clear suggestion is to apply \textit{ensemble forecasting} in a rolling window forecasting study framework.  This is to report multivariate forecast by providing a large ensemble from the forecasting model, which can be realized by Monte-Carlo simulation for statistical models.

Across several simulation studies, the strictly proper \textit{energy score} is the only evaluation measure which can clearly separate the true model from
all alternatives. This contradicts the conclusion from \cite{pinson2013discrimination} that the energy score is not suitable for evaluating differences in the dependency structure, 
especially correlations. A crucial element is that the \textit{Diebold-Mariono-test} is considered for checking the discrimination ability. 
Our simulation results allow the conclusion that the energy score should be preferred evaluation measure for multivariate predictions. Still, it should be combined in practice with 
other measures, esp. the \textit{CRPS and bivariate copula energy scores}, the first one for checking the marginal distribution fit, the latter one for the pairwise dependency fit.
Summarizing, our guideline for forecasters is:

 \begin{itemize}
\item[(1)] Do \textbf{ensemble forecasts}
% \item[(2)] Consider 
with a \textbf{large} Monte-Carlo / ensemble \textbf{sample size} 
\item[(2)] Evaluate on the full-dimensional \textbf{energy score}
\item[(3)] Apply the \textbf{Diebold-Mariano-test for significance} evaluation
\item[(4)] (\textit{optional but recommended}) back it up by other scores, esp. \newline \textbf{CRPS and bivariate copula energy score} across the forecasting horizon
\end{itemize}

From the practical point of view it is nice to remember that there is nothing beyond the multivariate forecasting, as the every characteristic can be derived from
a large ensemble. Thus, now further forecasting concept has to be developed. Still, there are a couple of open questions.
Concerning the marginal copula scores, this is for example to 
look for better isotonic functions which combine marginal scores and copula scores more efficiently than the considered multiplicative structure.
As it seems that the copula energy score is suitable for analyzing the dependency structure of the forecasts it might be worth to study more analytical properties, 
esp. better or even sharp bounds.

As the energy score seems to be the most promising known evaluation criterion so far, it might be worth to study it and the energy distance in more detail as well. For instance, 
the energy score definition \eqref{eq_energy_score}, depends on $\beta$ which we fixed essentially to $1$ in this paper, even though 
$\beta\in(0,2)$ could be analyzed. 
As pointed out by \cite{szekely2013energy}, the energy distance also works for more general function than just $x\mapsto |x|^{\beta}$ in its definition, such as $x\mapsto \log(1+|x|^2)$.
In fact, every function which is the negative logarithm of an infinitely divisible characteristic function works out and can be studied in this context.

Finally, we want to remark that there might exists other scoring rules suitable for forecasting evaluation that were not studied in literature so far. Promising candidates
could be derived from probability distances, as e.g. the Wasserstein distance or the Hellinger distance.
%  mention to look for better bounds, 
%  
%  mention to look for better isotonic functions $g$ that combine copula and marginal score.
 
% Just a cite... \cite{ziel2015efficient}
% 
% \section*{Acknowledgments}
% The authors thank Alfred M\"uller and Pierre Pinson for helpful comments concerning the copula energy score.

\section*{Appendix}

\subsection*{On bounds for the copula energy score}

% \KB{REST OF THE SECTION IN APPENDIX??} 

% \subsubsection{Simple bounds}
\begin{lemma} \label{lemma_1_appendix}
It holds:
$\frac{\sqrt{H}}{4} \le \E  \left\| \bsU_{\bsX} - \bsu_{\bsY} \right\|_2  \le \frac{\sqrt{H}}{\sqrt{3}}$. 
\end{lemma}

\begin{proof}
First we note that $ \E \left\| \bsU_{\bsX} - \bsu_{\bsY} \right\|_2 $ with is convex in $\bsu_{\bsY}$.
As convex functions on a compact domain take their maximum on an extremal point. Thus, $\E  \left\| \bsU_{\bsX} - \bsu_{\bsY} \right\|_2$ 
attains the maximum for $\bsu_{\bsY} \in \{0,1\}^H$.
Then we have with Jensens inequality and $ \E[U_{\bsX,h}^2]=\frac{1}{3}$ 
\begin{align*}
\E  \left\| \bsU_{\bsX} - \bsu_{\bsY} \right\|_2 = \E \sqrt{ \sum_{h=1}^H |U_{\bsX,h} - u_{\bsY, h}|^2 }
&= \E \sqrt{ \sum_{h=1}^H U_{\bsX,h}^2 } \\
&\leq \sqrt{ \E \left( \sum_{h=1}^H U_{\bsX,h}^2 \right) }
&= \sqrt{ \sum_{h=1}^H \E (U_{\bsX,h}^2) } = \sqrt{\frac{H}{3}}
\end{align*}

 For the lower bound we note first that 
 $ Z_{h,\bsu_{\bsY}}^2 = |U_{\bsX,h} - u_{\bsY, h}|^2  $ is stochastically larger than $ |U_{\bsX,h} - \frac{1}{2}|^2 $. 
  Let $(Z_{h,\bsu_{\bsY}}^+)^2$ for $h=1,\ldots,H$ comonotone random variables with the same distributions as $Z_{h,\bsu_{\bsY}}^2$. 
Notice that in general $\bsu_{\bsY}$ is not the corresponding $Z_{h,\bsu_{\bsY}}^2$ for $\bsC_{\bsX}=\bsM_H$. 
However, this holds true, if $u_{\bsY, 1}=u_{\bsY, 2}=\ldots = u_{\bsY, H}$, thus
in particular for $\bsu_{\bsY} = \frac{1}{2}\bsone$. Therefore, we get
 \begin{align*} 
   \E  \left\| \bsU_{\bsX} - \bsu_{\bsY} \right\|_2  =  \ \E\left( \sqrt{ \sum_{h=1}^H Z_{h,u_{\bsY, h}}^2 } \right) & \ge  \ \E\left( \sqrt{ \sum_{h=1}^H  (Z_{h,u_{\bsY, h}}^+)^2} \right)\\
  & \ge  \ \E\left( \sqrt{ \sum_{h=1}^H  (Z_{h,\frac{1}{2}\bsone}^+)^2} \right) \\
  &= 
  \ \E\left( \sqrt{ H  (U_{\bsX,1} - \frac{1}{2}\bsone )^2} \right) = \sqrt{H} \E |U_{\bsX,1} - \frac{1}{2}| =  \frac{\sqrt{H}}{4}.
 \end{align*}
\end{proof}

% and 
% \begin{align}

% second
% \end{align}

\begin{lemma} \label{lemma_2_appendix}
It holds:
$ \frac{\sqrt{H}}{3} \le \E\left( \| \bsU_{\bsX} - \wtilde{\bsU}_{\bsX} \|_2 \right) \le \frac{\sqrt{H}}{\sqrt{6}} $ 
\end{lemma}
\begin{proof}
The upper bound follows again by Jensens inequality
\begin{align}
 \E  \left\| \bsU_{\bsX} - \bsU_{\bsY} \right\|_2 
 = \E \sqrt{ \sum_{h=1}^H |U_{\bsX,h} - U_{\bsY, h}|^2 } %\\
&= \E \sqrt{ \sum_{h=1}^H |U_{\bsX,h} - U_{\bsY, h}|^2 } \\
&\leq \sqrt{ \E \left( \sum_{h=1}^H |U_{\bsX,h}^2 - U_{\bsY, h}|^2\right) } \\
&= \sqrt{ \sum_{h=1}^H \E |U_{\bsX,h} - U_{\bsY, h}|^2 } = \sqrt{\frac{H}{6}}
\end{align}
as $|U_{\bsX,h} - U_{\bsY, h}|^2$ has the density $(1-\sqrt{z})/\sqrt{z}$ on $(0,1)$ with yields $\E|U_{\bsX,h} - U_{\bsY, h}|^2 = \frac{1}{6}$.

For the lower bound we note again that 
$ \E \left\| \bsU_{\bsX} - \bsU_{\bsY}  \right\|_2 $ due to symmetry the minimum is attained for $\bsU_{\bsX} = \bsU_{\bsY}$.
Thus, we have
$ \E \left\| \bsU_{\bsX} - \bsU_{\bsY}  \right\|_2 =
\E \sqrt{ \sum_{h=1}^H Z_h }, $
where $Z_h$ has the density $(1-\sqrt{z})/\sqrt{z}$ on $(0,1)$.
Hence, $\E \sqrt{ \sum_{h=1}^H Z_h }$ can be written 
as $\E f(Z_1,\ldots, Z_H)$ for a submodular function $f$ (see e.g. \cite{muller2002comparison}).
It follows by \cite{tchen1980inequalities} that a lower bound 
is given by the upper Fr\'{e}chet-Hoeffding bound $\bsM_H$.
Hence,
$ \E \left\| \bsU_{\bsX} - \bsU_{\bsY}  \right\|_2 \geq \E \sqrt{ \sum_{h=1}^H \sqrt{Z_1} } = \sqrt{H} \E[\sqrt{Z_1}] = \sqrt{H}/3 .$

\end{proof}

\subsection*{DM-test tables}

% latex table generated in R 3.4.3 by xtable 1.8-2 package
% Mon Sep 10 15:25:32 2018
\begin{table}[ht]
\centering
\begin{tabular}{rlllllllll}
  \hline
 & \footnotesize AR(12) & \footnotesize AR(13) & \footnotesize AR(p) & \footnotesize AR(12)-M & \footnotesize AR(13)-M & \footnotesize AR(p)-M & \footnotesize AR(12)-W & \footnotesize AR(13)-W & \footnotesize AR(p)-W \\ 
  \hline
AR(12) & \cellcolor[rgb]{0.6,0.6,0.6} & \cellcolor[rgb]{1,0.6,0.6} $\underset{[>0.999]}{   5.45}$ & \cellcolor[rgb]{1,0.6,0.6} $\underset{[>0.999]}{   7.90}$ & \cellcolor[rgb]{0.6,1,0.6} $\underset{[<0.001]}{ -12.44}$ & \cellcolor[rgb]{1,0.671,0.6} $\underset{[0.997]}{   2.78}$ & \cellcolor[rgb]{1,0.6,0.6} $\underset{[>0.999]}{   5.65}$ & \cellcolor[rgb]{0.6,1,0.6} $\underset{[<0.001]}{ -11.83}$ & \cellcolor[rgb]{1,0.6,0.6} $\underset{[>0.999]}{   5.33}$ & \cellcolor[rgb]{1,0.6,0.6} $\underset{[>0.999]}{   8.03}$ \\ 
  AR(13) & \cellcolor[rgb]{0.6,1,0.6} $\underset{[<0.001]}{  -5.45}$ & \cellcolor[rgb]{0.6,0.6,0.6} & \cellcolor[rgb]{1,0.669,0.6} $\underset{[0.997]}{   2.79}$ & \cellcolor[rgb]{0.6,1,0.6} $\underset{[<0.001]}{ -11.30}$ & \cellcolor[rgb]{0.6,1,0.6} $\underset{[<0.001]}{  -4.74}$ & \cellcolor[rgb]{0.92,1,0.6} $\underset{[0.224]}{  -0.76}$ & \cellcolor[rgb]{0.6,1,0.6} $\underset{[<0.001]}{  -7.71}$ & \cellcolor[rgb]{0.601,1,0.6} $\underset{[<0.001]}{  -3.95}$ & \cellcolor[rgb]{1,0.834,0.6} $\underset{[0.962]}{   1.77}$ \\ 
  AR(p) & \cellcolor[rgb]{0.6,1,0.6} $\underset{[<0.001]}{  -7.90}$ & \cellcolor[rgb]{0.669,1,0.6} $\underset{[0.003]}{  -2.79}$ & \cellcolor[rgb]{0.6,0.6,0.6} & \cellcolor[rgb]{0.6,1,0.6} $\underset{[<0.001]}{ -12.49}$ & \cellcolor[rgb]{0.601,1,0.6} $\underset{[<0.001]}{  -3.87}$ & \cellcolor[rgb]{0.6,1,0.6} $\underset{[<0.001]}{  -5.73}$ & \cellcolor[rgb]{0.6,1,0.6} $\underset{[<0.001]}{ -10.16}$ & \cellcolor[rgb]{0.602,1,0.6} $\underset{[<0.001]}{  -3.86}$ & \cellcolor[rgb]{0.644,1,0.6} $\underset{[0.002]}{  -2.93}$ \\ 
  AR(12)-M & \cellcolor[rgb]{1,0.6,0.6} $\underset{[>0.999]}{  12.44}$ & \cellcolor[rgb]{1,0.6,0.6} $\underset{[>0.999]}{  11.30}$ & \cellcolor[rgb]{1,0.6,0.6} $\underset{[>0.999]}{  12.49}$ & \cellcolor[rgb]{0.6,0.6,0.6} & \cellcolor[rgb]{1,0.6,0.6} $\underset{[>0.999]}{   8.49}$ & \cellcolor[rgb]{1,0.6,0.6} $\underset{[>0.999]}{  11.33}$ & \cellcolor[rgb]{1,0.6,0.6} $\underset{[>0.999]}{   9.65}$ & \cellcolor[rgb]{1,0.6,0.6} $\underset{[>0.999]}{  11.35}$ & \cellcolor[rgb]{1,0.6,0.6} $\underset{[>0.999]}{  12.36}$ \\ 
  AR(13)-M & \cellcolor[rgb]{0.671,1,0.6} $\underset{[0.003]}{  -2.78}$ & \cellcolor[rgb]{1,0.6,0.6} $\underset{[>0.999]}{   4.74}$ & \cellcolor[rgb]{1,0.601,0.6} $\underset{[>0.999]}{   3.87}$ & \cellcolor[rgb]{0.6,1,0.6} $\underset{[<0.001]}{  -8.49}$ & \cellcolor[rgb]{0.6,0.6,0.6} & \cellcolor[rgb]{1,0.731,0.6} $\underset{[0.995]}{   2.55}$ & \cellcolor[rgb]{0.6,1,0.6} $\underset{[<0.001]}{  -4.69}$ & \cellcolor[rgb]{1,0.614,0.6} $\underset{[>0.999]}{   3.27}$ & \cellcolor[rgb]{1,0.618,0.6} $\underset{[>0.999]}{   3.21}$ \\ 
  AR(p)-M & \cellcolor[rgb]{0.6,1,0.6} $\underset{[<0.001]}{  -5.65}$ & \cellcolor[rgb]{1,0.92,0.6} $\underset{[0.776]}{   0.76}$ & \cellcolor[rgb]{1,0.6,0.6} $\underset{[>0.999]}{   5.73}$ & \cellcolor[rgb]{0.6,1,0.6} $\underset{[<0.001]}{ -11.33}$ & \cellcolor[rgb]{0.731,1,0.6} $\underset{[0.005]}{  -2.55}$ & \cellcolor[rgb]{0.6,0.6,0.6} & \cellcolor[rgb]{0.6,1,0.6} $\underset{[<0.001]}{  -7.94}$ & \cellcolor[rgb]{0.956,1,0.6} $\underset{[0.347]}{  -0.39}$ & \cellcolor[rgb]{1,0.603,0.6} $\underset{[>0.999]}{   3.73}$ \\ 
  AR(12)-W & \cellcolor[rgb]{1,0.6,0.6} $\underset{[>0.999]}{  11.83}$ & \cellcolor[rgb]{1,0.6,0.6} $\underset{[>0.999]}{   7.71}$ & \cellcolor[rgb]{1,0.6,0.6} $\underset{[>0.999]}{  10.16}$ & \cellcolor[rgb]{0.6,1,0.6} $\underset{[<0.001]}{  -9.65}$ & \cellcolor[rgb]{1,0.6,0.6} $\underset{[>0.999]}{   4.69}$ & \cellcolor[rgb]{1,0.6,0.6} $\underset{[>0.999]}{   7.94}$ & \cellcolor[rgb]{0.6,0.6,0.6} & \cellcolor[rgb]{1,0.6,0.6} $\underset{[>0.999]}{   7.79}$ & \cellcolor[rgb]{1,0.6,0.6} $\underset{[>0.999]}{  10.42}$ \\ 
  AR(13)-W & \cellcolor[rgb]{0.6,1,0.6} $\underset{[<0.001]}{  -5.33}$ & \cellcolor[rgb]{1,0.601,0.6} $\underset{[>0.999]}{   3.95}$ & \cellcolor[rgb]{1,0.602,0.6} $\underset{[>0.999]}{   3.86}$ & \cellcolor[rgb]{0.6,1,0.6} $\underset{[<0.001]}{ -11.35}$ & \cellcolor[rgb]{0.614,1,0.6} $\underset{[<0.001]}{  -3.27}$ & \cellcolor[rgb]{1,0.956,0.6} $\underset{[0.653]}{   0.39}$ & \cellcolor[rgb]{0.6,1,0.6} $\underset{[<0.001]}{  -7.79}$ & \cellcolor[rgb]{0.6,0.6,0.6} & \cellcolor[rgb]{1,0.666,0.6} $\underset{[0.997]}{   2.80}$ \\ 
  AR(p)-W & \cellcolor[rgb]{0.6,1,0.6} $\underset{[<0.001]}{  -8.03}$ & \cellcolor[rgb]{0.834,1,0.6} $\underset{[0.038]}{  -1.77}$ & \cellcolor[rgb]{1,0.644,0.6} $\underset{[0.998]}{   2.93}$ & \cellcolor[rgb]{0.6,1,0.6} $\underset{[<0.001]}{ -12.36}$ & \cellcolor[rgb]{0.618,1,0.6} $\underset{[<0.001]}{  -3.21}$ & \cellcolor[rgb]{0.603,1,0.6} $\underset{[<0.001]}{  -3.73}$ & \cellcolor[rgb]{0.6,1,0.6} $\underset{[<0.001]}{ -10.42}$ & \cellcolor[rgb]{0.666,1,0.6} $\underset{[0.003]}{  -2.80}$ & \cellcolor[rgb]{0.6,0.6,0.6} \\ 
   \hline
\end{tabular}
\caption{DM-test statistics with corresponding p-value given in squared brackets statistics for the variogram score (VS)} 
\label{tab_DM_VS}
\end{table}

% latex table generated in R 3.4.3 by xtable 1.8-2 package
% Mon Sep 10 15:25:32 2018
\begin{table}[ht]
\centering
\begin{tabular}{rlllllllll}
  \hline
 & \footnotesize AR(12) & \footnotesize AR(13) & \footnotesize AR(p) & \footnotesize AR(12)-M & \footnotesize AR(13)-M & \footnotesize AR(p)-M & \footnotesize AR(12)-W & \footnotesize AR(13)-W & \footnotesize AR(p)-W \\ 
  \hline
AR(12) & \cellcolor[rgb]{0.6,0.6,0.6} & \cellcolor[rgb]{1,0.6,0.6} $\underset{[>0.999]}{   7.01}$ & \cellcolor[rgb]{1,0.6,0.6} $\underset{[>0.999]}{   4.99}$ & \cellcolor[rgb]{0.6,1,0.6} $\underset{[<0.001]}{ -12.36}$ & \cellcolor[rgb]{0.6,1,0.6} $\underset{[<0.001]}{  -7.65}$ & \cellcolor[rgb]{0.6,1,0.6} $\underset{[<0.001]}{  -8.51}$ & \cellcolor[rgb]{0.6,1,0.6} $\underset{[<0.001]}{ -10.72}$ & \cellcolor[rgb]{0.6,1,0.6} $\underset{[<0.001]}{  -5.24}$ & \cellcolor[rgb]{0.6,1,0.6} $\underset{[<0.001]}{  -6.01}$ \\ 
  AR(13) & \cellcolor[rgb]{0.6,1,0.6} $\underset{[<0.001]}{  -7.01}$ & \cellcolor[rgb]{0.6,0.6,0.6} & \cellcolor[rgb]{0.748,1,0.6} $\underset{[0.011]}{  -2.30}$ & \cellcolor[rgb]{0.6,1,0.6} $\underset{[<0.001]}{ -12.36}$ & \cellcolor[rgb]{0.6,1,0.6} $\underset{[<0.001]}{  -7.65}$ & \cellcolor[rgb]{0.6,1,0.6} $\underset{[<0.001]}{  -8.51}$ & \cellcolor[rgb]{0.6,1,0.6} $\underset{[<0.001]}{ -10.72}$ & \cellcolor[rgb]{0.6,1,0.6} $\underset{[<0.001]}{  -5.24}$ & \cellcolor[rgb]{0.6,1,0.6} $\underset{[<0.001]}{  -6.01}$ \\ 
  AR(p) & \cellcolor[rgb]{0.6,1,0.6} $\underset{[<0.001]}{  -4.99}$ & \cellcolor[rgb]{1,0.748,0.6} $\underset{[0.989]}{   2.30}$ & \cellcolor[rgb]{0.6,0.6,0.6} & \cellcolor[rgb]{0.6,1,0.6} $\underset{[<0.001]}{ -12.36}$ & \cellcolor[rgb]{0.6,1,0.6} $\underset{[<0.001]}{  -7.65}$ & \cellcolor[rgb]{0.6,1,0.6} $\underset{[<0.001]}{  -8.51}$ & \cellcolor[rgb]{0.6,1,0.6} $\underset{[<0.001]}{ -10.72}$ & \cellcolor[rgb]{0.6,1,0.6} $\underset{[<0.001]}{  -5.24}$ & \cellcolor[rgb]{0.6,1,0.6} $\underset{[<0.001]}{  -6.01}$ \\ 
  AR(12)-M & \cellcolor[rgb]{1,0.6,0.6} $\underset{[>0.999]}{  12.36}$ & \cellcolor[rgb]{1,0.6,0.6} $\underset{[>0.999]}{  12.36}$ & \cellcolor[rgb]{1,0.6,0.6} $\underset{[>0.999]}{  12.36}$ & \cellcolor[rgb]{0.6,0.6,0.6} & \cellcolor[rgb]{1,0.6,0.6} $\underset{[>0.999]}{   4.68}$ & \cellcolor[rgb]{1,0.6,0.6} $\underset{[>0.999]}{   5.07}$ & \cellcolor[rgb]{0.612,1,0.6} $\underset{[<0.001]}{  -3.32}$ & \cellcolor[rgb]{0.908,1,0.6} $\underset{[0.181]}{  -0.91}$ & \cellcolor[rgb]{0.937,1,0.6} $\underset{[0.283]}{  -0.57}$ \\ 
  AR(13)-M & \cellcolor[rgb]{1,0.6,0.6} $\underset{[>0.999]}{   7.65}$ & \cellcolor[rgb]{1,0.6,0.6} $\underset{[>0.999]}{   7.65}$ & \cellcolor[rgb]{1,0.6,0.6} $\underset{[>0.999]}{   7.65}$ & \cellcolor[rgb]{0.6,1,0.6} $\underset{[<0.001]}{  -4.68}$ & \cellcolor[rgb]{0.6,0.6,0.6} & \cellcolor[rgb]{1,0.982,0.6} $\underset{[0.562]}{   0.16}$ & \cellcolor[rgb]{0.6,1,0.6} $\underset{[<0.001]}{  -7.68}$ & \cellcolor[rgb]{0.61,1,0.6} $\underset{[<0.001]}{  -3.35}$ & \cellcolor[rgb]{0.603,1,0.6} $\underset{[<0.001]}{  -3.72}$ \\ 
  AR(p)-M & \cellcolor[rgb]{1,0.6,0.6} $\underset{[>0.999]}{   8.51}$ & \cellcolor[rgb]{1,0.6,0.6} $\underset{[>0.999]}{   8.51}$ & \cellcolor[rgb]{1,0.6,0.6} $\underset{[>0.999]}{   8.51}$ & \cellcolor[rgb]{0.6,1,0.6} $\underset{[<0.001]}{  -5.07}$ & \cellcolor[rgb]{0.982,1,0.6} $\underset{[0.438]}{  -0.16}$ & \cellcolor[rgb]{0.6,0.6,0.6} & \cellcolor[rgb]{0.6,1,0.6} $\underset{[<0.001]}{  -7.73}$ & \cellcolor[rgb]{0.616,1,0.6} $\underset{[<0.001]}{  -3.24}$ & \cellcolor[rgb]{0.605,1,0.6} $\underset{[<0.001]}{  -3.53}$ \\ 
  AR(12)-W & \cellcolor[rgb]{1,0.6,0.6} $\underset{[>0.999]}{  10.72}$ & \cellcolor[rgb]{1,0.6,0.6} $\underset{[>0.999]}{  10.72}$ & \cellcolor[rgb]{1,0.6,0.6} $\underset{[>0.999]}{  10.72}$ & \cellcolor[rgb]{1,0.612,0.6} $\underset{[>0.999]}{   3.32}$ & \cellcolor[rgb]{1,0.6,0.6} $\underset{[>0.999]}{   7.68}$ & \cellcolor[rgb]{1,0.6,0.6} $\underset{[>0.999]}{   7.73}$ & \cellcolor[rgb]{0.6,0.6,0.6} & \cellcolor[rgb]{1,0.888,0.6} $\underset{[0.889]}{   1.22}$ & \cellcolor[rgb]{1,0.745,0.6} $\underset{[0.990]}{   2.33}$ \\ 
  AR(13)-W & \cellcolor[rgb]{1,0.6,0.6} $\underset{[>0.999]}{   5.24}$ & \cellcolor[rgb]{1,0.6,0.6} $\underset{[>0.999]}{   5.24}$ & \cellcolor[rgb]{1,0.6,0.6} $\underset{[>0.999]}{   5.24}$ & \cellcolor[rgb]{1,0.908,0.6} $\underset{[0.819]}{   0.91}$ & \cellcolor[rgb]{1,0.61,0.6} $\underset{[>0.999]}{   3.35}$ & \cellcolor[rgb]{1,0.616,0.6} $\underset{[>0.999]}{   3.24}$ & \cellcolor[rgb]{0.888,1,0.6} $\underset{[0.111]}{  -1.22}$ & \cellcolor[rgb]{0.6,0.6,0.6} & \cellcolor[rgb]{1,0.903,0.6} $\underset{[0.836]}{   0.98}$ \\ 
  AR(p)-W & \cellcolor[rgb]{1,0.6,0.6} $\underset{[>0.999]}{   6.01}$ & \cellcolor[rgb]{1,0.6,0.6} $\underset{[>0.999]}{   6.01}$ & \cellcolor[rgb]{1,0.6,0.6} $\underset{[>0.999]}{   6.01}$ & \cellcolor[rgb]{1,0.937,0.6} $\underset{[0.717]}{   0.57}$ & \cellcolor[rgb]{1,0.603,0.6} $\underset{[>0.999]}{   3.72}$ & \cellcolor[rgb]{1,0.605,0.6} $\underset{[>0.999]}{   3.53}$ & \cellcolor[rgb]{0.745,1,0.6} $\underset{[0.010]}{  -2.33}$ & \cellcolor[rgb]{0.903,1,0.6} $\underset{[0.164]}{  -0.98}$ & \cellcolor[rgb]{0.6,0.6,0.6} \\ 
   \hline
\end{tabular}
\caption{DM-test statistics with corresponding p-value given in squared brackets statistics for the Dawid-Sebastiani score (DSS)} 
\label{tab_DM_DSS}
\end{table}

% latex table generated in R 3.4.3 by xtable 1.8-2 package
% Mon Sep 10 15:25:32 2018
\begin{table}[ht]
\centering
\begin{tabular}{rlllllllll}
  \hline
 & \footnotesize AR(12) & \footnotesize AR(13) & \footnotesize AR(p) & \footnotesize AR(12)-M & \footnotesize AR(13)-M & \footnotesize AR(p)-M & \footnotesize AR(12)-W & \footnotesize AR(13)-W & \footnotesize AR(p)-W \\ 
  \hline
AR(12) & \cellcolor[rgb]{0.6,0.6,0.6} & \cellcolor[rgb]{0.635,1,0.6} $\underset{[0.001]}{  -3.00}$ & \cellcolor[rgb]{0.652,1,0.6} $\underset{[0.002]}{  -2.88}$ & \cellcolor[rgb]{0.6,1,0.6} $\underset{[<0.001]}{ -22.52}$ & \cellcolor[rgb]{0.6,1,0.6} $\underset{[<0.001]}{ -14.49}$ & \cellcolor[rgb]{0.6,1,0.6} $\underset{[<0.001]}{ -13.76}$ & \cellcolor[rgb]{0.6,1,0.6} $\underset{[<0.001]}{ -22.37}$ & \cellcolor[rgb]{0.6,1,0.6} $\underset{[<0.001]}{ -14.84}$ & \cellcolor[rgb]{0.6,1,0.6} $\underset{[<0.001]}{ -14.00}$ \\ 
  AR(13) & \cellcolor[rgb]{1,0.635,0.6} $\underset{[0.999]}{   3.00}$ & \cellcolor[rgb]{0.6,0.6,0.6} & \cellcolor[rgb]{1,0.963,0.6} $\underset{[0.627]}{   0.32}$ & \cellcolor[rgb]{0.6,1,0.6} $\underset{[<0.001]}{ -16.04}$ & \cellcolor[rgb]{0.6,1,0.6} $\underset{[<0.001]}{ -19.80}$ & \cellcolor[rgb]{0.6,1,0.6} $\underset{[<0.001]}{ -14.46}$ & \cellcolor[rgb]{0.6,1,0.6} $\underset{[<0.001]}{ -15.24}$ & \cellcolor[rgb]{0.6,1,0.6} $\underset{[<0.001]}{ -20.63}$ & \cellcolor[rgb]{0.6,1,0.6} $\underset{[<0.001]}{ -14.29}$ \\ 
  AR(p) & \cellcolor[rgb]{1,0.652,0.6} $\underset{[0.998]}{   2.88}$ & \cellcolor[rgb]{0.963,1,0.6} $\underset{[0.373]}{  -0.32}$ & \cellcolor[rgb]{0.6,0.6,0.6} & \cellcolor[rgb]{0.6,1,0.6} $\underset{[<0.001]}{ -14.88}$ & \cellcolor[rgb]{0.6,1,0.6} $\underset{[<0.001]}{ -16.89}$ & \cellcolor[rgb]{0.6,1,0.6} $\underset{[<0.001]}{ -18.82}$ & \cellcolor[rgb]{0.6,1,0.6} $\underset{[<0.001]}{ -14.54}$ & \cellcolor[rgb]{0.6,1,0.6} $\underset{[<0.001]}{ -17.23}$ & \cellcolor[rgb]{0.6,1,0.6} $\underset{[<0.001]}{ -19.10}$ \\ 
  AR(12)-M & \cellcolor[rgb]{1,0.6,0.6} $\underset{[>0.999]}{  22.52}$ & \cellcolor[rgb]{1,0.6,0.6} $\underset{[>0.999]}{  16.04}$ & \cellcolor[rgb]{1,0.6,0.6} $\underset{[>0.999]}{  14.88}$ & \cellcolor[rgb]{0.6,0.6,0.6} & \cellcolor[rgb]{0.602,1,0.6} $\underset{[<0.001]}{  -3.75}$ & \cellcolor[rgb]{0.642,1,0.6} $\underset{[0.002]}{  -2.95}$ & \cellcolor[rgb]{0.603,1,0.6} $\underset{[<0.001]}{  -3.69}$ & \cellcolor[rgb]{0.6,1,0.6} $\underset{[<0.001]}{  -4.36}$ & \cellcolor[rgb]{0.608,1,0.6} $\underset{[<0.001]}{  -3.42}$ \\ 
  AR(13)-M & \cellcolor[rgb]{1,0.6,0.6} $\underset{[>0.999]}{  14.49}$ & \cellcolor[rgb]{1,0.6,0.6} $\underset{[>0.999]}{  19.80}$ & \cellcolor[rgb]{1,0.6,0.6} $\underset{[>0.999]}{  16.89}$ & \cellcolor[rgb]{1,0.602,0.6} $\underset{[>0.999]}{   3.75}$ & \cellcolor[rgb]{0.6,0.6,0.6} & \cellcolor[rgb]{1,0.934,0.6} $\underset{[0.727]}{   0.60}$ & \cellcolor[rgb]{1,0.627,0.6} $\underset{[0.999]}{   3.08}$ & \cellcolor[rgb]{0.6,1,0.6} $\underset{[<0.001]}{  -4.17}$ & \cellcolor[rgb]{0.962,1,0.6} $\underset{[0.369]}{  -0.33}$ \\ 
  AR(p)-M & \cellcolor[rgb]{1,0.6,0.6} $\underset{[>0.999]}{  13.76}$ & \cellcolor[rgb]{1,0.6,0.6} $\underset{[>0.999]}{  14.46}$ & \cellcolor[rgb]{1,0.6,0.6} $\underset{[>0.999]}{  18.82}$ & \cellcolor[rgb]{1,0.642,0.6} $\underset{[0.998]}{   2.95}$ & \cellcolor[rgb]{0.934,1,0.6} $\underset{[0.273]}{  -0.60}$ & \cellcolor[rgb]{0.6,0.6,0.6} & \cellcolor[rgb]{1,0.738,0.6} $\underset{[0.993]}{   2.43}$ & \cellcolor[rgb]{0.877,1,0.6} $\underset{[0.075]}{  -1.44}$ & \cellcolor[rgb]{0.609,1,0.6} $\underset{[<0.001]}{  -3.41}$ \\ 
  AR(12)-W & \cellcolor[rgb]{1,0.6,0.6} $\underset{[>0.999]}{  22.37}$ & \cellcolor[rgb]{1,0.6,0.6} $\underset{[>0.999]}{  15.24}$ & \cellcolor[rgb]{1,0.6,0.6} $\underset{[>0.999]}{  14.54}$ & \cellcolor[rgb]{1,0.603,0.6} $\underset{[>0.999]}{   3.69}$ & \cellcolor[rgb]{0.627,1,0.6} $\underset{[0.001]}{  -3.08}$ & \cellcolor[rgb]{0.738,1,0.6} $\underset{[0.007]}{  -2.43}$ & \cellcolor[rgb]{0.6,0.6,0.6} & \cellcolor[rgb]{0.603,1,0.6} $\underset{[<0.001]}{  -3.68}$ & \cellcolor[rgb]{0.642,1,0.6} $\underset{[0.002]}{  -2.95}$ \\ 
  AR(13)-W & \cellcolor[rgb]{1,0.6,0.6} $\underset{[>0.999]}{  14.84}$ & \cellcolor[rgb]{1,0.6,0.6} $\underset{[>0.999]}{  20.63}$ & \cellcolor[rgb]{1,0.6,0.6} $\underset{[>0.999]}{  17.23}$ & \cellcolor[rgb]{1,0.6,0.6} $\underset{[>0.999]}{   4.36}$ & \cellcolor[rgb]{1,0.6,0.6} $\underset{[>0.999]}{   4.17}$ & \cellcolor[rgb]{1,0.877,0.6} $\underset{[0.925]}{   1.44}$ & \cellcolor[rgb]{1,0.603,0.6} $\underset{[>0.999]}{   3.68}$ & \cellcolor[rgb]{0.6,0.6,0.6} & \cellcolor[rgb]{1,0.951,0.6} $\underset{[0.670]}{   0.44}$ \\ 
  AR(p)-W & \cellcolor[rgb]{1,0.6,0.6} $\underset{[>0.999]}{  14.00}$ & \cellcolor[rgb]{1,0.6,0.6} $\underset{[>0.999]}{  14.29}$ & \cellcolor[rgb]{1,0.6,0.6} $\underset{[>0.999]}{  19.10}$ & \cellcolor[rgb]{1,0.608,0.6} $\underset{[>0.999]}{   3.42}$ & \cellcolor[rgb]{1,0.962,0.6} $\underset{[0.631]}{   0.33}$ & \cellcolor[rgb]{1,0.609,0.6} $\underset{[>0.999]}{   3.41}$ & \cellcolor[rgb]{1,0.642,0.6} $\underset{[0.998]}{   2.95}$ & \cellcolor[rgb]{0.951,1,0.6} $\underset{[0.330]}{  -0.44}$ & \cellcolor[rgb]{0.6,0.6,0.6} \\ 
   \hline
\end{tabular}
\caption{DM-test statistics with corresponding p-value given in squared brackets statistics for the CRPS-copula energy score (CRPS-CES)} 
\label{tab_DM_CRPS-CES}
\end{table}

% latex table generated in R 3.4.3 by xtable 1.8-2 package
% Mon Sep 10 15:25:32 2018
\begin{table}[ht]
\centering
\begin{tabular}{rlllllllll}
  \hline
 & \footnotesize AR(12) & \footnotesize AR(13) & \footnotesize AR(p) & \footnotesize AR(12)-M & \footnotesize AR(13)-M & \footnotesize AR(p)-M & \footnotesize AR(12)-W & \footnotesize AR(13)-W & \footnotesize AR(p)-W \\ 
  \hline
AR(12) & \cellcolor[rgb]{0.6,0.6,0.6} & \cellcolor[rgb]{0.777,1,0.6} $\underset{[0.020]}{  -2.05}$ & \cellcolor[rgb]{0.744,1,0.6} $\underset{[0.010]}{  -2.34}$ & \cellcolor[rgb]{0.6,1,0.6} $\underset{[<0.001]}{  -6.28}$ & \cellcolor[rgb]{0.601,1,0.6} $\underset{[<0.001]}{  -4.11}$ & \cellcolor[rgb]{0.6,1,0.6} $\underset{[<0.001]}{  -4.62}$ & \cellcolor[rgb]{0.6,1,0.6} $\underset{[<0.001]}{ -17.10}$ & \cellcolor[rgb]{0.6,1,0.6} $\underset{[<0.001]}{ -11.38}$ & \cellcolor[rgb]{0.6,1,0.6} $\underset{[<0.001]}{ -13.02}$ \\ 
  AR(13) & \cellcolor[rgb]{1,0.777,0.6} $\underset{[0.980]}{   2.05}$ & \cellcolor[rgb]{0.6,0.6,0.6} & \cellcolor[rgb]{1,0.917,0.6} $\underset{[0.786]}{   0.79}$ & \cellcolor[rgb]{0.659,1,0.6} $\underset{[0.002]}{  -2.84}$ & \cellcolor[rgb]{0.6,1,0.6} $\underset{[<0.001]}{  -6.09}$ & \cellcolor[rgb]{0.607,1,0.6} $\underset{[<0.001]}{  -3.45}$ & \cellcolor[rgb]{0.6,1,0.6} $\underset{[<0.001]}{  -5.32}$ & \cellcolor[rgb]{0.6,1,0.6} $\underset{[<0.001]}{ -15.18}$ & \cellcolor[rgb]{0.6,1,0.6} $\underset{[<0.001]}{  -6.46}$ \\ 
  AR(p) & \cellcolor[rgb]{1,0.744,0.6} $\underset{[0.990]}{   2.34}$ & \cellcolor[rgb]{0.917,1,0.6} $\underset{[0.214]}{  -0.79}$ & \cellcolor[rgb]{0.6,0.6,0.6} & \cellcolor[rgb]{0.6,1,0.6} $\underset{[<0.001]}{  -5.08}$ & \cellcolor[rgb]{0.602,1,0.6} $\underset{[<0.001]}{  -3.79}$ & \cellcolor[rgb]{0.6,1,0.6} $\underset{[<0.001]}{  -5.60}$ & \cellcolor[rgb]{0.6,1,0.6} $\underset{[<0.001]}{ -14.82}$ & \cellcolor[rgb]{0.6,1,0.6} $\underset{[<0.001]}{ -11.75}$ & \cellcolor[rgb]{0.6,1,0.6} $\underset{[<0.001]}{ -14.65}$ \\ 
  AR(12)-M & \cellcolor[rgb]{1,0.6,0.6} $\underset{[>0.999]}{   6.28}$ & \cellcolor[rgb]{1,0.659,0.6} $\underset{[0.998]}{   2.84}$ & \cellcolor[rgb]{1,0.6,0.6} $\underset{[>0.999]}{   5.08}$ & \cellcolor[rgb]{0.6,0.6,0.6} & \cellcolor[rgb]{0.871,1,0.6} $\underset{[0.055]}{  -1.60}$ & \cellcolor[rgb]{0.878,1,0.6} $\underset{[0.077]}{  -1.43}$ & \cellcolor[rgb]{0.603,1,0.6} $\underset{[<0.001]}{  -3.69}$ & \cellcolor[rgb]{0.6,1,0.6} $\underset{[<0.001]}{  -5.30}$ & \cellcolor[rgb]{0.6,1,0.6} $\underset{[<0.001]}{  -4.58}$ \\ 
  AR(13)-M & \cellcolor[rgb]{1,0.601,0.6} $\underset{[>0.999]}{   4.11}$ & \cellcolor[rgb]{1,0.6,0.6} $\underset{[>0.999]}{   6.09}$ & \cellcolor[rgb]{1,0.602,0.6} $\underset{[>0.999]}{   3.79}$ & \cellcolor[rgb]{1,0.871,0.6} $\underset{[0.945]}{   1.60}$ & \cellcolor[rgb]{0.6,0.6,0.6} & \cellcolor[rgb]{1,0.935,0.6} $\underset{[0.725]}{   0.60}$ & \cellcolor[rgb]{0.969,1,0.6} $\underset{[0.393]}{  -0.27}$ & \cellcolor[rgb]{0.615,1,0.6} $\underset{[<0.001]}{  -3.24}$ & \cellcolor[rgb]{0.871,1,0.6} $\underset{[0.054]}{  -1.61}$ \\ 
  AR(p)-M & \cellcolor[rgb]{1,0.6,0.6} $\underset{[>0.999]}{   4.62}$ & \cellcolor[rgb]{1,0.607,0.6} $\underset{[>0.999]}{   3.45}$ & \cellcolor[rgb]{1,0.6,0.6} $\underset{[>0.999]}{   5.60}$ & \cellcolor[rgb]{1,0.878,0.6} $\underset{[0.923]}{   1.43}$ & \cellcolor[rgb]{0.935,1,0.6} $\underset{[0.275]}{  -0.60}$ & \cellcolor[rgb]{0.6,0.6,0.6} & \cellcolor[rgb]{0.899,1,0.6} $\underset{[0.149]}{  -1.04}$ & \cellcolor[rgb]{0.601,1,0.6} $\underset{[<0.001]}{  -3.88}$ & \cellcolor[rgb]{0.627,1,0.6} $\underset{[0.001]}{  -3.08}$ \\ 
  AR(12)-W & \cellcolor[rgb]{1,0.6,0.6} $\underset{[>0.999]}{  17.10}$ & \cellcolor[rgb]{1,0.6,0.6} $\underset{[>0.999]}{   5.32}$ & \cellcolor[rgb]{1,0.6,0.6} $\underset{[>0.999]}{  14.82}$ & \cellcolor[rgb]{1,0.603,0.6} $\underset{[>0.999]}{   3.69}$ & \cellcolor[rgb]{1,0.969,0.6} $\underset{[0.607]}{   0.27}$ & \cellcolor[rgb]{1,0.899,0.6} $\underset{[0.851]}{   1.04}$ & \cellcolor[rgb]{0.6,0.6,0.6} & \cellcolor[rgb]{0.604,1,0.6} $\underset{[<0.001]}{  -3.62}$ & \cellcolor[rgb]{0.601,1,0.6} $\underset{[<0.001]}{  -3.89}$ \\ 
  AR(13)-W & \cellcolor[rgb]{1,0.6,0.6} $\underset{[>0.999]}{  11.38}$ & \cellcolor[rgb]{1,0.6,0.6} $\underset{[>0.999]}{  15.18}$ & \cellcolor[rgb]{1,0.6,0.6} $\underset{[>0.999]}{  11.75}$ & \cellcolor[rgb]{1,0.6,0.6} $\underset{[>0.999]}{   5.30}$ & \cellcolor[rgb]{1,0.615,0.6} $\underset{[>0.999]}{   3.24}$ & \cellcolor[rgb]{1,0.601,0.6} $\underset{[>0.999]}{   3.88}$ & \cellcolor[rgb]{1,0.604,0.6} $\underset{[>0.999]}{   3.62}$ & \cellcolor[rgb]{0.6,0.6,0.6} & \cellcolor[rgb]{1,0.903,0.6} $\underset{[0.835]}{   0.97}$ \\ 
  AR(p)-W & \cellcolor[rgb]{1,0.6,0.6} $\underset{[>0.999]}{  13.02}$ & \cellcolor[rgb]{1,0.6,0.6} $\underset{[>0.999]}{   6.46}$ & \cellcolor[rgb]{1,0.6,0.6} $\underset{[>0.999]}{  14.65}$ & \cellcolor[rgb]{1,0.6,0.6} $\underset{[>0.999]}{   4.58}$ & \cellcolor[rgb]{1,0.871,0.6} $\underset{[0.946]}{   1.61}$ & \cellcolor[rgb]{1,0.627,0.6} $\underset{[0.999]}{   3.08}$ & \cellcolor[rgb]{1,0.601,0.6} $\underset{[>0.999]}{   3.89}$ & \cellcolor[rgb]{0.903,1,0.6} $\underset{[0.165]}{  -0.97}$ & \cellcolor[rgb]{0.6,0.6,0.6} \\ 
   \hline
\end{tabular}
\caption{DM-test statistics with corresponding p-value given in squared brackets statistics for the CRPS-copula variogram score (CRPS-CVS)} 
\label{tab_DM_CRPS-CVS}
\end{table}

% latex table generated in R 3.4.3 by xtable 1.8-2 package
% Mon Sep 10 15:25:32 2018
\begin{table}[ht]
\centering
\begin{tabular}{rlllllllll}
  \hline
 & \footnotesize AR(12) & \footnotesize AR(13) & \footnotesize AR(p) & \footnotesize AR(12)-M & \footnotesize AR(13)-M & \footnotesize AR(p)-M & \footnotesize AR(12)-W & \footnotesize AR(13)-W & \footnotesize AR(p)-W \\ 
  \hline
AR(12) & \cellcolor[rgb]{0.6,0.6,0.6} & \cellcolor[rgb]{0.601,1,0.6} $\underset{[<0.001]}{  -4.00}$ & \cellcolor[rgb]{0.636,1,0.6} $\underset{[0.001]}{  -3.00}$ & \cellcolor[rgb]{0.736,1,0.6} $\underset{[0.007]}{  -2.47}$ & \cellcolor[rgb]{0.601,1,0.6} $\underset{[<0.001]}{  -3.98}$ & \cellcolor[rgb]{0.635,1,0.6} $\underset{[0.001]}{  -3.00}$ & \cellcolor[rgb]{0.916,1,0.6} $\underset{[0.208]}{  -0.81}$ & \cellcolor[rgb]{0.601,1,0.6} $\underset{[<0.001]}{  -3.99}$ & \cellcolor[rgb]{0.636,1,0.6} $\underset{[0.001]}{  -2.99}$ \\ 
  AR(13) & \cellcolor[rgb]{1,0.601,0.6} $\underset{[>0.999]}{   4.00}$ & \cellcolor[rgb]{0.6,0.6,0.6} & \cellcolor[rgb]{1,0.934,0.6} $\underset{[0.728]}{   0.61}$ & \cellcolor[rgb]{1,0.601,0.6} $\underset{[>0.999]}{   3.96}$ & \cellcolor[rgb]{1,0.786,0.6} $\underset{[0.977]}{   2.00}$ & \cellcolor[rgb]{1,0.934,0.6} $\underset{[0.729]}{   0.61}$ & \cellcolor[rgb]{1,0.601,0.6} $\underset{[>0.999]}{   3.97}$ & \cellcolor[rgb]{1,0.916,0.6} $\underset{[0.789]}{   0.80}$ & \cellcolor[rgb]{1,0.934,0.6} $\underset{[0.730]}{   0.61}$ \\ 
  AR(p) & \cellcolor[rgb]{1,0.636,0.6} $\underset{[0.999]}{   3.00}$ & \cellcolor[rgb]{0.934,1,0.6} $\underset{[0.272]}{  -0.61}$ & \cellcolor[rgb]{0.6,0.6,0.6} & \cellcolor[rgb]{1,0.64,0.6} $\underset{[0.998]}{   2.96}$ & \cellcolor[rgb]{0.938,1,0.6} $\underset{[0.285]}{  -0.57}$ & \cellcolor[rgb]{1,0.988,0.6} $\underset{[0.543]}{   0.11}$ & \cellcolor[rgb]{1,0.638,0.6} $\underset{[0.999]}{   2.98}$ & \cellcolor[rgb]{0.936,1,0.6} $\underset{[0.279]}{  -0.58}$ & \cellcolor[rgb]{1,0.924,0.6} $\underset{[0.764]}{   0.72}$ \\ 
  AR(12)-M & \cellcolor[rgb]{1,0.736,0.6} $\underset{[0.993]}{   2.47}$ & \cellcolor[rgb]{0.601,1,0.6} $\underset{[<0.001]}{  -3.96}$ & \cellcolor[rgb]{0.64,1,0.6} $\underset{[0.002]}{  -2.96}$ & \cellcolor[rgb]{0.6,0.6,0.6} & \cellcolor[rgb]{0.601,1,0.6} $\underset{[<0.001]}{  -3.95}$ & \cellcolor[rgb]{0.64,1,0.6} $\underset{[0.002]}{  -2.96}$ & \cellcolor[rgb]{1,0.802,0.6} $\underset{[0.972]}{   1.91}$ & \cellcolor[rgb]{0.601,1,0.6} $\underset{[<0.001]}{  -3.95}$ & \cellcolor[rgb]{0.641,1,0.6} $\underset{[0.002]}{  -2.95}$ \\ 
  AR(13)-M & \cellcolor[rgb]{1,0.601,0.6} $\underset{[>0.999]}{   3.98}$ & \cellcolor[rgb]{0.786,1,0.6} $\underset{[0.023]}{  -2.00}$ & \cellcolor[rgb]{1,0.938,0.6} $\underset{[0.715]}{   0.57}$ & \cellcolor[rgb]{1,0.601,0.6} $\underset{[>0.999]}{   3.95}$ & \cellcolor[rgb]{0.6,0.6,0.6} & \cellcolor[rgb]{1,0.938,0.6} $\underset{[0.716]}{   0.57}$ & \cellcolor[rgb]{1,0.601,0.6} $\underset{[>0.999]}{   3.96}$ & \cellcolor[rgb]{0.909,1,0.6} $\underset{[0.186]}{  -0.89}$ & \cellcolor[rgb]{1,0.937,0.6} $\underset{[0.717]}{   0.57}$ \\ 
  AR(p)-M & \cellcolor[rgb]{1,0.635,0.6} $\underset{[0.999]}{   3.00}$ & \cellcolor[rgb]{0.934,1,0.6} $\underset{[0.271]}{  -0.61}$ & \cellcolor[rgb]{0.988,1,0.6} $\underset{[0.457]}{  -0.11}$ & \cellcolor[rgb]{1,0.64,0.6} $\underset{[0.998]}{   2.96}$ & \cellcolor[rgb]{0.938,1,0.6} $\underset{[0.284]}{  -0.57}$ & \cellcolor[rgb]{0.6,0.6,0.6} & \cellcolor[rgb]{1,0.638,0.6} $\underset{[0.999]}{   2.98}$ & \cellcolor[rgb]{0.936,1,0.6} $\underset{[0.278]}{  -0.59}$ & \cellcolor[rgb]{1,0.953,0.6} $\underset{[0.664]}{   0.42}$ \\ 
  AR(12)-W & \cellcolor[rgb]{1,0.916,0.6} $\underset{[0.792]}{   0.81}$ & \cellcolor[rgb]{0.601,1,0.6} $\underset{[<0.001]}{  -3.97}$ & \cellcolor[rgb]{0.638,1,0.6} $\underset{[0.001]}{  -2.98}$ & \cellcolor[rgb]{0.802,1,0.6} $\underset{[0.028]}{  -1.91}$ & \cellcolor[rgb]{0.601,1,0.6} $\underset{[<0.001]}{  -3.96}$ & \cellcolor[rgb]{0.638,1,0.6} $\underset{[0.001]}{  -2.98}$ & \cellcolor[rgb]{0.6,0.6,0.6} & \cellcolor[rgb]{0.601,1,0.6} $\underset{[<0.001]}{  -3.96}$ & \cellcolor[rgb]{0.639,1,0.6} $\underset{[0.001]}{  -2.97}$ \\ 
  AR(13)-W & \cellcolor[rgb]{1,0.601,0.6} $\underset{[>0.999]}{   3.99}$ & \cellcolor[rgb]{0.916,1,0.6} $\underset{[0.211]}{  -0.80}$ & \cellcolor[rgb]{1,0.936,0.6} $\underset{[0.721]}{   0.58}$ & \cellcolor[rgb]{1,0.601,0.6} $\underset{[>0.999]}{   3.95}$ & \cellcolor[rgb]{1,0.909,0.6} $\underset{[0.814]}{   0.89}$ & \cellcolor[rgb]{1,0.936,0.6} $\underset{[0.722]}{   0.59}$ & \cellcolor[rgb]{1,0.601,0.6} $\underset{[>0.999]}{   3.96}$ & \cellcolor[rgb]{0.6,0.6,0.6} & \cellcolor[rgb]{1,0.936,0.6} $\underset{[0.723]}{   0.59}$ \\ 
  AR(p)-W & \cellcolor[rgb]{1,0.636,0.6} $\underset{[0.999]}{   2.99}$ & \cellcolor[rgb]{0.934,1,0.6} $\underset{[0.270]}{  -0.61}$ & \cellcolor[rgb]{0.924,1,0.6} $\underset{[0.236]}{  -0.72}$ & \cellcolor[rgb]{1,0.641,0.6} $\underset{[0.998]}{   2.95}$ & \cellcolor[rgb]{0.937,1,0.6} $\underset{[0.283]}{  -0.57}$ & \cellcolor[rgb]{0.953,1,0.6} $\underset{[0.336]}{  -0.42}$ & \cellcolor[rgb]{1,0.639,0.6} $\underset{[0.999]}{   2.97}$ & \cellcolor[rgb]{0.936,1,0.6} $\underset{[0.277]}{  -0.59}$ & \cellcolor[rgb]{0.6,0.6,0.6} \\ 
   \hline
\end{tabular}
\caption{DM-test statistics with corresponding p-value given in squared brackets statistics for the CRPS} 
\label{tab_DM_CRPS}
\end{table}

% latex table generated in R 3.4.3 by xtable 1.8-2 package
% Mon Sep 10 15:25:32 2018
\begin{table}[ht]
\centering
\begin{tabular}{rlllllllll}
  \hline
 & \footnotesize AR(12) & \footnotesize AR(13) & \footnotesize AR(p) & \footnotesize AR(12)-M & \footnotesize AR(13)-M & \footnotesize AR(p)-M & \footnotesize AR(12)-W & \footnotesize AR(13)-W & \footnotesize AR(p)-W \\ 
  \hline
AR(12) & \cellcolor[rgb]{0.6,0.6,0.6} & \cellcolor[rgb]{0.995,1,0.6} $\underset{[0.482]}{  -0.05}$ & \cellcolor[rgb]{0.99,1,0.6} $\underset{[0.466]}{  -0.08}$ & \cellcolor[rgb]{0.6,1,0.6} $\underset{[<0.001]}{-130.97}$ & \cellcolor[rgb]{0.6,1,0.6} $\underset{[<0.001]}{ -31.51}$ & \cellcolor[rgb]{0.6,1,0.6} $\underset{[<0.001]}{ -31.91}$ & \cellcolor[rgb]{0.6,1,0.6} $\underset{[<0.001]}{-158.85}$ & \cellcolor[rgb]{0.6,1,0.6} $\underset{[<0.001]}{ -28.16}$ & \cellcolor[rgb]{0.6,1,0.6} $\underset{[<0.001]}{ -34.35}$ \\ 
  AR(13) & \cellcolor[rgb]{1,0.995,0.6} $\underset{[0.518]}{   0.05}$ & \cellcolor[rgb]{0.6,0.6,0.6} & \cellcolor[rgb]{0.993,1,0.6} $\underset{[0.474]}{  -0.06}$ & \cellcolor[rgb]{0.6,1,0.6} $\underset{[<0.001]}{ -35.50}$ & \cellcolor[rgb]{0.6,1,0.6} $\underset{[<0.001]}{-121.30}$ & \cellcolor[rgb]{0.6,1,0.6} $\underset{[<0.001]}{ -59.88}$ & \cellcolor[rgb]{0.6,1,0.6} $\underset{[<0.001]}{ -34.93}$ & \cellcolor[rgb]{0.6,1,0.6} $\underset{[<0.001]}{ -99.72}$ & \cellcolor[rgb]{0.6,1,0.6} $\underset{[<0.001]}{ -73.32}$ \\ 
  AR(p) & \cellcolor[rgb]{1,0.99,0.6} $\underset{[0.534]}{   0.08}$ & \cellcolor[rgb]{1,0.993,0.6} $\underset{[0.526]}{   0.06}$ & \cellcolor[rgb]{0.6,0.6,0.6} & \cellcolor[rgb]{0.6,1,0.6} $\underset{[<0.001]}{ -38.59}$ & \cellcolor[rgb]{0.6,1,0.6} $\underset{[<0.001]}{ -61.15}$ & \cellcolor[rgb]{0.6,1,0.6} $\underset{[<0.001]}{ -79.79}$ & \cellcolor[rgb]{0.6,1,0.6} $\underset{[<0.001]}{ -37.71}$ & \cellcolor[rgb]{0.6,1,0.6} $\underset{[<0.001]}{ -49.46}$ & \cellcolor[rgb]{0.6,1,0.6} $\underset{[<0.001]}{-126.75}$ \\ 
  AR(12)-M & \cellcolor[rgb]{1,0.6,0.6} $\underset{[>0.999]}{ 130.97}$ & \cellcolor[rgb]{1,0.6,0.6} $\underset{[>0.999]}{  35.50}$ & \cellcolor[rgb]{1,0.6,0.6} $\underset{[>0.999]}{  38.59}$ & \cellcolor[rgb]{0.6,0.6,0.6} & \cellcolor[rgb]{1,0.951,0.6} $\underset{[0.668]}{   0.44}$ & \cellcolor[rgb]{1,0.933,0.6} $\underset{[0.732]}{   0.62}$ & \cellcolor[rgb]{0.601,1,0.6} $\underset{[<0.001]}{  -4.01}$ & \cellcolor[rgb]{0.907,1,0.6} $\underset{[0.179]}{  -0.92}$ & \cellcolor[rgb]{0.895,1,0.6} $\underset{[0.137]}{  -1.10}$ \\ 
  AR(13)-M & \cellcolor[rgb]{1,0.6,0.6} $\underset{[>0.999]}{  31.51}$ & \cellcolor[rgb]{1,0.6,0.6} $\underset{[>0.999]}{ 121.30}$ & \cellcolor[rgb]{1,0.6,0.6} $\underset{[>0.999]}{  61.15}$ & \cellcolor[rgb]{0.951,1,0.6} $\underset{[0.332]}{  -0.44}$ & \cellcolor[rgb]{0.6,0.6,0.6} & \cellcolor[rgb]{1,0.958,0.6} $\underset{[0.645]}{   0.37}$ & \cellcolor[rgb]{0.87,1,0.6} $\underset{[0.052]}{  -1.63}$ & \cellcolor[rgb]{0.61,1,0.6} $\underset{[<0.001]}{  -3.38}$ & \cellcolor[rgb]{0.711,1,0.6} $\underset{[0.004]}{  -2.63}$ \\ 
  AR(p)-M & \cellcolor[rgb]{1,0.6,0.6} $\underset{[>0.999]}{  31.91}$ & \cellcolor[rgb]{1,0.6,0.6} $\underset{[>0.999]}{  59.88}$ & \cellcolor[rgb]{1,0.6,0.6} $\underset{[>0.999]}{  79.79}$ & \cellcolor[rgb]{0.933,1,0.6} $\underset{[0.268]}{  -0.62}$ & \cellcolor[rgb]{0.958,1,0.6} $\underset{[0.355]}{  -0.37}$ & \cellcolor[rgb]{0.6,0.6,0.6} & \cellcolor[rgb]{0.833,1,0.6} $\underset{[0.038]}{  -1.77}$ & \cellcolor[rgb]{0.753,1,0.6} $\underset{[0.013]}{  -2.24}$ & \cellcolor[rgb]{0.62,1,0.6} $\underset{[<0.001]}{  -3.17}$ \\ 
  AR(12)-W & \cellcolor[rgb]{1,0.6,0.6} $\underset{[>0.999]}{ 158.85}$ & \cellcolor[rgb]{1,0.6,0.6} $\underset{[>0.999]}{  34.93}$ & \cellcolor[rgb]{1,0.6,0.6} $\underset{[>0.999]}{  37.71}$ & \cellcolor[rgb]{1,0.601,0.6} $\underset{[>0.999]}{   4.01}$ & \cellcolor[rgb]{1,0.87,0.6} $\underset{[0.948]}{   1.63}$ & \cellcolor[rgb]{1,0.833,0.6} $\underset{[0.962]}{   1.77}$ & \cellcolor[rgb]{0.6,0.6,0.6} & \cellcolor[rgb]{1,0.98,0.6} $\underset{[0.571]}{   0.18}$ & \cellcolor[rgb]{1,0.977,0.6} $\underset{[0.580]}{   0.20}$ \\ 
  AR(13)-W & \cellcolor[rgb]{1,0.6,0.6} $\underset{[>0.999]}{  28.16}$ & \cellcolor[rgb]{1,0.6,0.6} $\underset{[>0.999]}{  99.72}$ & \cellcolor[rgb]{1,0.6,0.6} $\underset{[>0.999]}{  49.46}$ & \cellcolor[rgb]{1,0.907,0.6} $\underset{[0.821]}{   0.92}$ & \cellcolor[rgb]{1,0.61,0.6} $\underset{[>0.999]}{   3.38}$ & \cellcolor[rgb]{1,0.753,0.6} $\underset{[0.987]}{   2.24}$ & \cellcolor[rgb]{0.98,1,0.6} $\underset{[0.429]}{  -0.18}$ & \cellcolor[rgb]{0.6,0.6,0.6} & \cellcolor[rgb]{0.997,1,0.6} $\underset{[0.491]}{  -0.02}$ \\ 
  AR(p)-W & \cellcolor[rgb]{1,0.6,0.6} $\underset{[>0.999]}{  34.35}$ & \cellcolor[rgb]{1,0.6,0.6} $\underset{[>0.999]}{  73.32}$ & \cellcolor[rgb]{1,0.6,0.6} $\underset{[>0.999]}{ 126.75}$ & \cellcolor[rgb]{1,0.895,0.6} $\underset{[0.863]}{   1.10}$ & \cellcolor[rgb]{1,0.711,0.6} $\underset{[0.996]}{   2.63}$ & \cellcolor[rgb]{1,0.62,0.6} $\underset{[>0.999]}{   3.17}$ & \cellcolor[rgb]{0.977,1,0.6} $\underset{[0.420]}{  -0.20}$ & \cellcolor[rgb]{1,0.997,0.6} $\underset{[0.509]}{   0.02}$ & \cellcolor[rgb]{0.6,0.6,0.6} \\ 
   \hline
\end{tabular}
\caption{DM-test statistics with corresponding p-value given in squared brackets statistics for the copula energy score (CES)} 
\label{tab_DM_CES}
\end{table}

% latex table generated in R 3.4.3 by xtable 1.8-2 package
% Mon Sep 10 15:25:32 2018
\begin{table}[ht]
\centering
\begin{tabular}{rlllllllll}
  \hline
 & \footnotesize AR(12) & \footnotesize AR(13) & \footnotesize AR(p) & \footnotesize AR(12)-M & \footnotesize AR(13)-M & \footnotesize AR(p)-M & \footnotesize AR(12)-W & \footnotesize AR(13)-W & \footnotesize AR(p)-W \\ 
  \hline
AR(12) & \cellcolor[rgb]{0.6,0.6,0.6} & \cellcolor[rgb]{0.9,1,0.6} $\underset{[0.152]}{  -1.03}$ & \cellcolor[rgb]{0.949,1,0.6} $\underset{[0.322]}{  -0.46}$ & \cellcolor[rgb]{0.6,1,0.6} $\underset{[<0.001]}{  -7.32}$ & \cellcolor[rgb]{0.602,1,0.6} $\underset{[<0.001]}{  -3.81}$ & \cellcolor[rgb]{0.604,1,0.6} $\underset{[<0.001]}{  -3.62}$ & \cellcolor[rgb]{0.6,1,0.6} $\underset{[<0.001]}{ -45.83}$ & \cellcolor[rgb]{0.6,1,0.6} $\underset{[<0.001]}{ -13.77}$ & \cellcolor[rgb]{0.6,1,0.6} $\underset{[<0.001]}{ -14.87}$ \\ 
  AR(13) & \cellcolor[rgb]{1,0.9,0.6} $\underset{[0.848]}{   1.03}$ & \cellcolor[rgb]{0.6,0.6,0.6} & \cellcolor[rgb]{1,0.919,0.6} $\underset{[0.781]}{   0.77}$ & \cellcolor[rgb]{0.6,1,0.6} $\underset{[<0.001]}{  -4.70}$ & \cellcolor[rgb]{0.6,1,0.6} $\underset{[<0.001]}{  -6.63}$ & \cellcolor[rgb]{0.602,1,0.6} $\underset{[<0.001]}{  -3.75}$ & \cellcolor[rgb]{0.6,1,0.6} $\underset{[<0.001]}{  -8.51}$ & \cellcolor[rgb]{0.6,1,0.6} $\underset{[<0.001]}{ -21.12}$ & \cellcolor[rgb]{0.6,1,0.6} $\underset{[<0.001]}{  -8.83}$ \\ 
  AR(p) & \cellcolor[rgb]{1,0.949,0.6} $\underset{[0.678]}{   0.46}$ & \cellcolor[rgb]{0.919,1,0.6} $\underset{[0.219]}{  -0.77}$ & \cellcolor[rgb]{0.6,0.6,0.6} & \cellcolor[rgb]{0.6,1,0.6} $\underset{[<0.001]}{  -5.96}$ & \cellcolor[rgb]{0.6,1,0.6} $\underset{[<0.001]}{  -4.29}$ & \cellcolor[rgb]{0.6,1,0.6} $\underset{[<0.001]}{  -5.71}$ & \cellcolor[rgb]{0.6,1,0.6} $\underset{[<0.001]}{ -13.38}$ & \cellcolor[rgb]{0.6,1,0.6} $\underset{[<0.001]}{ -14.13}$ & \cellcolor[rgb]{0.6,1,0.6} $\underset{[<0.001]}{ -32.64}$ \\ 
  AR(12)-M & \cellcolor[rgb]{1,0.6,0.6} $\underset{[>0.999]}{   7.32}$ & \cellcolor[rgb]{1,0.6,0.6} $\underset{[>0.999]}{   4.70}$ & \cellcolor[rgb]{1,0.6,0.6} $\underset{[>0.999]}{   5.96}$ & \cellcolor[rgb]{0.6,0.6,0.6} & \cellcolor[rgb]{0.985,1,0.6} $\underset{[0.449]}{  -0.13}$ & \cellcolor[rgb]{1,0.961,0.6} $\underset{[0.633]}{   0.34}$ & \cellcolor[rgb]{0.603,1,0.6} $\underset{[<0.001]}{  -3.72}$ & \cellcolor[rgb]{0.602,1,0.6} $\underset{[<0.001]}{  -3.82}$ & \cellcolor[rgb]{0.612,1,0.6} $\underset{[<0.001]}{  -3.32}$ \\ 
  AR(13)-M & \cellcolor[rgb]{1,0.602,0.6} $\underset{[>0.999]}{   3.81}$ & \cellcolor[rgb]{1,0.6,0.6} $\underset{[>0.999]}{   6.63}$ & \cellcolor[rgb]{1,0.6,0.6} $\underset{[>0.999]}{   4.29}$ & \cellcolor[rgb]{1,0.985,0.6} $\underset{[0.551]}{   0.13}$ & \cellcolor[rgb]{0.6,0.6,0.6} & \cellcolor[rgb]{1,0.942,0.6} $\underset{[0.702]}{   0.53}$ & \cellcolor[rgb]{0.798,1,0.6} $\underset{[0.027]}{  -1.93}$ & \cellcolor[rgb]{0.612,1,0.6} $\underset{[<0.001]}{  -3.32}$ & \cellcolor[rgb]{0.771,1,0.6} $\underset{[0.018]}{  -2.09}$ \\ 
  AR(p)-M & \cellcolor[rgb]{1,0.604,0.6} $\underset{[>0.999]}{   3.62}$ & \cellcolor[rgb]{1,0.602,0.6} $\underset{[>0.999]}{   3.75}$ & \cellcolor[rgb]{1,0.6,0.6} $\underset{[>0.999]}{   5.71}$ & \cellcolor[rgb]{0.961,1,0.6} $\underset{[0.367]}{  -0.34}$ & \cellcolor[rgb]{0.942,1,0.6} $\underset{[0.298]}{  -0.53}$ & \cellcolor[rgb]{0.6,0.6,0.6} & \cellcolor[rgb]{0.734,1,0.6} $\underset{[0.006]}{  -2.50}$ & \cellcolor[rgb]{0.609,1,0.6} $\underset{[<0.001]}{  -3.39}$ & \cellcolor[rgb]{0.613,1,0.6} $\underset{[<0.001]}{  -3.29}$ \\ 
  AR(12)-W & \cellcolor[rgb]{1,0.6,0.6} $\underset{[>0.999]}{  45.83}$ & \cellcolor[rgb]{1,0.6,0.6} $\underset{[>0.999]}{   8.51}$ & \cellcolor[rgb]{1,0.6,0.6} $\underset{[>0.999]}{  13.38}$ & \cellcolor[rgb]{1,0.603,0.6} $\underset{[>0.999]}{   3.72}$ & \cellcolor[rgb]{1,0.798,0.6} $\underset{[0.973]}{   1.93}$ & \cellcolor[rgb]{1,0.734,0.6} $\underset{[0.994]}{   2.50}$ & \cellcolor[rgb]{0.6,0.6,0.6} & \cellcolor[rgb]{0.904,1,0.6} $\underset{[0.169]}{  -0.96}$ & \cellcolor[rgb]{0.986,1,0.6} $\underset{[0.451]}{  -0.12}$ \\ 
  AR(13)-W & \cellcolor[rgb]{1,0.6,0.6} $\underset{[>0.999]}{  13.77}$ & \cellcolor[rgb]{1,0.6,0.6} $\underset{[>0.999]}{  21.12}$ & \cellcolor[rgb]{1,0.6,0.6} $\underset{[>0.999]}{  14.13}$ & \cellcolor[rgb]{1,0.602,0.6} $\underset{[>0.999]}{   3.82}$ & \cellcolor[rgb]{1,0.612,0.6} $\underset{[>0.999]}{   3.32}$ & \cellcolor[rgb]{1,0.609,0.6} $\underset{[>0.999]}{   3.39}$ & \cellcolor[rgb]{1,0.904,0.6} $\underset{[0.831]}{   0.96}$ & \cellcolor[rgb]{0.6,0.6,0.6} & \cellcolor[rgb]{1,0.911,0.6} $\underset{[0.808]}{   0.87}$ \\ 
  AR(p)-W & \cellcolor[rgb]{1,0.6,0.6} $\underset{[>0.999]}{  14.87}$ & \cellcolor[rgb]{1,0.6,0.6} $\underset{[>0.999]}{   8.83}$ & \cellcolor[rgb]{1,0.6,0.6} $\underset{[>0.999]}{  32.64}$ & \cellcolor[rgb]{1,0.612,0.6} $\underset{[>0.999]}{   3.32}$ & \cellcolor[rgb]{1,0.771,0.6} $\underset{[0.982]}{   2.09}$ & \cellcolor[rgb]{1,0.613,0.6} $\underset{[>0.999]}{   3.29}$ & \cellcolor[rgb]{1,0.986,0.6} $\underset{[0.549]}{   0.12}$ & \cellcolor[rgb]{0.911,1,0.6} $\underset{[0.192]}{  -0.87}$ & \cellcolor[rgb]{0.6,0.6,0.6} \\ 
   \hline
\end{tabular}
\caption{DM-test statistics with corresponding p-value given in squared brackets statistics for copula variogram score (CVS)} 
\label{tab_DM_CVS}
\end{table}

% latex table generated in R 3.4.3 by xtable 1.8-2 package
% Mon Sep 10 15:25:32 2018
\begin{table}[ht]
\centering
\begin{tabular}{rlllllllll}
  \hline
 & \footnotesize AR(12) & \footnotesize AR(13) & \footnotesize AR(p) & \footnotesize AR(12)-M & \footnotesize AR(13)-M & \footnotesize AR(p)-M & \footnotesize AR(12)-W & \footnotesize AR(13)-W & \footnotesize AR(p)-W \\ 
  \hline
AR(12) & \cellcolor[rgb]{0.6,0.6,0.6} & \cellcolor[rgb]{0.901,1,0.6} $\underset{[0.157]}{  -1.01}$ & \cellcolor[rgb]{0.889,1,0.6} $\underset{[0.115]}{  -1.20}$ & \cellcolor[rgb]{0.6,0.6,0.6} & \cellcolor[rgb]{0.6,0.6,0.6} & \cellcolor[rgb]{0.6,0.6,0.6} & \cellcolor[rgb]{0.6,0.6,0.6} & \cellcolor[rgb]{0.6,0.6,0.6} & \cellcolor[rgb]{0.6,0.6,0.6} \\ 
  AR(13) & \cellcolor[rgb]{1,0.901,0.6} $\underset{[0.843]}{   1.01}$ & \cellcolor[rgb]{0.6,0.6,0.6} & \cellcolor[rgb]{0.934,1,0.6} $\underset{[0.270]}{  -0.61}$ & \cellcolor[rgb]{0.6,0.6,0.6} & \cellcolor[rgb]{0.6,0.6,0.6} & \cellcolor[rgb]{0.6,0.6,0.6} & \cellcolor[rgb]{0.6,0.6,0.6} & \cellcolor[rgb]{0.6,0.6,0.6} & \cellcolor[rgb]{0.6,0.6,0.6} \\ 
  AR(p) & \cellcolor[rgb]{1,0.889,0.6} $\underset{[0.885]}{   1.20}$ & \cellcolor[rgb]{1,0.934,0.6} $\underset{[0.730]}{   0.61}$ & \cellcolor[rgb]{0.6,0.6,0.6} & \cellcolor[rgb]{0.6,0.6,0.6} & \cellcolor[rgb]{0.6,0.6,0.6} & \cellcolor[rgb]{0.6,0.6,0.6} & \cellcolor[rgb]{0.6,0.6,0.6} & \cellcolor[rgb]{0.6,0.6,0.6} & \cellcolor[rgb]{0.6,0.6,0.6} \\ 
  AR(12)-M & \cellcolor[rgb]{0.6,0.6,0.6} & \cellcolor[rgb]{0.6,0.6,0.6} & \cellcolor[rgb]{0.6,0.6,0.6} & \cellcolor[rgb]{0.6,0.6,0.6} & \cellcolor[rgb]{0.6,0.6,0.6} & \cellcolor[rgb]{0.6,0.6,0.6} & \cellcolor[rgb]{0.6,0.6,0.6} & \cellcolor[rgb]{0.6,0.6,0.6} & \cellcolor[rgb]{0.6,0.6,0.6} \\ 
  AR(13)-M & \cellcolor[rgb]{0.6,0.6,0.6} & \cellcolor[rgb]{0.6,0.6,0.6} & \cellcolor[rgb]{0.6,0.6,0.6} & \cellcolor[rgb]{0.6,0.6,0.6} & \cellcolor[rgb]{0.6,0.6,0.6} & \cellcolor[rgb]{0.6,0.6,0.6} & \cellcolor[rgb]{0.6,0.6,0.6} & \cellcolor[rgb]{0.6,0.6,0.6} & \cellcolor[rgb]{0.6,0.6,0.6} \\ 
  AR(p)-M & \cellcolor[rgb]{0.6,0.6,0.6} & \cellcolor[rgb]{0.6,0.6,0.6} & \cellcolor[rgb]{0.6,0.6,0.6} & \cellcolor[rgb]{0.6,0.6,0.6} & \cellcolor[rgb]{0.6,0.6,0.6} & \cellcolor[rgb]{0.6,0.6,0.6} & \cellcolor[rgb]{0.6,0.6,0.6} & \cellcolor[rgb]{0.6,0.6,0.6} & \cellcolor[rgb]{0.6,0.6,0.6} \\ 
  AR(12)-W & \cellcolor[rgb]{0.6,0.6,0.6} & \cellcolor[rgb]{0.6,0.6,0.6} & \cellcolor[rgb]{0.6,0.6,0.6} & \cellcolor[rgb]{0.6,0.6,0.6} & \cellcolor[rgb]{0.6,0.6,0.6} & \cellcolor[rgb]{0.6,0.6,0.6} & \cellcolor[rgb]{0.6,0.6,0.6} & \cellcolor[rgb]{0.6,0.6,0.6} & \cellcolor[rgb]{0.6,0.6,0.6} \\ 
  AR(13)-W & \cellcolor[rgb]{0.6,0.6,0.6} & \cellcolor[rgb]{0.6,0.6,0.6} & \cellcolor[rgb]{0.6,0.6,0.6} & \cellcolor[rgb]{0.6,0.6,0.6} & \cellcolor[rgb]{0.6,0.6,0.6} & \cellcolor[rgb]{0.6,0.6,0.6} & \cellcolor[rgb]{0.6,0.6,0.6} & \cellcolor[rgb]{0.6,0.6,0.6} & \cellcolor[rgb]{0.6,0.6,0.6} \\ 
  AR(p)-W & \cellcolor[rgb]{0.6,0.6,0.6} & \cellcolor[rgb]{0.6,0.6,0.6} & \cellcolor[rgb]{0.6,0.6,0.6} & \cellcolor[rgb]{0.6,0.6,0.6} & \cellcolor[rgb]{0.6,0.6,0.6} & \cellcolor[rgb]{0.6,0.6,0.6} & \cellcolor[rgb]{0.6,0.6,0.6} & \cellcolor[rgb]{0.6,0.6,0.6} & \cellcolor[rgb]{0.6,0.6,0.6} \\ 
   \hline
\end{tabular}
\caption{DM-test statistics with corresponding p-value given in squared brackets statistics for copula Dawid-Sebastiani score (CDSS)} 
\label{tab_DM_CDSS}
\end{table}

 \clearpage
\bibliographystyle{apalike}
\bibliography{eval}

\end{document}